\documentclass[nonblindrev,copyedit]{informs1} % for review, nonblinded

\RequirePackage[OT1]{fontenc}
\RequirePackage{amsmath}
\allowdisplaybreaks[4]
\usepackage{amssymb}
\usepackage{bbm}
\usepackage{bm}
\usepackage{float}
\usepackage{graphicx}
\usepackage[dvipsnames]{xcolor}
 \usepackage{tikz}
\usetikzlibrary{arrows.meta}
\usetikzlibrary{decorations.pathreplacing}
\usepackage{hyperref}[]
\hypersetup{
	colorlinks=true,
	linkcolor=blue,
	filecolor=magenta,      
	urlcolor=cyan,
	citecolor=blue,
}
\usepackage{cleveref}
\usepackage{subcaption}
\usepackage[dvipsnames]{xcolor}
\usepackage{booktabs}
\usepackage{array}
\newcolumntype{Z}{>{\setbox0=\hbox\bgroup}c<{\egroup}@{\hspace*{-\tabcolsep}}}
\usepackage{xr}
\usepackage{setspace}
\usepackage{diagbox}

\usepackage{natbib}
 \bibpunct[, ]{(}{)}{,}{a}{}{,}%
 \def\bibfont{\small}%
 \def\bibsep{\smallskipamount}%

\theoremstyle{TH}%
\newtheorem{thm}{Theorem}
\newtheorem{lemma}{Lemma}
\newtheorem{pro}{Proposition}
\newtheorem{cor}{Corollary}
\newtheorem{ass}{Assumption}
\newenvironment{thmbis}[1]
{\renewcommand{\thelemma}{\ref{#1}$^*$}%
	\addtocounter{lemma}{-1}%
	\begin{lemma}}
	{\end{lemma}}

\usepackage{algorithm}% http://ctan.org/pkg/algorithms
\usepackage{algpseudocode}% http://ctan.org/pkg/algorithmicx
\algnewcommand\algorithmicinput{\textbf{INPUT:}}
\algnewcommand\INPUT{\item[\algorithmicinput]}
\algnewcommand\algorithmicoutput{\textbf{OUTPUT:}}
\algnewcommand\OUTPUT{\item[\algorithmicoutput]}

\usepackage{algorithm,algpseudocode}% http://ctan.org/pkg/{algorithms,algorithmx}
\algnewcommand{\Inputs}[1]{%
  \State \textbf{Input:}
  \Statex \hspace*{\algorithmicindent}\parbox[t]{.8\linewidth}{\raggedright #1}
}
\algnewcommand{\Initialize}[1]{%
  \State \textbf{Initialization:}
  \Statex \hspace*{\algorithmicindent}\parbox[t]{.8\linewidth}{\raggedright #1}
}
\algnewcommand{\Outputs}[1]{%
  \State \textbf{Output:} #1
}

\algdef{SE}[SUBALG]{Indent}{EndIndent}{}{\algorithmicend\ }%
\algtext*{Indent}
\algtext*{EndIndent}
\usepackage{xurl}

\begin{document}

\TITLE{\Large High-Dimensional Dynamic Pricing under Non-Stationarity: Learning and Earning with Change-Point Detection}

\ARTICLEAUTHORS{

\AUTHOR{Zifeng Zhao}
\AFF{Mendoza College of Business, University of Notre Dame, zifeng.zhao@nd.edu}

\AUTHOR{Feiyu Jiang}
\AFF{School of Management, Fudan University, jiangfy@fudan.edu.cn}

\AUTHOR{Yi Yu}
\AFF{Department of Statistics, University of Warwick, yi.yu.2@warwick.ac.uk}

\AUTHOR{Xi Chen}
\AFF{Stern School of Business, New York University, xc13@stern.nyu.edu}

}

\ABSTRACT{
We consider a high-dimensional dynamic pricing problem under non-stationarity, where a firm sells products to $T$ sequentially arriving consumers that behave according to an unknown demand model with potential changes at unknown times. The demand model is assumed to be a high-dimensional generalized linear model~(GLM), allowing for a feature vector in $\mathbb R^d$ that encodes products and consumer information. To achieve optimal revenue~(i.e.,\ least regret), the firm needs to learn and exploit the unknown GLMs while monitoring for potential change-points.  To tackle such a problem, we first design a novel penalized likelihood-based online change-point detection algorithm for high-dimensional GLMs, which is the first algorithm in the change-point literature that achieves optimal minimax localization error rate for high-dimensional GLMs.  A change-point detection assisted dynamic pricing (CPDP) policy is further proposed and achieves a near-optimal regret of order $O(s\sqrt{\Upsilon_T T}\log(Td))$, where $s$ is the sparsity level and $\Upsilon_T$ is the number of change-points. This regret is accompanied with a minimax lower bound, demonstrating the optimality of CPDP (up to logarithmic factors). In particular, the optimality with respect to $\Upsilon_T$ is seen for the first time in the dynamic pricing literature, and is achieved via a novel accelerated exploration mechanism. Extensive simulation experiments and a real data application on online lending illustrate the efficiency of the proposed policy and the importance and practical value of handling non-stationarity in dynamic pricing.
}


\maketitle

\setlength{\abovedisplayskip}{6pt}
\setlength{\belowdisplayskip}{6pt}
\setlength{\abovedisplayshortskip}{4pt}
\setlength{\belowdisplayshortskip}{4pt}

\vspace{-0.5cm}
\section{Introduction}\label{sec-intro}

With the technological advances and prevalence of online marketplaces, many firms can now dynamically make pricing decisions while having access to an abundance of contextual information such as consumer characteristics, product features and economic environment. On the other hand, in practice, the demand model is unknown and firms need to dynamically learn how the contextual information impacts consumer demand. Thus, to maximize its revenue, the firm needs to implement dynamic pricing, which aims to optimally balance the trade-off between learning the unknown demand function and earning revenues by exploiting the estimated demand model. 

Due to its importance in revenue management, dynamic pricing has been extensively studied in the literature under various settings. The majority of studies on dynamic pricing, however, focus on the case where the demand model is unknown but \textit{stationary}. In other words, it assumes that the way consumers react to prices, product features and economic environment remains unchanged over time. While this can be a reasonable assumption over a short period of time, empirical evidence suggests that consumer behavior \textit{changes} over time due to various reasons. For example, unforeseen external shocks such as financial crisis and the Covid-19 pandemic can greatly impact the demand of certain products, such as personal care and health products, home improvement, fitness equipment and electronics~\citep[see e.g.,][]{bauer2020semiconductor, Tarlton2020, Whitten2020}. 

As a concrete example, we examine a dataset popular in the dynamic pricing literature, which consists of consumer-level transaction data collected by an online auto loan lender. \cite{besbes2011minimax} manually partition the dataset into two 6-month periods. They find that the estimated logistic regression models on the relationship between consumer decisions (to accept or reject) and prices (interest rate of the loan) differ significantly between the two periods~(see Figure 1 therein). We formalize their analysis by running an offline change-point estimation procedure and identify three change-points. We document the estimated logistic regression models on each detected segment in \Cref{tab:loan}~(\Cref{subsec:autoloan}), which shows that the demand function exhibits notable changes over time and provides further evidence of the non-stationarity in consumer behavior. %For more intuition, we plot the price sensitivity estimated via a moving window of size 10000 in Figure \textcolor{red}{1} \textcolor{red}{(where is the pic?)} along with the detected change-points, 

Motivated by these observations, in this paper we study a high-dimensional dynamic pricing problem under non-stationarity. To be specific, we consider a firm selling products to $T$ sequentially arriving consumers, where the $t$th consumer arrives in time period $t$, $t \in \{1, \ldots, T\}$. For each time period $t$, the contextual information is featurized by a vector $z_t\in \mathbb R^{d-1}$, $d \geq 1$.  Conditional on the covariate $z_t$ and the price $p_t$, the consumer demand $y_t$ follows a generalized linear model~(GLM) with parameter $\theta_t\in \mathbb R^d.$ We allow the dimension $d$ to diverge as $T$ grows unbounded but assume the consumer demand is only impacted by $s\leq d$ unknown covariates. Thus, $s$ is the sparsity level. Due to non-stationarity, the demand model, i.e.,\ the parameter sequence $\{\theta_t\}_{t=1}^T$, may change abruptly at unknown $\Upsilon_T$ number of change-points. The firm initially has no information about the parameter $\{\theta_t\}_{t=1}^T$ nor the number or locations of change-points. We aim to design a pricing policy that achieves near-optimal revenue performance, measured by the firm’s $T$-period expected regret. That is, the revenue loss compared to a clairvoyant who has perfect knowledge of the underlying demand model.

\subsection{List of Contributions}
This paper makes the following main contributions:

\begin{itemize}
    \item We design a novel \textit{change-point detection assisted dynamic pricing}~(CPDP) algorithm, detailed in \Cref{algorithm:cpd2}, which achieves a near-optimal regret of order $O(s\sqrt{\Upsilon_T T}\log(Td))$ as shown in \Cref{thm:regret}. To our best knowledge, CPDP is the first algorithm in the dynamic pricing literature that can handle non-stationary demand models with covariates. In addition, unlike existing algorithms for dynamic pricing under non-stationarity, CPDP does not require the knowledge of the minimal change size and further allows vanishing change sizes. Similar to existing works, CPDP runs in cycles that consists of both exploration and exploitation components. However, we carefully \textit{accelerate} the exploration to exploitation ratio upon the detection of every change-point. Thanks to this novel accelerated exploration mechanism, the regret of CPDP scales optimally with the number of change-points $\Upsilon_T$, while existing algorithms in the literature scale linearly. Furthermore, under mild conditions, we derive a new high-probability error bound for the Lasso estimator~\citep{tibshirani1996regression} under a mixture of two high-dimensional GLMs. Based on this, we show somewhat surprisingly, the regret of CPDP can be independent of the change size as stated in \Cref{cor_regret}, which is seen for the first time in the literature. 
    
    \item As a key ingredient as well as an important byproduct, we propose a novel penalized likelihood-based online change-point detection algorithm for high-dimensional GLMs, detailed in \Cref{algorithm:cp} with theoretical guarantees in \Cref{prop:cp}. The high-dimensionality and non-linearity of the GLM pose great challenges to change-point detection. In particular, due to loss of linearity, the pseudo-true model parameter of a non-stationary high-dimensional GLM process may not be (approximately) sparse. Thus, neither the method nor theory of existing works for change-point detection in high-dimensional linear models can be borrowed. To overcome these difficulties, we propose a change-point detection procedure equipped with a newly designed $\ell_1$-difference penalty and further develop new technical arguments to establish its optimality. Together, it provides the first algorithm in the change-point literature that can achieve an optimal minimax localization error rate for high-dimensional GLMs. 
    
    \item We show that, for dynamic pricing under non-stationarity, the regret lower bound is $\Omega(\sqrt{\Upsilon_T T})$, where $T$ is the total number of consumers and $\Upsilon_T$ is the number of change-points. In particular, the dependence on $\Upsilon_T$ is seen for the first time in the dynamic pricing literature. To reveal the problem difficulty in terms of $\Upsilon_T$, new technical arguments are developed in the proof, which is of independent interest, as we need to handle the case of multiple change-points with $\Upsilon_T$ allowed to diverge as $T$ grows unbounded. This result is collected in \Cref{sec-cpdp-lower}.
\end{itemize}

The theoretical findings are further supported with extensive numerical studies on both synthetic and real data in \Cref{sec:num}, illustrating the importance of handling non-stationarity in dynamic pricing.  Our analysis of a U.S.\ online auto loan dataset reveals the key managerial insight that significant revenue loss can occur if we ignore the potential non-stationarity and simply rely on a dynamic pricing algorithm designed for a stationary demand model.

The rest of this paper is organized as follows. \Cref{subsec:lit} provides a detailed literature review and further highlights our contributions and differences from existing works. We formulate the problem rigorously in \Cref{sec:problem_formulation} and propose the CPDP algorithm in \Cref{sec:alg}. Theoretical results including upper and lower bounds on the regret, along with a sketch of proofs, are presented in \Cref{sec:theory}. Numerical studies on both synthetic and real data are given in \Cref{sec:num} to illustrate the efficiency of CPDP. We conclude with discussions in \Cref{sec:conclusion}.  All technical arguments and additional simulation studies can be found in the Appendix. %, which also includes the description of the subroutine of change-point detection.  

\subsection{Related Literature}\label{subsec:lit}

%In view of the high-dimensional dynamic pricing under non-stationarity problem considered in our paper, we identify three closely-related research areas, elaborated below.  

Three streams of literature are closely related to our paper: dynamic pricing with demand learning, statistical change-point detection, and the switching bandit problem.

\noindent\textbf{Dynamic pricing with demand learning.} There is a growing body of literature on dynamic pricing under various settings, where a firm aims to maximize revenue by balancing the trade-off between learning and earning~(i.e.,\ exploration and exploitation) when faced with an unknown (but stationary) demand function. See, for example, \cite{araman2009dynamic}, \cite{besbes2009dynamic}, \cite{farias2010dynamic}, \cite{broder2012dynamic}, \cite{harrison2012bayesian}, \cite{keskin2014dynamic}, \cite{den2015dynamic},  \cite{cheung2017dynamic}, \cite{chen2019welfare,chen2021nonparametric}, \cite{nambiar2019dynamic}, \cite{wang2021multimodal}, \cite{chen2022differential},  \cite{bastani2022meta} and \cite{jia2022online}. More recently, \cite{javanmard2019dynamic} and \cite{ban2021personalized} design near-optimal dynamic pricing policies for demand functions with high-dimensional contextual features (i.e.,\ covariates). In particular, \cite{ban2021personalized} assume the stationary demand follows a high-dimensional GLM and show that the regret lower bound is $\Omega(s\sqrt{T})$, where recall $s$ is the sparsity level.
%, where $T$ is the number of consumers and .

While the majority of studies on dynamic pricing focus on the case where the demand model stays unchanged across time, there are also works tackling non-stationarity. \cite{besbes2011minimax} study a setting where the demand function may change once in an unknown time period, while assuming that the firm has the knowledge of the pre- and post-change demand functions. \cite{keskin2017chasing} study a setting where the demand is a linear function of price and the model parameter may be time-varying.  Using a notion of ``budget", which measures the total variation of parameter change, \cite{keskin2017chasing} study two scenarios depending on whether the parameter change is smooth or abrupt, and show that the problem complexity is different for the two cases. \cite{keskin2022data} consider a setting where the seller needs to make joint pricing and inventory ordering decisions under possible changes in a demand function with a given non-linear form. \cite{chen2019dynamic} investigate the setting where the firm needs to make pricing decisions among a finite number of prices for a single product on multiple local markets with evolving market sizes and possible changes in the demand function. \cite{chen2019dynamic} consider both zeroth-order change (abrupt change in demand) and first-order change (abrupt change in slope of demand) and design novel algorithms to solve both cases simultaneously.

All the aforementioned works are concerned with dynamic pricing in a  covariate-free  setting, where the demand is a function of only price without other contextual information. This allows one to handle the non-stationarity, to be specific, to design a change-point detection component in a relatively simple manner. In such case, the change-point detection component typically boils down to comparing average demand at two or three pre-specified price points.  In addition, the aforementioned works assume the knowledge of a minimal change size. The technicality for analyzing change-point detection is, therefore, to control the deviation of a sub-Gaussian or sub-Exponential sum with respect to its mean.  Our setting, in contrast, is much more difficult in terms of change-point detection, as we allow the demand function to be a GLM with high-dimensional covariates, without assuming a minimal change size.  The high-dimensionality also poses challenges for the analysis of regret on each stationary segment. The technical arguments needed are fundamentally different as we need to analyze Lasso estimators~\citep{tibshirani1996regression} sequentially obtained based on mixtures of high-dimensional GLMs. In addition, thanks to a novel accelerated exploration mechanism, our regret scales sublinearly instead of linearly with the number of change-points.
\medskip

\noindent\textbf{Statistical change-point detection.}  Change-point analysis is an active and important area in statistics, and is witnessing a renaissance in recent years. Due to the dynamic nature of our problem, our work relies on an online change-point detection component, where one aims to flag changes in data that come sequentially, with minimum delay and a control on false alarms. Existing works have studied online change-point detection in univariate or multivariate observations \citep[e.g.,][]{siegmund1985sequential, brodsky1993nonparametric, lai1995sequential, lai1998information, chen2012parametric, tartakovsky2014sequential, maillard2019sequential, yu2020note}, as well as high-dimensional sequences \citep[e.g.,][]{keshavarz2020sequential}, vector autoregressive models \citep[e.g.,][]{safikhani2022joint}, dynamic networks \citep[e.g.,][]{dubey2021online} and more general structures \citep[e.g.,][]{he2018sequential, chen2019sequential}, among many others. However, we are the first to study the problem of online change-point detection for a sequence of high-dimensional GLMs, accompanied with its theoretical guarantees. This is in fact an important ingredient and also a significant contribution of this paper.

%In this paper, we are concerned with monitoring changes in a sequence of high-dimensional GLMs where the pre- and post-change parameters are unknown.  To handle the high-dimensionality, we propose a novel penalized likelihood based online change-point detection algorithm, which is the first in the change-point literature that achieves optimal minimax localization error rate for high-dimensional GLM. 

%To go beyond the linear models, some attempts on high-dimensional GLMs have also been made.  \cite{wang2020detecting} considered a high-dimensional self-exciting Poisson process change-point localization problem and identified the difficulty of handling the non-linearity in the model assumption, when a penalized estimator is deployed to conquer the high-dimensionality but also facing the mixture distributions.  \cite{li2022detecting} studied the change-point problems in a sequence of high-dimensional Bradley--Terry--Luce models.  The non-linearity again imposes fundamental difficulties and demonstrated different problem complexity from that in the linear regression models.  The aforementioned works focus on online change-point detection. 

In view of the high-dimensional GLM component in our problem, it is worth mentioning the literature of offline change-point detection for the high-dimensional \textit{linear} model. See for example \cite{lee2016lasso}, \cite{kaul2019efficient}, \cite{wang2021statistically} and \cite{rinaldo2021localizing}, where different aspects of offline change-point analysis such as estimation and inference are studied. However, due to the online nature of our problem and the non-linearity of GLM, both the methodology and technical arguments developed in our work are significantly different (and in fact more challenging) than the ones used in offline change-point detection for the high-dimensional linear model. 
\medskip

%We would like to emphasize that the non-linearity imposes different technical challenges for different non-linear models.  This require us to develop novel techniques from those seen in \cite{wang2020detecting} and \cite{li2022detecting}.

\noindent\textbf{Switching bandit problem.}  When the candidate prices form a finite set and when there is no contextual information, the dynamic pricing problem can be seen as a multi-armed bandit (MAB) problem.  MAB problems with non-stationarity are often called the ``switching bandit problem", where the basic setting is to achieve optimal regret for an MAB whose reward functions are piecewise stationary \citep{auer2002nonstochastic}. To handle non-stationarity, two types of strategies are studied in the literature. The passive strategy involves a simple mechanism to forget the past that consists in either discounting rewards far into the history or only considering recent rewards in a sliding window, see e.g.,\ \cite{kocsis2006bandit}, \cite{garivier2011upper}. \cite{besbes2014stochastic} study the passive strategy for MAB under a drifting environment setting where the reward function is allowed to change smoothly under a total variation constraint and \cite{cheung2019learning,cheung2022hedging} further extend the passive strategy to fixed-dimensional non-stationary linear and generalized linear bandit under the drifting environment setting. The active strategy involves a change-point detection component, where one forgets all of the past history and restarts once a change-point is detected \citep[e.g.,][]{liu2018change, cao2019nearly, besson2022efficient}.  

Compared to the switching bandit setting where rewards are from finite number of arms with different means, our setting involves rewards generated based on a high-dimensional GLM and a continuous action space (i.e.,\ price), and thus requires significantly different methodology and technical arguments. We further implement the passive strategy in the numerical studies, where it is seen that in general the active strategy with change-point detection is more adaptive to dynamic pricing under piecewise non-stationarity and achieves better regret.

\medskip
\noindent \textbf{Notation.} For any vector $v \in \mathbb{R}^d$, denote its support as $S(v) = \{j: \, v_j \neq 0\} \subseteq \{1, \ldots, d\}$, its $\ell_1$-, $\ell_2$- and its supremum norms as $\|v\|_1$, $\|v\|_2$ and $\|v\|_{\infty}$. When no ambiguity arises, $\|v\|$ denotes the $\ell_2$-norm. For any set $S$, denote $|S|$ as its cardinality. For any matrix $M \in \mathbb{R}^{d \times d}$, denote its smallest eigenvalue as $\lambda_{\min}(M)$. Let $\mathbb{N}_+$ be the collection of positive integers. Let $a(T)$ and $b(T)$ be two quantities depending on $T$, we denote $a(T)\gtrsim b(T)$ if there exists $T_0\in\mathbb{N}_+$ and $C>0$ such that for any $T\geq T_0$, $a(T)\geq C b(T)$. For $a\in \mathbb{R}$, denote $\lceil a \rceil$ as the smallest integer value  greater than or equal to $a$. For $a,b\in\mathbb{R}$,  denote $a\vee b=\max\{a,b\}$ and $a\wedge b=\min\{a,b\}$.

The (implicit) asymptotic regime in this paper is driven by the number of time periods $T$.  The dimension $d$, sparsity $s$ and number of change-points $\Upsilon_T$ are allowed to grow with $T.$  We refer to a quantity as an absolute constant, if it does not depend on $T$. 

%We denote $a(T)\asymp b(T)$ if  $a(T)\gtrsim b(T)$ and  $b(T)\gtrsim a(T)$.  $\lfloor a\rfloor$  as greatest integer value smaller than or equal to $a$, and 

%\textcolor{red}{$\ell_1$, $\ell_2$ norms, $\lambda_{\min}$}  what are absoulte constants.

%\zifeng{There are a few absolute constants in our paper. $c_m, c_\lambda, c_\gamma, c_{\mathrm{snr}}, c_d$. Do we want to clearly state what they depend on? Basically all the constants on \Cref{assum_model} and \Cref{assum_moment}?} 

\section{Problem Formulation} \label{sec:problem_formulation}

As introduced in \Cref{sec-intro}, in this paper, we consider a firm, hereafter referred to as the \textit{seller}, that sells a product to $T \in \mathbb{N}_+$ sequentially arriving consumers. In each time period $t \in \{1, \ldots, T\}$, the seller observes contextual information featurized as a vector $z_t\in\mathbb{R}^{d-1}$, which may include consumer personal information and product characteristics.  Upon observing $z_t$, the seller offers a price $p_t$ from a bounded price range $[p_l, p_u] \subseteq \mathbb{R}$ to the consumer and observes a demand $y_t \in \mathbb{R}$.

We present the detailed model setup and assumptions in \Cref{sec-basic-model} and define the class of pricing policies considered and its performance metrics (i.e.,\ regret) in \Cref{sec-pricing-policies}.

\subsection{Model Setup}\label{sec-basic-model}

For any $t \in \{1, \ldots, T\}$, we assume that the demand $y_t$ follows a generalized linear model (GLM), conditional on $z_t$ and $p_t$. In particular, given the covariate $x_t = (z_t^{\top}, p_t)^{\top} \in \mathbb{R}^d$, the demand $y_t$ follows the probability distribution
%
%Upon observing $z_t$, the seller offers a price $p_t$ from a bounded price range $[p_l,p_u] \subset \mathbb{R}_+$ to the consumer and observes a demand $y_t \in \mathbb{R}$, which follows a generalized linear model~(GLM). Specifically, given the $d$-dimensional covariate $x_t=(z_t^{\top},p_t)^{\top}$, the demand $y_t$ follows the probability distribution 
\begin{equation}\label{model_GLM}
    f(y_t=y|p_t,z_t)= \exp\left\{\frac{y  x_t^{\top}\theta_t-\psi(x_t^{\top}\theta_t)}{a(\phi)}+h(y)\right\},
\end{equation}
where $\theta_t = (\alpha_t^{\top},\beta_t)^{\top}\in\mathbb{R}^{d}$ is the unknown model parameter associated with time period $t$, $\psi(\cdot):\mathbb {R}\mapsto \mathbb{R}$ is a known function with derivative $\psi'(\cdot)$, $a(\phi)$ is a fixed and known scale parameter and $h(y)$ is a known normalizing function of the distribution family.  This setting covers important GLMs, including Gaussian, logistic and Poisson regression models.  

Note that, conditioning on $x_t$, following \eqref{model_GLM}, it holds that $\mathbb E[y_t|p_t,z_t]=\psi'(x_t^{\top}\theta_t)=\psi'(z_t^\top \alpha_t+\beta_tp_t)$, where the inverse function of $\psi'(\cdot)$ is commonly known as the link function of a GLM. In time period $t$, the consumer demand thus depends on (i) the intrinsic utility $z_t^\top\alpha_t$ and (ii) the pricing effect $\beta_t p_t$. The seller's expected revenue can be written as 
\begin{equation}\label{eq-r-defi}
    r(p_t, \theta_t, z_t)= p_t \mathbb E[y_t|p_t, z_t]= p_t\psi'(z_t^\top \alpha_t+\beta_tp_t).
\end{equation}
Denote $p_t^*$ as the optimal price that maximizes $r(p_t,\theta_t,z_t)$. Throughout this paper, we impose the following assumptions on the GLM \eqref{model_GLM}, covariate $z_t$ and model parameter $\theta_t$. %{\color{red} FY:impose i.i.d. condition somewhere?} \textcolor{purple}{(Yi: good point!  I added Assumption 2.1(v) for this.)}% in \eqref{model_GLM}. %Denote $\Theta=\{\theta\in\mathbb{R}^{d}:\|\theta\|_1\leq C_\theta\}$ as the model parameter space, where $C_\theta>0$ is an absolute constant. 

\begin{ass}\label{assum_model}
For any $t \in \{1, \ldots, T\}$, the following holds.
    (i) The covariate $z_t$ is a random vector with $\|z_t\|_\infty \leq C_b$, where $C_b>0$ is an absolute constant. (ii) The model parameter $\theta_t \in \Theta = \{\theta\in\mathbb{R}^{d}:\|\theta\|_1\leq C_\theta\}$, where $C_\theta>0$ is an absolute constant.  Its support $S(\theta_t)$ satisfies that $|S(\theta_t)|\leq s.$ (iii) The function $\psi(\cdot)$ is infinitely differentiable and $\psi''(\cdot)>0$ is strictly positive. (iv) The optimal price $p_t^*$ is unique and falls into the bounded price range $[p_l,p_u]$. %\textcolor{purple}{(Yi: not sure if ``bounded'' is enough to say $p_l$ and $p_u$ are absolute constants..)} (v) Given $\{(p_t, z_t)\}_{t = 1}^T$, the sequence $\{y_t\}_{t = 1}^T$ is independent and distributed following \eqref{model_GLM}. %,  with $p_l, p_u \geq 0$ being absolute constants.
\end{ass}

%In particular, for $\psi''_u=\sup_{|x|\leq (C_{\theta}+1)C_b} \psi''(x) $, and $\psi''_l=\inf_{|x|\leq C_bC_{\theta}} \psi''(x)$ such that $0<\psi''_l\leq \psi''_u<\infty$.

\Cref{assum_model}(i) requires the covariate $z_t$ is entry-wise upper bounded and \Cref{assum_model}(ii) requires that the $\ell_1$-norm of $\theta_t$ is bounded, which are commonly used in the dynamic pricing literature~\citep[e.g.,][]{javanmard2019dynamic,luo2021distribution,ban2021personalized,chen2022differential}. These two assumptions imply that the intrinsic utility is bounded as $|z_t^\top \alpha_t| \leq \|z_t\|_\infty \|\alpha_t\|_1 \leq C_b C_\theta$.  \Cref{assum_model}(iii) is standard and holds for commonly used GLMs.

\Cref{assum_model}(iv) is standard in the dynamic pricing literature and imposes the uniqueness of $p_t^*$, i.e.,~omitting the dependence on $\theta_t$ and $z_t$ in the notation,
\begin{equation}\label{eq-p-t-star-def}
    p_t^*=\varphi( z_t^\top \alpha_t,\beta_t)= \argmax\limits_{p_t\in[p_l,p_u]}r(p_t,\theta_t,z_t),
\end{equation}
where $\varphi(\cdot,\cdot)$ is a bivariate function that maps the intrinsic utility $z_t^\top \alpha_t$ and price sensitivity $\beta_t$ to the optimal price $p_t^*$. To provide an upper bound on the estimation error of the optimal prices, in the existing literature~(e.g., \cite{broder2012dynamic}, \cite{ban2021personalized}), a Lipschitz condition is explicitly or implicitly imposed on $\varphi(\cdot,\cdot)$. In this paper, we show (see \Cref{lem_punique} in the Appendix) by Berge's maximum theorem \citep{berge1957two} that $\varphi(\cdot,\cdot)$ is Lipschitz under \Cref{assum_model}(iv), i.e.,~there exists an absolute constant $C_\varphi>0$ such that
\begin{equation}\label{C_phi}
    \left|\varphi( z^\top \alpha,\beta)-\varphi( z^\top \alpha',\beta')\right|\leq C_{\varphi}\left\{\left|(\alpha-\alpha')^{\top}z\right|+\left|\beta-\beta'\right|\right\},
\end{equation}
for any $\theta=(\alpha^{\top},\beta)^{\top},\theta'=(\alpha^{'\top},\beta')^{\top} \in \mathbb{R}^d$ and $z \in \mathbb{R}^{d-1}$ that satisfy \Cref{assum_model}. %\textcolor{blue}{Note that the Lipschitz condition in \eqref{C_phi} establishes a bound between estimation error of optimal prices and model parameters.} \textcolor{red}{(do we need this sentence here?)}  This uniqueness condition holds in popular GLMs as demonstrated below. %

We give some common GLMs that satisfy all imposed conditions.  (I) Gaussian GLM, where $\psi(x)=x^2/2$ and $p_t^*=-{z_t^\top \alpha_t}/{(2\beta_t)}$. (II) Poisson GLM, where $\psi(x)=e^{x}$ and $p_t^*=-{1}/{\beta_t}$. (III) Logistic GLM, where $\psi(x)=\log(1+e^x)$ and $p_t^*$ is the unique solution to $1+\exp(z_t^\top\alpha_t +\beta_t p)+\beta_t p=0$. Furthermore, it is easy to verify that if the price sensitivity $\beta_t$ is upper bounded by a negative constant, for all three GLMs in (I)-(III), there exists a bounded price range $[p_l,p_u]$ that covers the optimal price $p_t^*$ for all $t \in \{1, \ldots, T\}$. 
%the intrinsic utility $|z_t^\top \alpha_t|$ is upper bounded and 

\textbf{Market environment evolution processes.}  The marginal model assumptions are collected in \Cref{assum_model}.  Given a sequence of GLMs, we further allow the market environment, i.e.,~the model parameter sequence $\bm\theta_T = \{\theta_t\}_{t = 1}^T$ to evolve over time. In particular, we assume that there exist $\Upsilon_T$ unknown time periods $1\leq\tau_1<\tau_2<\cdots <\tau_{\Upsilon_T} <T$ such that 
\begin{equation}\label{eq-cpd-environment}
    \theta_t \neq \theta_{t + 1} \quad \mbox{if and only if} \quad t \in \{\tau_1, \ldots, \tau_{\Upsilon_T}\}.    
\end{equation}
Denote $\tau_0=0$ and $\tau_{\Upsilon_T+1}=T$. We refer to $\{\tau_k\}_{k = 1}^{\Upsilon_T}$ as change-points, which partition the whole time course into $\Upsilon_T + 1$ stationary segments. We not only allow the model parameter $\theta_t$ to vary between segments, but also the covariate distribution. This is formalized in \Cref{assum_moment} below.

%$\Upsilon_T\geq 0$ CPs $1\leq\tau_1<\tau_2<\cdots <\tau_{\Upsilon_T} <T$ such that
%\begin{align*}
%    \theta_1=\cdots=\theta_{\tau_1} \neq \theta_{\tau_1+1} =\cdots =\theta_{\tau_2}\neq \cdots \neq \theta_{\tau_{\Upsilon_T}+1}=\cdots= \theta_{T}.
%\end{align*}
%In other words, the $\Upsilon_T$ CPs partition the entire horizon $\{1, \ldots,T\}$ into $\Upsilon_T+1$ stationary segments $\{\tau_{k}+1,\tau_{k}+2,\dots,\tau_{k+1}\}$ for $k=0,1,\cdots,\Upsilon_T$, where $\tau_0=0$ and $\tau_{\Upsilon_T+1}=T.$ On the $k$th segment, all time points share the same model parameter $\theta_{\tau_k+1}.$

\begin{ass}\label{assum_moment}
For each $k \in \{0,1,\ldots,\Upsilon_T\}$, the covariates $\{z_t\}_{t = \tau_k+1}^{\tau_{k+1}}$ are i.i.d.\ random vectors generated from an unknown distribution $P_z^{(k)}$ and the second-order moment matrix $\mathbb E(z_{\tau_k+1}z_{\tau_k+1}^\top)=\Sigma_k$ satisfies that $\lambda_{\min}(\Sigma_k)\geq\sigma_l>0$. In addition, there exists $\delta >0$ such that 
\[
    \sup_{\|\Delta\|_2=1, \, \Delta \in \mathbb{R}^{d-1}} \mathbb{E}\{|\Delta^\top z_{\tau_k+1}|^{2+\delta}\} \leq\sigma_u <\infty.
\]
Here, $\sigma_l, \sigma_u, \delta > 0$ are absolute constants.
\end{ass}

\Cref{assum_moment} imposes a mild moment condition on the $\{z_t\}$ sequence and importantly allows a non-stationary covariate process. By Jensen's inequality, it further implies that the maximum eigenvalue of $\mathbb E(z_tz_t^\top)$ is upper bounded by $\sigma_u^{2/(2+\delta)}.$ \Cref{assum_moment} is used to establish a restricted strong convexity condition~\citep[see e.g.,][]{negahban2012unified} and upper bound the estimation error of the Lasso estimator later. Note that we only require $z_t$ to be \textit{entry-wise} bounded~(\Cref{assum_model}(i)) and satisfy a $(2+\delta)$-moment condition. This is weaker than (and thus can be implied by) the assumption that $z_t$ has a bounded support in $\mathbb R^{d-1}$ such that $\|z_t\|_2 <C$ for some absolute constant $C > 0$~\citep[see e.g.,][]{ban2021personalized}.

%Add an explanation with $\delta=0$ will inflate by a $\log^3$ factor, see Section 4 in \cite{rudelson2012reconstruction}.

%Remark that this also implies that  the minimum eigenvalue of $E(x_tx_t^{\top})$ is lower bounded, and the maximum eigenvalue of $E(x_tx_t^{\top})$ is upper bounded for $p_t$ uniformly and independently sampled.

\subsection{Pricing Policies and Performance Metric}\label{sec-pricing-policies}

In a dynamic pricing problem, the key objective is to design a pricing policy that maximizes the revenue.  In this paper, we assume that the seller has no knowledge of any market environmental parameter before the start of the selling horizon.  To be specific, the seller does not know (1) the number of change-points $\Upsilon_T$, (2) the locations of the change-points $\{\tau_1, \ldots, \tau_{\Upsilon_T}\}$, or (3) the model parameters $\bm\theta_T=\{\theta_t\}_{t=1}^T$. %or (4) the covariate probability distributions $\{P_z^{(0)}, \ldots, P_z^{(\Upsilon_T)} \}$.

For each $t \in \{0, \ldots, T-1\}$, denote $\mathcal{F}_t = \sigma (p_1,\cdots,p_t,y_1,\cdots,y_t,z_1,\cdots,z_{t+1})$ as the natural filtration generated by the demand, price and covariate history up to time $t$, together with the covariate $z_{t+1}$.  We let $p_0 = y_0 = 0$ for notational completeness.  Denote by $\Pi$ the family of all price processes $\{p_1, \ldots, p_T\}$ satisfying the condition that $p_t$ is $\mathcal{F}_{t-1}$-measurable for all $t\in\{1, \ldots, T\}.$ In other words, we require the pricing policy to be non-anticipating. 

Given a pricing policy $\pi\in\Pi$, we evaluate its performance using the common notion of regret: the expected revenue loss compared with a clairvoyant that has the perfect knowledge of the model parameter $\bm\theta_T=\{\theta_t\}_{t=1}^T$ (and thus sets the prices at $\{p_t^*\}_{t=1}^T$). In particular, with $r(\cdot, \cdot, \cdot)$ defined in~\eqref{eq-r-defi}, let the regret be
\begin{align}\label{eq:regret}
    R_T^\pi(\bm \theta_T)= \sum_{k=0}^{\Upsilon_T}\sum_{t=\tau_{k}+1}^{\tau_{k+1}} \mathbb E\left\{ r(p_t^*,\theta_t,z_t)-r(p_t^\pi,\theta_t,z_t) \right\},
\end{align}
where the expectation is with respect to the distribution of demands $y_t$ and covariates $z_t$, and $\{p^{\pi}_t\}$ is the sequence of prices under the pricing policy $\pi$.

\section{The Change-Point Assisted Dynamic Pricing Algorithm}\label{sec:alg}

To tackle the non-stationarity as well as the high-dimensionality in the dynamic pricing problem introduced in \Cref{sec:problem_formulation}, we propose the change-point detection assisted dynamic pricing~(CPDP) algorithm, which is shown later to achieve a near-optimal regret.   

There are three key ingredients in the CPDP algorithm: (i) high-dimensional GLM estimation, (ii) exploration and exploitation in the dynamic pricing procedure and (iii) a change-point detection subroutine.   % for the non-stationary high-dimensional dynamic pricing problem defined in \Cref{sec:problem_formulation}. %
As for (i), we adopt the Lasso estimator \citep{tibshirani1996regression}. Given an interval $I\subseteq \{1, \ldots,T\}$, the negative log-likelihood function of the GLM based on \eqref{model_GLM} is $L(\theta, I)= \sum_{t\in I} \{\psi(x^{\top}_t\theta) - y_t x^{\top}_t\theta\}$.
The Lasso estimator on $I$ is then defined as
\begin{align}\label{eq:lasso}
    \widehat \theta_I = \argmin_{\theta \in \Theta} \big\{L(\theta, I)+ \lambda\sqrt{|I|}\|\theta\|_1\big\} = \argmin_{\theta \in \Theta} \Big\{\sum_{t\in I} \{\psi(x^{\top}_t\theta) - y_t x^{\top}_t\theta\}+ \lambda\sqrt{|I|}\|\theta\|_1\Big\},
\end{align}
where $\lambda > 0$ is a pre-specified tuning parameter and $\Theta$ is defined in \Cref{assum_model}(ii).  As for (ii) and (iii), we elaborate in Sections~\ref{sec-cpdp} and \ref{subsec:cp}. %\zifeng{I set the log-likelihood here are true \textit{not} negative loglikelihood. We need to mention this in the proof.}

\subsection{Algorithm Description}\label{sec-cpdp}

We kick off this subsection with some high-level discussion before formally presenting the algorithm.

For a dynamic pricing problem under stationarity, due to the presence of the price sensitivity~$\beta$, it is well-known that there may exist an ``uninformative" price which makes a pure-greedy policy fail~\citep{broder2012dynamic}. To achieve an optimal regret, a pricing policy thus needs to balance between exploration (i.e.,\ price experiment) and exploitation. In particular, the pricing policy is designed such that the ratio of the number of exploration steps to that of exploitation steps decreases with time~\citep[e.g.,][]{broder2012dynamic, ban2021personalized}. This makes intuitive sense as little is known about the model parameter in the early time periods (thus more exploration), while estimation quality improves in the later time periods, as \textit{all} price exploration steps accumulate information under stationarity.

Moving on to the non-stationarity case, the policy needs to further monitor the pricing process, in order to detect any potential change-point with minimum delay and to avoid false alarms.  We therefore need to design an efficient online change-point detection algorithm for high-dimensional GLMs and integrate it with the exploration and exploitation components. Above all, we need to redistribute the exploration and exploitation steps along $\{1, \ldots, T\}$ and in fact \textit{accelerate} the exploration frequency to achieve an optimal regret.  This is because that due to the non-stationarity, price exploration steps prior to a change-point no longer accumulate information.

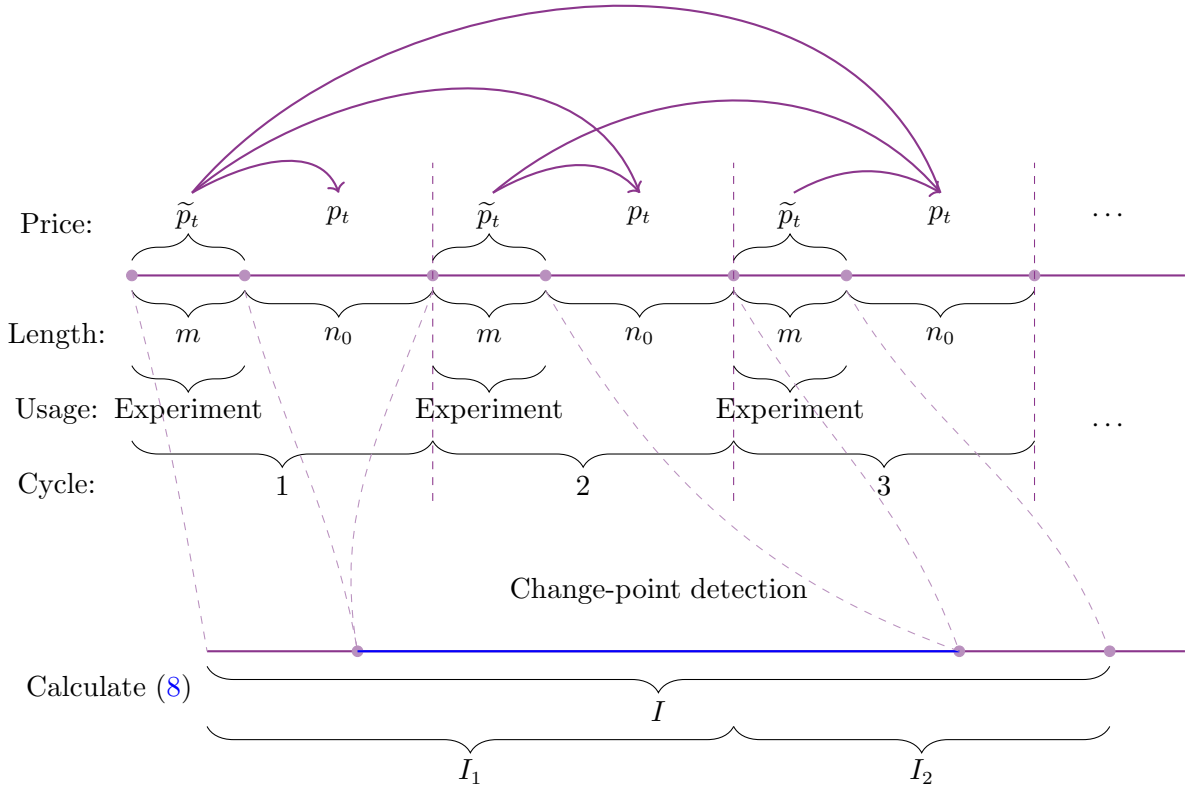
\begin{figure}[ht]
\begin{center}
\begin{tikzpicture}
\draw[Fuchsia!100, thick]  (0, 0) -- (14, 0);
\draw (-1, 0.7) node {Price:};
\draw (-1, -0.8) node {Length:};
\draw (-1, -1.8) node {Usage:};
\draw (-1, -2.8) node {Cycle:};

\filldraw [Fuchsia!50] (0,0) circle (2pt);	
\filldraw [Fuchsia!50] (1.5,0) circle (2pt);	
\filldraw [Fuchsia!50] (4,0) circle (2pt);	

\draw [decorate,decoration={brace,amplitude=10pt},xshift=0pt,yshift=0pt]
(0,0.2) -- (1.5, 0.2) node [black,midway,xshift=0cm, yshift = 0.6cm] {$\widetilde{p}_t$};
\draw (2.75, 0.8) node {$p_t$}; 
\draw [Fuchsia!100, ->, thick] (0.8, 1.1) to [out = 30, in = 110] (2.75, 1.1);

\draw [decorate,decoration={brace,amplitude=10pt, mirror},xshift=0pt,yshift=0pt]
(0, -0.2) -- (1.5, -0.2) node [black,midway,xshift=0cm, yshift = -0.6cm] {$m$};
\draw [decorate,decoration={brace,amplitude=10pt, mirror},xshift=0pt,yshift=0pt]
(1.5, -0.2) -- (4, -0.2) node [black,midway,xshift=0cm, yshift = -0.6cm] {$n_0$};
\draw [decorate,decoration={brace,amplitude=10pt, mirror},xshift=0pt,yshift=0pt]
(0, -1.2) -- (1.5, -1.2) node [black,midway,xshift=0cm, yshift = -0.6cm] {Experiment};
\draw [decorate,decoration={brace,amplitude=10pt, mirror},xshift=0pt,yshift=0pt]
(0, -2.2) -- (4, -2.2) node [black,midway,xshift=0cm, yshift = -0.6cm] {$1$};

\draw[Fuchsia!100, dashed]  (4, -3) -- (4, 1.5);

\filldraw [Fuchsia!50] (5.5,0) circle (2pt);	
\filldraw [Fuchsia!50] (8,0) circle (2pt);	

\draw [decorate,decoration={brace,amplitude=10pt},xshift=0pt,yshift=0pt]
(4,0.2) -- (5.5, 0.2) node [black,midway,xshift=0cm, yshift = 0.6cm] {$\widetilde{p}_t$};
\draw (6.75, 0.8) node {$p_t$}; 
\draw [Fuchsia!100, ->, thick] (4.8, 1.1) to [out = 30, in = 130] (6.75, 1.1);
\draw [Fuchsia!100, thick] (0.8, 1.1) to [out = 40, in = 110] (6.75, 1.1);

\draw [decorate,decoration={brace,amplitude=10pt, mirror},xshift=0pt,yshift=0pt]
(4, -0.2) -- (5.5, -0.2) node [black,midway,xshift=0cm, yshift = -0.6cm] {$m$};
\draw [decorate,decoration={brace,amplitude=10pt, mirror},xshift=0pt,yshift=0pt]
(5.5, -0.2) -- (8, -0.2) node [black,midway,xshift=0cm, yshift = -0.6cm] {$n_0$};
\draw [decorate,decoration={brace,amplitude=10pt, mirror},xshift=0pt,yshift=0pt]
(4, -1.2) -- (5.5, -1.2) node [black,midway,xshift=0cm, yshift = -0.6cm] {Experiment};
\draw [decorate,decoration={brace,amplitude=10pt, mirror},xshift=0pt,yshift=0pt]
(4, -2.2) -- (8, -2.2) node [black,midway,xshift=0cm, yshift = -0.6cm] {$2$};

\draw[Fuchsia!100, dashed]  (8, -3) -- (8, 1.5);

\filldraw [Fuchsia!50] (9.5,0) circle (2pt);	
\filldraw [Fuchsia!50] (12,0) circle (2pt);	

\draw [decorate,decoration={brace,amplitude=10pt},xshift=0pt,yshift=0pt]
(8,0.2) -- (9.5, 0.2) node [black,midway,xshift=0cm, yshift = 0.6cm] {$\widetilde{p}_t$};
\draw (10.75, 0.8) node {$p_t$}; 
\draw [Fuchsia!100, ->, thick] (8.8, 1.1) to [out = 30, in = 150] (10.75, 1.1);
\draw [Fuchsia!100, thick] (4.8, 1.1) to [out = 40, in = 130] (10.75, 1.1);
\draw [Fuchsia!100, thick] (0.8, 1.1) to [out = 50, in = 110] (10.75, 1.1);

\draw [decorate,decoration={brace,amplitude=10pt, mirror},xshift=0pt,yshift=0pt]
(8, -0.2) -- (9.5, -0.2) node [black,midway,xshift=0cm, yshift = -0.6cm] {$m$};
\draw [decorate,decoration={brace,amplitude=10pt, mirror},xshift=0pt,yshift=0pt]
(9.5, -0.2) -- (12, -0.2) node [black,midway,xshift=0cm, yshift = -0.6cm] {$n_0$};
\draw [decorate,decoration={brace,amplitude=10pt, mirror},xshift=0pt,yshift=0pt]
(8, -1.2) -- (9.5, -1.2) node [black,midway,xshift=0cm, yshift = -0.6cm] {Experiment};
\draw [decorate,decoration={brace,amplitude=10pt, mirror},xshift=0pt,yshift=0pt]
(8, -2.2) -- (12, -2.2) node [black,midway,xshift=0cm, yshift = -0.6cm] {$3$};

\draw[Fuchsia!100, dashed]  (12, -3) -- (12, 1.5);

\draw (13, 0.8) node {$\cdots$}; 
\draw (13, -2) node {$\cdots$}; 

\draw (7, -4.2) node {Change-point detection};
\draw[Fuchsia!100, thick]  (1, -5) -- (14, -5);
\filldraw [Fuchsia!50] (3,-5) circle (2pt);	
\filldraw [Fuchsia!50] (11,-5) circle (2pt);	
\filldraw [Fuchsia!50] (13,-5) circle (2pt);	
\draw[blue!100, thick]  (3, -5) -- (11, -5);

\draw [Fuchsia!50, dashed] (0, -0.2) to [out = 285, in = 100] (1, -5);
\draw [Fuchsia!50, dashed] (1.5, -0.2) to [out = 285, in = 100] (3, -5);
\draw [Fuchsia!50, dashed] (4, -0.2) to [out = 250, in = 100] (3, -5);
\draw [Fuchsia!50, dashed] (5.5, -0.2) to [out = 300, in = 160] (11, -5);
\draw [Fuchsia!50, dashed] (8, -0.2) to [out = 300, in = 110] (11, -5);
\draw [Fuchsia!50, dashed] (9.5, -0.2) to [out = 300, in = 110] (13, -5);

\draw [decorate,decoration={brace,amplitude=10pt, mirror},xshift=0pt,yshift=0pt]
(1, -5.2) -- (13, -5.2) node [black,midway,xshift=0cm, yshift = -0.6cm] {$I$};
\draw [decorate,decoration={brace,amplitude=10pt, mirror},xshift=0pt,yshift=0pt]
(1, -6) -- (8, -6) node [black,midway,xshift=0cm, yshift = -0.6cm] {$I_1$};
\draw [decorate,decoration={brace,amplitude=10pt, mirror},xshift=0pt,yshift=0pt]
(8, -6) -- (13, -6) node [black,midway,xshift=0cm, yshift = -0.6cm] {$I_2$};

\draw (-0.3, -5.5) node {Calculate \eqref{eq:pLRT}};

\end{tikzpicture}
\end{center}
\caption{Illustration of the CPDP algorithm.}\label{fig:alg}
\end{figure}

\Cref{algorithm:cpd2} formally presents the CPDP algorithm, accompanied by a graphic illustration in \Cref{fig:alg}. The algorithm runs in cycles, each of which includes a number of consecutive time periods and includes both exploration and exploitation components. For any $e \in \mathbb{N}_+$, we denote $\mathcal{E}_e$ as the collection of time periods in the $e$th cycle and denote $N(e) = \sum_{i=1}^{e}|\mathcal{E}_i|$ as the number of total consumers up to the $e$th cycle.

\begin{algorithm}[ht]
\begin{algorithmic}
    %\Inputs{Horizon $T$. CP detection threshold  $\gamma.$ Price experiment length $m$. Price exploitation length $n_k$ (decreasing with $k$), for $k\in \mathbb N$.   Two prespecified price $(p_1,p_2).$ \textcolor{red}{maybe use $\widetilde{p}_1$ and $\widetilde{p}_2$ here to avoid confusing of $p_t$ later?}}

    \Inputs{Horizon $T$. Lasso tuning parameter $\lambda$. Change-point detection threshold  $\gamma$. Price experiment length $m=c_ms\log(Td)$. Price experiment set $\widetilde{\mathcal{P}}$.}
    \medskip
    
    \Initialize{set $k\gets0$, $\widehat{\tau}_k\gets 0$,  $\mathcal{M}_k\gets\varnothing$}
    
    \For {each cycle $e=1,2,3\cdots $~(until $T$ is reached)}
      \State [\textbf{A.\ Exploration}] 
      \Indent
        \State (A1) for the first $m$ instances, uniformly sample price $p_t$ from $\widetilde{\mathcal{P}}$ and observe 
          \begin{align*}
              \mathcal{M}^{(e)}=\left\{y_t,z_t,p_t\right\}_{t=N(e-1)+1}^{N(e-1)+m};
          \end{align*}
          
      \State (A2) append $\mathcal{M}^{(e)}$ into $\mathcal{M}_k$, i.e., $\mathcal{M}_k\gets \mathcal{M}_k\cup \mathcal{M}^{(e)}.$ 
      \EndIndent
      
      \State [\textbf{B.\ Change-point detection}] Run the change-point detection procedure $\mathrm{CPT}(\mathcal{M}_k, \lambda, \gamma, m)$  \\
      \hspace{1.1cm}  in \Cref{algorithm:cp}. If there is a change, record the change-point and restart, i.e.,\ set
      $$k\gets k+1,~ \widehat{\tau}_k\gets N(e-1)+m,~ \mathcal{M}_k\gets\varnothing, ~
      \text{and go to next cycle}.$$

      \State [\textbf{C.\ Exploitation}] Otherwise,
      \Indent
         \State (C1) update the estimator $\widehat{\theta}_{\mathcal{M}_k}$ based on ${\mathcal{M}_k}$ via Lasso as in \eqref{eq:lasso};
         \State (C2) for the next $n_k=\sqrt{T/(k+1)}$ time periods, set price at $p_t=\varphi(\widehat{\theta}_{\mathcal{M}_k}, z_t)$.
      \EndIndent
    \EndFor	
\caption{The CPDP algorithm}
\label{algorithm:cpd2}
\end{algorithmic}
\end{algorithm}

The CPDP algorithm consists of three components: exploration, change-point detection and exploitation. We discuss each component in the following. We refer to \Cref{sec-cpdp-upper-bound} for theoretical guidance and \Cref{subsec:generalsetting} for practical guidance on the choice of tuning parameters $(\lambda,\gamma,m)$ used in the CPDP algorithm. 

\textbf{A.\ Exploration.} In each cycle $e$, the first $m=c_ms\log(Td)$ time periods are used for price experiments, where the seller uniformly samples a price $p_t$ from a set $\widetilde{\mathcal{P}}$. Here, $\widetilde{\mathcal{P}} \subseteq [p_l,p_u]$ denotes the set where the seller prefers to conduct price experiments and is thus specified by the seller. For example, $\widetilde{\mathcal{P}}$ can be an interval $[\widetilde p_l, \widetilde p_u]$ or a collection of finite price points. The only requirement is that $|\widetilde{\mathcal{P}}|\geq 2$. The data observed in price experiments are then appended together, serving as the foundation for change-point detection and price exploitation. Note that with $c_m > 0$ being an absolute constant, we require $m$ to be of the order $O(s\log(Td))$,  which is the minimum sample size needed to establish optimal error bounds of the Lasso estimator and subsequentially the optimality in the context of change-point detection. 

\textbf{B.\ Change-point detection.} Denote the accumulated data observed in price experiments as $\mathcal{M}_k$, where~$k$ is the number of detected change-points so far. We run the proposed online change-point detection algorithm (\Cref{algorithm:cp}) on $\mathcal{M}_k$. If no change-point is flagged, we proceed to price exploitation, otherwise we restart the system.  Specifically, we discard the accumulated price experiments $\mathcal{M}_k$ and further update the number of change-points detected.  We remark that the change-point detection is challenging due to the high-dimensionality and non-linearity of the GLM. We refer to \Cref{subsec:cp} for more discussion on the novelty and optimality of \Cref{algorithm:cp}.

%as will be seen, $k$ dictates the exploration-exploitation ratio for the upcoming cycles. 

\textbf{C.\ Exploitation.} Once no change-point is detected in the accumulated price experiments $\mathcal{M}_k$, the seller updates the Lasso estimator~$\widehat\theta_{\mathcal{M}_k}$ and then prices at the estimated optimal price $p_t=\varphi(\widehat{\theta}_{\mathcal{M}_k}, z_t)$ for the next $n_k=\sqrt{T/(k+1)}$ consumers, with some abuse of notation on the bivariate function $\varphi(\cdot, \cdot)$ defined in \eqref{eq-p-t-star-def}.  Note that the exploration-exploitation ratio for cycles occurring between the $k$th and $(k+1)$th detected change-points, is $m/n_k=m \sqrt{(k+1)/T}$.  This is an increasing function of $k$, the number of detected change-points. In other words, we \textit{accelerate} the price experiments frequency with more change-points being detected. This is in contrast with existing algorithms for dynamic pricing under non-stationarity~\citep{keskin2017chasing, chen2019dynamic}, where the exploration-exploitation ratio remains constant regardless of the number of change-points detected. As will be made clear in \Cref{sec:theory}, this accelerated exploration mechanism helps CPDP achieve the optimal scaling $\sqrt{\Upsilon_T}$ with respect to the number of change-points.

%\textcolor{red}{remarks: 1, $p^*$ is a function of two unknown real values, so two candidates are enough.  but these two require some knowledge of $p_l$ and $p_u$,  discuss. remark that  $[\widetilde{p}_1,\widetilde{p}_2]$ can be replaced by two numbers.}
%\zifeng{For \Cref{algorithm:cpd2}, I think we can put step 4 in front of step 2.}

%For simplicity of presentation, we use a uniform sampling scheme on an interval $[\widetilde{p}_1,\widetilde{p}_2]$ for price experiments in \Cref{algorithm:cpd2}. We remark that our analysis remains valid as long as two or more experimental prices are used.

\subsection{An Optimal Change-Point Detection Subroutine} \label{subsec:cp}

A key component of \Cref{algorithm:cpd2} is the change-point detection subroutine, where it is applied to the price experiments data denoted as $\mathcal{M}_k$, with tuning parameters $\lambda$, $\gamma$ and $m$.  For notational simplicity, in this section, for any fixed $k \in \mathbb{N}$, we reindex all the data points in $\mathcal{M}_k$ as $\{y_t,z_t,p_t\}_{t=1}^n$ with $|\mathcal{M}_k|=n.$ Note that by design, we have $n<T$ and importantly $\{p_t\}_{t=1}^n$ are i.i.d.\ random variables uniformly drawn from the price experiment set $\widetilde{\mathcal{P}}$.

Given a time period $t\in \{1, \ldots, n\}$, we define the following penalized likelihood ratio test~(LRT), which serves as the building block of our change-point detection.  For any $t \in \{1, \ldots, n-1\}$, let
\begin{align}\label{eq:pLRT}
    \mathcal{D}(t,n)= L(\widehat{\theta}_I, I) - L(\widehat\theta_{I_1}, I_1) - L(\widehat{\theta}_{I_2}, I_2) + \lambda \sqrt{|I_1|} \|\widehat{\theta}_I-\widehat{\theta}_{I_1}\|_1 + \lambda \sqrt{|I_2|} \|\widehat{\theta}_I-\widehat{\theta}_{I_2}\|_1,
\end{align}
where $I = \{1, \ldots, n\}$, $I_1 = \{1, \ldots, t\}$ and $I_2 = \{t + 1, \ldots, n\}$. As defined in \eqref{eq:lasso}, $L(\cdot, \cdot)$ is the negative log-likelihood function, $\lambda > 0$ is the Lasso tuning parameter and $\widehat\theta$ is the Lasso estimator. %\zifeng{I think we need to highlight the novelty/difference of \eqref{eq:pLRT} with existing literature. i.e., highlight something like this is not a direct result from change-point detection in high-dimensional linear model.}

\Cref{algorithm:cp} presents the penalized LRT based change-point detection procedure. Given the detection threshold $\gamma > 0$ and trimming parameter $m = c_m s\log(Td)$, we declare non-stationarity in the data $\{y_t,z_t,p_t\}_{t=1}^n$ if
\begin{align*}
    \max_{t=m,m+1,\cdots,n-m} \mathcal{D}(t,n) >\gamma.
\end{align*}
We set the trimming parameter $m$ the same as the length of price experiments within each cycle in \Cref{algorithm:cpd2}. As discussed above, the parameter $m$ is the minimum sample size needed to upper bound the estimation error of all the Lasso estimators $\widehat{\theta}_{I_1},\widehat{\theta}_{I_2}$ and $\widehat{\theta}_{I}$.  Note that change-point detection starts only when $\mathcal{M}_k$ has at least two cycles of price experiments.

\begin{algorithm}[ht]
\begin{algorithmic}
    \Inputs{Data $\{y_t,z_t,p_t\}_{t=1}^n$. Lasso tuning parameter $\lambda$.\\ Change-point detection threshold  $\gamma$. Trimming parameter $m$.}
    \medskip
    
    \Initialize{set $t\gets m$, $\mathcal{T}_n\gets 0$}    		
    
    \While {$\mathcal{T}_n= 0$ and $t\leq n-m$}			
    \State  $\mathcal{T}_n \leftarrow \mathbb I \{\mathcal{D}(t,n) >\gamma \}$
    \State $t \gets t+1$
    \EndWhile
    
    % \If{FLAG = 0}
    %    \State $t \leftarrow \varnothing$ 
    %\EndIf
    \Outputs{$\mathcal{T}_n$}
    \caption{The CPT algorithm.  CPT$(\text{Data}, \lambda, \gamma, m)$}
    \label{algorithm:cp}
\end{algorithmic}
\end{algorithm} 

The penalized LRT in \eqref{eq:pLRT} consists of two parts: the gain of log-likelihood due to partitioning $L(\widehat{\theta}_I, I) - L(\widehat\theta_{I_1}, I_1) - L(\widehat{\theta}_{I_2}, I_2)$ and the $\ell_1$-penalization $\lambda \sqrt{|I_1|} \|\widehat{\theta}_I-\widehat{\theta}_{I_1}\|_1 + \lambda \sqrt{|I_2|} \|\widehat{\theta}_I-\widehat{\theta}_{I_2}\|_1$. We note that various penalized LRT have been proposed in the change-point literature \citep[e.g.,][]{lai2010sequential, wang2020detecting}. However, we would like to highlight the novelty and subtlety of~\eqref{eq:pLRT}, which is new in the literature.

The imposition of an $\ell_1$ penalty is to overcome the high-dimensionality of the underlying model. In the existing literature, such penalty is often of the form 
\begin{equation}\label{eq-sub-optimal-penalisation}
    \lambda \sqrt{|I|} \|\widehat{\theta}_I\|_1 - \lambda \sqrt{|I_1|} \|\widehat{\theta}_{I_1}\|_1 - \lambda \sqrt{|I_2|} \|\widehat{\theta}_{I_2}\|_1.
\end{equation}
This penalty is easier to analyze using the definition of Lasso estimators, but unfortunately often leads to sub-optimal change-point localization error rates \citep[e.g.,][]{wang2020detecting}.  As demonstrated in the proofs, we conjecture this sub-optimality is partially due to the under-penalization of \eqref{eq-sub-optimal-penalisation}.  To be specific, the penalization term in \eqref{eq:pLRT} is always larger than that in \eqref{eq-sub-optimal-penalisation}, i.e.,
\begin{align*}
    \sqrt{|I_1|} \|\widehat{\theta}_I - \widehat{\theta}_{I_1}\|_1 + \sqrt{|I_2|} \|\widehat{\theta}_I - \widehat{\theta}_{I_2}\|_1  & \geq \sqrt{|I_1|} \|\widehat{\theta}_I\|_1 - \sqrt{|I_1|} \widehat{\theta}_{I_1}\|_1 + \sqrt{|I_2|} \|\widehat{\theta}_I\|_1 - \sqrt{|I_2|} \widehat{\theta}_{I_2}\|_1 \\
    & \geq  \sqrt{|I|} \|\widehat{\theta}_I\|_1 -  \sqrt{|I_1|} \|\widehat{\theta}_{I_1}\|_1 -  \sqrt{|I_2|} \|\widehat{\theta}_{I_2}\|_1,
\end{align*}
due to repeated use of the triangle inequality. This makes the penalized LRT in \eqref{eq:pLRT} more powerful against non-stationarity. As will be shown in \Cref{prop:cp} and discussions afterwards, this novel penalization term delivers optimal online change-point detection.  We remark that, unlike \eqref{eq-sub-optimal-penalisation}, this novel form of penalization prevents a direct use of the basic inequality for Lasso estimators \citep[e.g.,][]{buhlmann2011statistics} and therefore requires development of new technical arguments.

We establish theoretical guarantees of \Cref{algorithm:cp} in terms of both false alarm control and maximum detection delay, in \Cref{prop:cp} below. %This provides the first systematic treatment and optimal result for online change-point detection in high-dimensional GLMs.

\begin{pro}\label{prop:cp}
%Let $\widehat{t}$ be the output of the change-point detection algorithm detailed in \Cref{algorithm:cp}, with the following inputs: (1) data $\{y_t, z_t, p_t\}_{i = 1}^n$ generated from the GLMs defined in \eqref{model_GLM}, under Assumptions~\ref{assum_model} and \ref{assum_moment}, with $\{p_t\}_{t = 1}^n$ \textcolor{red}{(i.i.d.~from a subset of $[\widetilde{p}_l, \widetilde{p}_u]$?)}, (2) Lasso tuning parameter $\lambda=c_{\lambda}\sqrt{\log(Td)}$, (3) detection threshold $\gamma=c_{\gamma}s\lambda^2$ and (4) the trimming parameter $m=c_ms\log(Td)$, where $c_{\lambda}$, $c_{\gamma}$ and $c_m$ are all absolute positive constants.

Let $\mathcal{T}_n$ be the output of \Cref{algorithm:cp}, with $m=c_ms\log(Td)$, $\lambda=c_{\lambda}\sqrt{\log(Td)}$ and $\gamma=c_{\gamma}s\lambda^2$, where $c_{\lambda}$, $c_{\gamma}$, $c_m>0$ are absolute constants. Under Assumptions~\ref{assum_model} and \ref{assum_moment}, we have
    %\vspace{-4mm}
    \begin{itemize}
        \item [(i)] If there is no change-point in $\{1, \ldots, n\}$, then 
        \[
            \mathbb{P}\left(\mbox{no change-point is detected} \right) =\mathbb{P}(\mathcal{T}_n = 0) = \mathbb{P}\left( \max_{t=m,\cdots,n-m} \mathcal{D}(t,n) \leq \gamma \right) \geq 1 - 3T^{-3}.
        \]
        \item [(ii)] If there exists a single change-point $\tau \in \{m, \ldots, n-m\}$, assuming that 
        \begin{equation}\label{eq-snr-prop-3.1}
            \min(\tau,n-\tau)\kappa^2>c_{\mathrm{snr}}s\log(Td),   
        \end{equation}
        where $\kappa=\|\theta_{\tau+1}-\theta_\tau\|_2$ is the change size and $c_{\mathrm{snr}}>0$ is an absolute constant, it holds that
        \[
            \mathbb{P}\left(\mbox{a change-point is detected}\right)=\mathbb{P}\left(\mathcal{T}_n=1 \right)   \geq \mathbb {P} \left(\mathcal{D}(\tau,n)>\gamma \right)  \geq 1 - 3T^{-4}.
        \]%\tau \leq 
        %\[
        %\mathbb{P}\left(\max_{t=m,m+1,\cdots,\tau} \mathcal{D}(t,n) >\gamma\right)\leq 3T^{-3} \text{ and } \mathbb{P}\left(\max_{t=\tau+1,\cdots,\tau+\varepsilon} \mathcal{D}(t,n) >\gamma\right)> 1-3T^{-4},
        %\]
        %where $\varepsilon=c_{\mathrm{snr}}s\log(Td)/\kappa^2$ is the detection delay.
    \end{itemize}
    %\begin{align*}
     %   \mathbb{P}\left(\max_{t=m,m+1,\cdots,\tau} \mathcal{D}(t,n) >\gamma\right)\leq 3T^{-3} \text{ and } \mathbb{P}\left(\max_{t=\tau+1,\cdots,\tau+\varepsilon} \mathcal{D}(t,n) >\gamma\right)> 1-3T^{-3}.
    %\end{align*}
\end{pro}

\Cref{prop:cp} provides non-asymptotic probability bounds for the performance of \Cref{algorithm:cp}. We remark that \Cref{prop:cp} holds for any sufficiently large absolute constants $c_\lambda, c_\gamma, c_m$ and $c_{\text{snr}}$. In particular, the requirement for $c_\lambda, c_\gamma, c_m$ only depend on the absolute quantities in Assumptions \ref{assum_model} and \ref{assum_moment}, and the requirement for $c_{\text{snr}}$ only depend on $c_\lambda, c_\gamma$ and Assumptions \ref{assum_model} and \ref{assum_moment}. We refer to the proof of \Cref{prop:cp} for their detailed characterization. 

\Cref{prop:cp} shows that, when there is no change-point, with high probability, no false alarm will be raised. When there indeed exists a change-point, satisfying \eqref{eq-snr-prop-3.1}, the proposed algorithm can detect its presence with high probability. The condition \eqref{eq-snr-prop-3.1} is equivalent to $\min(\tau,n-\tau)>c_{\mathrm{snr}}s\log(Td)/\kappa^2$. In other words, the change-point $\tau$ can be detected with high probability as long as there are sufficient amount of data before and after $\tau$.

Note that we are in fact in an online setting where $n$ grows. Define the detection delay for $\tau$ as the number of time periods between its occurrence~(i.e., $\tau$) and its detection~(i.e., the smallest $n$ such that $\mathcal{T}_n=1$). \Cref{prop:cp}(ii) states that the detection delay, measured by $n-\tau$, will be upper bounded by $c_{\mathrm{snr}}s\log(Td)/\kappa^2$ with high probability~(assuming we have a sufficient pre-change sample size with $\tau>c_{\mathrm{snr}}s\log(Td)/\kappa^2$). Recall that the price experiment length $m=c_ms\log(Td)$ in the CPDP algorithm.  An important implication of \Cref{prop:cp}(ii) is thus that under mild conditions~(to be specified in \Cref{sec-cpdp-upper-bound}), once a change occurs, we can detect its presence within $O(1/\kappa^2)$ number of cycles with high probability.  This is a key ingredient for establishing the optimal regret bound for the CPDP algorithm. %no false alarm will be raised and 

\textbf{Optimality of \Cref{algorithm:cp}}: The condition \eqref{eq-snr-prop-3.1} is often referred to as the signal-to-noise ratio~(SNR) condition in the change-point analysis literature.  It is of the form
\[
    \underbrace{\mbox{minimal spacing} \times \mbox{change size}^2}_{\text{signal}} \gtrsim \underbrace{\mbox{model complexity} \times \mbox{a logarithmic factor}}_{\text{noise}},
\]
where the model complexity term is the sparsity level $s$ in this paper. This general form is standard and shown to be optimal in various change-point problems, online or offline \citep[e.g.,][]{yu2020review}. The condition \eqref{eq-snr-prop-3.1} matches the minimax optimal SNR condition for change-point detection in high-dimensional linear model~\citep{rinaldo2021localizing,wang2021statistically}. Considering our GLM setup covers the linear model, \Cref{prop:cp} shows that the proposed \Cref{algorithm:cp} works under a minimal condition, and is thus minimax rate-optimal in terms of the required SNR condition.  %Intuitively, it requires the signal (i.e.,\ $\min(\tau,n-\tau)\kappa^2$) to dominate the noise from Lasso estimation (i.e.,\ $s\log(Td)$). For the low dimensional case when $d$ is fixed, \eqref{eq-snr-prop-3.1} further reduces to $\min(\tau,n-\tau)\kappa^2>c_{\mathrm{snr}}\log(T)$,  the information theoretic lower bound (up to constants and logarithmic terms) for the univariate mean change-point detection problem (see e.g.,\ Chan and Walther, 2013; Frick et al., 2014).

Another key quantity in \Cref{prop:cp} is the high-probability detection delay, which is of order $s\log(Td)/\kappa^2$ as discussed above. This is of the form 
\[
    \frac{\mbox{model complexity} \times \mbox{a logarithmic factor}}{\mbox{change size}^2},
\]
which is the standard and optimal localization error in the change-point literature, regardless of the online or offline nature. For the high-dimensional linear model, \cite{rinaldo2021localizing} and \cite{wang2021statistically} derive a minimax lower bound of order $s\kappa^{-2}$. Thus, up to a logarithmic factor, \Cref{prop:cp} shows that \Cref{algorithm:cp} achieves a minimax rate-optimal detection delay.

\section{Theoretical Guarantees of the CPDP Algorithm}\label{sec:theory}

In this section, we analyze the theoretical performance of the proposed CPDP algorithm. In particular, \Cref{sec-cpdp-upper-bound} establishes a near-optimal upper bound on the regret of CPDP and  \Cref{sec-cpdp-lower} further provides a minimax lower bound.

\subsection{Regret Analysis of CPDP} \label{sec-cpdp-upper-bound}

%To analyze the regret bounds on the output of the CPDP Algorithm \ref{algorithm:cpd2}, we first introduce some additional assumptions on the CPs.  %the assumption on the minimal spacing between consecutive change-points, which reflects the difficulty of change-point detection in terms of the change size $\kappa_k$. %Define $l_k=m+n_k$ \textcolor{red}{(referring to the def of $n$ and $m$.  give $l_k$ here an eq.~no.)} and set $d_0=d_{\Upsilon_T+1}=0.$

%Recall that the main objective of this paper is to derive an upper bound on the regret, defined in \eqref{eq:regret}, in a non-stationary environment with the presence of change-points. 

As discussed in \Cref{subsec:cp}, performance of the change-point detection subroutine requires some form of signal-to-noise ratio~(SNR) condition. Before we present an upper bound on the regret, we first introduce an assumption, which is a generalization of the SNR condition \eqref{eq-snr-prop-3.1} in \Cref{prop:cp}. 

Recall that there exist $\Upsilon_T$ unknown change-points as defined in \eqref{eq-cpd-environment}. For $k \in \{1, \ldots, \Upsilon_T\}$, let $\kappa_k = \|\theta_{\tau_k} - \theta_{\tau_k + 1}\|_2$ be the change size at the $k$th change-point and let $\kappa_{\min} = \min_{k = 1, \ldots, \Upsilon_T} \kappa_k$. Recall that for cycles occurring between the $k$th and $(k+1)$th detected change-points, $\mathrm{CPDP}$ (\Cref{algorithm:cpd2}) sets $m = c_m s\log(Td)$ time periods for price experiments and $n_k = \sqrt{T/(k+1)}$ periods for price exploitation. Denote $l_k = m+n_k$ as its cycle length. %Recall $\tau_0=0$ and $\tau_{\Upsilon_T+1}=T.$

%\begin{ass}[SNR]\label{ass_spacing_1}
%  For $k \in \{1, \ldots, \Upsilon_T\}$, let $d_k=(\lceil {c_d}/{\kappa_k^2}\rceil+1)  l_{k-1}$, where $c_d > 0$ is an absolute constant, with $d_0 = d_{\Upsilon_{T+1}} = 0$. Assume that $\tau_{k+1}-\tau_{k}\geq 2(d_k\vee d_{k+1})$, for $k \in \{0, \ldots, \Upsilon_T\}$. , assume that $\tau_{k+1}-\tau_{k}\geq 2(d_k\vee d_{k+1})$, where $d_k=(\lceil {C_d}/{\kappa_k^2}\rceil+1)  l_{k-1}$, $l_k = m+n_k$, $d_0 = d_{\Upsilon_{T+1}} = 0$ and $C_d > 0$ is an absolute constant.
%\end{ass}

\begin{ass}[SNR] \label{ass_spacing}
For all $k \in \{1, \ldots, \Upsilon_T\}$, we have that
\[
    \min\left(\frac{\tau_k-\tau_{k-1}}{l_{k-1}},\frac{\tau_{k+1}-\tau_{k}}{l_{k-1}}\right)\kappa_k^2>c_{\mathrm{snr}}^*,
\] 
where $c_{\mathrm{snr}}^* > 0$ is a sufficiently large absolute constant.
%There exists a sufficiently large absolute constant $c_{\mathrm{snr}}^*>0$ such that for $k \in \{1, \ldots, \Upsilon_T\}$, we have that
%$$\min\left(\frac{\tau_k-\tau_{k-1}}{l_{k-1}},\frac{\tau_{k+1}-\tau_{k}}{l_{k-1}}\right)\kappa_k^2>c_{\mathrm{snr}}^*.$$
\end{ass}

\Cref{ass_spacing} is closely related to the SNR condition \eqref{eq-snr-prop-3.1} for the online change-point detector in \Cref{prop:cp}. In particular, for the $k$th change-point $\tau_k$, conditional on the event that $\tau_{k-1}$ is accurately detected, the quantities $(\tau_k-\tau_{k-1})/l_{k-1}$ and $(\tau_{k+1}-\tau_{k})/{l_{k-1}}$ roughly measure the maximum number of available cycles before and after $\tau_k$, prior to its detection. Due to the exploration-exploitation nature of the CPDP algorithm, the effective sample size for change-point detection is then $(\tau_k-\tau_{k-1})m/l_{k-1}$ and $(\tau_{k+1}-\tau_{k})m/{l_{k-1}}$. Plugging this into \eqref{eq-snr-prop-3.1}, we recover the general form of \Cref{ass_spacing}. The constant $c_{\mathrm{snr}}^*$ is different than $c_{\mathrm{snr}}$ as we need to further account for possible detection delay resulted from change-points prior to $\tau_k$. We refer to the sketch of proofs for \Cref{thm:regret} later for the detailed characterization of $c_{\mathrm{snr}}^*$.

%\textcolor{red}{(recall the def of $n_k$ here.  be careful with $m$ here - we have multiple $m$ which are not necessarily the same..)}

%For the $k$th CP, with the jump size $\kappa_k$ measured in the $L_2$ norm and the spacing $\tau_{k+1} - \tau_k$, \Cref{ass_spacing} is a more general form of the signal-to-noise ratio condition \eqref{eq-snr-prop-3.1} used in \Cref{prop:cp}.

%Let $\kappa_{\min} = \min_k \kappa_k$ and
Plugging in the order of all related quantities, \Cref{ass_spacing} reads as
\begin{equation} \label{eq-ass-4.1-inter}
    \{\kappa_k \wedge \kappa_{k+1}\}^2(\tau_{k+1} - \tau_k) \gtrsim \sqrt{T/(k+1)} + \{s\log(Td)\}, \text{ for } k\in \{0,1,\cdots,\Upsilon_T\},
\end{equation}
where we define $\kappa_0=\kappa_{\Upsilon_T+1}=\infty.$ A few remarks are in order.
\begin{itemize}
%    \item In the CP detection literature, the counterpart of \eqref{eq-ass-4.1-inter} is usually of the form $\kappa_{\min}^2 (\tau_{k+1} - \tau_k) \gtrsim \{s\log(Td)\}$. \textcolor{red}{(refs)}  The condition \eqref{eq-ass-4.1-inter} however inflates this condition by a factor of $\sqrt{T/k}$, due to the inevitable exploration and exploitation nature of the dynamic pricing problems.
    \item As we stated, a close cousin of dynamic pricing under non-stationarity is the switching bandit problem, the counterpart of \eqref{eq-ass-4.1-inter} there is usually of the form $\{\kappa_k \wedge \kappa_{k+1}\}^2(\tau_{k+1} - \tau_k) \gtrsim \sqrt{T}$ \citep[e.g.,][]{cao2019nearly, besson2022efficient}, which matches \eqref{eq-ass-4.1-inter} when $d$ is considered as fixed.

    \item Unlike existing works for dynamic pricing under non-stationarity~\citep[e.g.,][]{keskin2017chasing,chen2019dynamic}, CPDP does not require the knowledge of a \textit{fixed} lower bound on the minimum change size $\kappa_{\min}$.  In fact, we allow a \textit{diminishing} $\kappa_{\min}$ with $T$ growing unbounded, provided that the change-points are well separated, i.e.,~$(\tau_{k+1}-\tau_k)$'s are sufficiently large. When $\kappa_{\min}$ is lower bounded and $d$ is fixed, \eqref{eq-ass-4.1-inter} reduces to $(\tau_{k+1} - \tau_k) \gtrsim \sqrt{T}$, which recovers the SNR condition in \cite{chen2019dynamic}.

    \item Note that the number of change-points $\Upsilon_T$ is allowed to grow under \eqref{eq-ass-4.1-inter}. In particular, when $\kappa_{\min}$ is lowered bounded, it is easy to see that $\Upsilon_T$ can be of maximum order $O(\sqrt{T}/\log(Td))$ under the mild condition that the sparsity $s=O(\sqrt{T})$. In addition, thanks to the accelerated exploration mechanism of CPDP, \eqref{eq-ass-4.1-inter} in fact becomes \textit{weaker} for later occurred change-points.
    
    %\item Similar assumption has been adopted by  for nonstationary MAB problems and \cite{chen2019dynamic} for nonstationary dynamic pricing problems.  We note that in \cite{keskin2017chasing} a constant lower bound on $\kappa_k$'s is imposed whereas our treatment  implicitly allows $\kappa_k$'s to be diminishing. Under their setting, a sufficient condition for \Cref{ass_spacing} is that  $\tau_{k+1}-\tau_k\geq C(s\log (Td)+\sqrt{T})$ for some large $C>0$.
\end{itemize}

%\textcolor{blue}{Clearly, Assumption \ref{ass_spacing} is related to the signal-to-noise ratio condition in \Cref{prop:cp}, it requires that the spacing between change-points scales  with the change-point magnitude.  Similar assumption has been adopted by \cite{cao2019nearly} and \cite{besson2022efficient} for nonstationary MAB problems.}

%\textcolor{blue}{We note that in \cite{keskin2017chasing} a constant lower bound on $\kappa_k$'s is imposed whereas our treatment  implicitly allows $\kappa_k$'s to be diminishing. Under their setting, a sufficient condition for \Cref{ass_spacing} is that  $\tau_{k+1}-\tau_k\geq C(s\log (Td)+\sqrt{T})$ for some large $C>0$.}

With \Cref{ass_spacing} in hand, we are able to provide a regret upper bound for CPDP.  

\begin{thm}\label{thm:regret}
%Under Assumptions~\ref{assum_model} and \ref{assum_moment}, there exist sufficiently large absolute constants $c_m$, $c_{\lambda}$, $c_{\gamma}>0$, such that with $m=c_ms\log(Td),$ $\lambda=c_{\lambda}\sqrt{\log(Td)}$, $\gamma=c_{\gamma}s\lambda^2$, and $\pi$ being the CPDP algorithm, we have that
%\begin{align*}
 %   R_T^\pi(\bm \theta_T)\leq C \sqrt{\Upsilon_TT}\left(\kappa_{\min}^{-2}\vee {s\log (Td)}\right),
%\end{align*}
%for any $\bm\theta_T = \{\theta_1, \ldots, \theta_T\}$ sequence that possesses the change-point environment \eqref{eq-cpd-environment} and satisfies \Cref{ass_spacing}, where $C>0$ is an absolute constant.
%
%For the GLM demand model defined in \eqref{model_GLM}, assume that 

Let $\pi$ be the CPDP algorithm with $m=c_ms\log(Td)$, $\lambda=c_{\lambda}\sqrt{\log(Td)}$, and $\gamma=c_{\gamma}s\lambda^2$, where $c_{\lambda}, c_{\gamma}, c_m > 0$ are absolute constants. Under Assumptions~\ref{assum_model} and \ref{assum_moment}, it holds that
\begin{equation}\label{eq-regret-upper-bound}
    R_T^\pi(\bm \theta_T)\leq C \sqrt{\Upsilon_TT}\left(\kappa_{\min}^{-2}\vee {s\log (Td)}\right),
\end{equation}
for any $\bm\theta_T$ that satisfies \Cref{ass_spacing}, where $C > 0$ is an absolute constant. 
\end{thm}

The upper bound in \Cref{thm:regret} is further dominated by the order $O(\kappa_{\min}^{-2}s\sqrt{\Upsilon_TT}\log (Td))$, as $\kappa_{\min}$ is upper bounded by an absolute constant due to \Cref{assum_model}(ii). This resembles the best available upper bound in the switching bandit problem, where it takes the form $O(\kappa_{\min}^{-2}\sqrt{K\Upsilon_T T\log (T)})$, with $K$ being the number of arms~\citep[e.g.,][]{cao2019nearly, besson2022efficient}.

\textbf{Near-optimal performance of CPDP.} Consider the case where $\Upsilon_T$ is upper bounded and $\kappa_{\min}$ is lower bounded, which is the setting used in existing works~\citep[e.g.,][]{keskin2017chasing,chen2019dynamic} to establish optimality of the proposed algorithm for dynamic pricing under non-stationarity. The upper bound \eqref{eq-regret-upper-bound} reduces to $C s\sqrt{T}\log (Td)$, which matches the lower bound $\Omega(s\sqrt{T})$ (up to logarithmic terms) derived in \cite{ban2021personalized} for dynamic pricing under a stationary high-dimensional GLM~(see Theorem 3 therein). Thus, CPDP achieves near-optimal regret. In other words, the existence of (finite number of) change-points do not inflate the order of regret, which agrees with the observation in \cite{keskin2017chasing} and \cite{chen2019dynamic}.

We further explicitly characterize the impact of the number of change-points $\Upsilon_T$ in the regret upper bound. In particular, owing to the accelerated exploration mechanism, \eqref{eq-regret-upper-bound} scales with $\sqrt{\Upsilon_T}$, which is later shown to be optimal with a minimax lower bound in \Cref{sec-cpdp-lower}. To our best knowledge, this is new in the dynamic pricing literature. Without accelerated exploration, the regret scales linearly with $\Upsilon_T$, see e.g.,\ \cite{chen2019dynamic} and Corollary 1 therein. We refer to the sketch of proofs below for a more detailed discussion.

%\Cref{thm:regret} explicitly characterizes the determining factors in the  regret  upper bound.  We emphasize that the rate in terms of $\Upsilon_T$ is optimal in view of \Cref{thm:lowerbound}, %Furthermore, CPDP does not require the knowledge of an upper bound on the number of change-points.  This is in sharp contrast to \cite{keskin2017chasing}, where an upper bound is required.   In fact, if $\Upsilon_T$ is known, one can simply run CPDP by setting $n_0=\cdots=n_{\Upsilon_T}=\sqrt{T/(\Upsilon_T+1)}$, and  obtain the same regret upper bound. 

% Intuitively, when more change-points are present, more exploration should be forced to reduce the possibility of failure in detection.%For constant changes,  i.e.,\ $\kappa_{\min}^{-1}=O(1)$, \Cref{thm:regret} matches the lower bound derived in \Cref{thm:lowerbound} up to a logarithmic factor, suggesting the optimality of CPDP.  \textcolor{red}{(what about $s$?)} For vanishing changes, i.e.,\ $\kappa_{\min}\to 0$, \textcolor{red}{(what is the asymptotic regime here?)} our result is also consistent with  \cite{keskin2017chasing}  \textcolor{red}{(what is their result?)}, see Remark 3 therein.  

The regret upper bound in \Cref{thm:regret} also involves the minimal change size $\kappa_{\min}$. This is intuitive as a smaller change leads to a larger detection delay $s\log(Td)/\kappa_{\min}^2$ as in \Cref{prop:cp}, and thus results in more ``mis-priced" time periods. On the other hand, a smaller $\kappa_{\min}$ means that the pre- and post-change model parameters are closer to each other, and thus suggests that the regret on the ``mis-priced" time periods due to detection delay can be better controlled. Note that due to detection delay, the price experiment set $\mathcal{M}_k$ will consist of both pre- and post-change data. Thus, to materialize the above intuition, we need a finer analysis of the estimator $\widehat\theta_{\mathcal{M}_k}$ when it is based on a mixture of two high-dimensional GLMs. 

To proceed, we introduce a mild assumption on the mixture of two high-dimensional GLMs. Let $\alpha\in (0,1)$ be the mixture proportion. For $k \in \{1, \ldots, \Upsilon_T\}$, define
\begin{align}\label{eq-pseudo}
    \theta^{(k)}_\alpha=\argmin_{\theta\in\Theta} \mathbb E\left[ \alpha\left\{\psi(x^{\top}_{\tau_k}\theta) - y_{\tau_k} x^{\top}_{\tau_k}\theta\right\} + (1-\alpha)\left\{\psi(x^{\top}_{\tau_{k}+1}\theta) - y_{\tau_{k}+1} x^{\top}_{\tau_{k}+1}\theta\right\} \right].
\end{align}
In other words, for the $k$th change-point, $\theta^{(k)}_\alpha$ denotes the pseudo true parameter that maximizes the expected log-likelihood of a mixture of two high-dimensional GLMs, where $\alpha$-proportion is from the pre-change model $\theta_{\tau_k}$ and $(1-\alpha)$-proportion is from the post-change model $\theta_{\tau_{k+1}}.$ We then impose the following assumption on $\theta^{(k)}_\alpha$. 

\begin{ass}\label{assum:mixture}
For any $k \in \{1, \ldots, \Upsilon_T\}$ and any $\alpha \in (0, 1)$, the pseudo true parameter $\theta^{(k)}_\alpha$ satisfies that (i) it is in the interior of $\Theta$, defined in \Cref{assum_model}; and (ii) there exists an absolute constant $C_M>0$, such that 
\[
    \max\left\{\big\|\theta^{(k)}_\alpha-\theta_{\tau_k}\big\|_1, \big\|\theta^{(k)}_\alpha-\theta_{\tau_{k+1}}\big\|_1\right\}\leq C_M\|\theta_{\tau_k}-\theta_{\tau_{k+1}}\|_1.
\]

%There exists an absolute constant $C_M>0$ such that for any $\alpha\in (0,1)$ and $k=1,2,\cdots,\Upsilon_T$, the pseudo true parameter $\theta^{(k)}_\alpha$, defined in \eqref{eq-pseudo}, lies in the interior of $\Theta$, defined in \Cref{assum_model}(ii) and $$\max\left\{\big\|\theta^{(k)}_\alpha-\theta_{\tau_k}\big\|_1, \big\|\theta^{(k)}_\alpha-\theta_{\tau_{k+1}}\big\|_1\right\}\leq C_M\|\theta_{\tau_k}-\theta_{\tau_{k+1}}\|_1.$$
\end{ass}
\Cref{assum:mixture} essentially requires that the pseudo true model parameter $\theta^{(k)}_\alpha$ is in-between the two model parameters $\theta_{\tau_{k}}$ and $\theta_{\tau_{k+1}}$ that form the mixture, which is intuitive. For an i.i.d.\ covariate process $\{x_t\}_{t=1}^T,$ \Cref{assum:mixture} holds for the Gaussian GLM~(i.e.,\ linear model) automatically with $C_M=1$, as we have $\theta^{(k)}_\alpha=\alpha\theta_{\tau_k}+(1-\alpha)\theta_{\tau_{k+1}}.$ We refer to the sketch of proofs later for a more detailed discussion of \Cref{assum:mixture} and its implication. 

With \Cref{assum:mixture}, the following corollary shows that the regret upper bound of the CPDP algorithm is independent of $\kappa_{\min}$, which is new in the literature. This result is significant as it indicates that once the change-points are detectable (i.e.,\ the SNR condition in \Cref{ass_spacing} holds), the actual change size does \textit{not} impact the order of the regret.

\begin{cor}\label{cor_regret}
In addition to all the conditions in \Cref{thm:regret}, suppose that Assumption  \ref{assum:mixture} also holds, we have that
\begin{align*}
    R_T^\pi(\bm \theta_T)\leq C s\sqrt{\Upsilon_T T}\log (Td),
\end{align*}
where $C>0$ is an absolute constant.
\end{cor}

%Under the mild condition that $s^2\Upsilon_T=O(T)$, which holds if $s=O(T^{1/4})$ since  $\Upsilon_T=o(\sqrt{T})$ under \Cref{ass_spacing} as discussed before, the result in \Cref{cor_regret} implies that $R_T^\pi(\bm \theta_T)$ is upper bounded by $Cs\sqrt{\Upsilon_TT}\log (Td)$, which is the upper bound in \Cref{thm:regret} but free of $\kappa_{\min}.$

\medskip

\noindent\textbf{A Sketch of Proofs:} We conclude this subsection with a sketch of proofs for \Cref{thm:regret} and \Cref{cor_regret}. For readability, we focus on the high-level intuition and refer to the Appendix for the detailed and rigorous technical arguments. 

 %except the last cycle, which is of length $m.$ 

 At a high-level, the proofs are conducted in a high-probability event, mainly driven by the performances of the change-point detection algorithm.  The proofs can then be modularized into a few steps: (\textbf{Step 1}) deriving a high-probability bound for accurate change-point detection; (\textbf{Step 2}) decomposing the regret into regrets originated from different sources; and (\textbf{Step 3}) upper-bounding regrets from different sources separately.

\medskip
\noindent \textbf{Step 1: }\textbf{High-probability bound for accurate change-point detection.} Let $\{\widehat{\tau}_k\}_{k=1}^{\widehat{\Upsilon}_T}$ be the estimated change-points with $\widehat{\tau}_0=0$ and $\widehat{\tau}_{\widehat\Upsilon_T+1}=T$. By design, each of the cycles between $\widehat{\tau}_k$ and $\widehat{\tau}_{k+1}$ is of length $l_k=n_k+m$. By \Cref{prop:cp}, to detect $\tau_k$ with high-probability, it requires 
\begin{align*}
    \left\lceil \frac{c_{\mathrm{snr}}s\log(Td)}{\kappa_k^2} \Big / m \right\rceil =\left\lceil \frac{c_{\mathrm{snr}}}{c_m\kappa_k^2}\right\rceil
\end{align*}
number of cycles both before and after $\tau_k$. Define $d_k=\lceil c_{\mathrm{snr}}/(c_m\kappa_k^2)\rceil l_{k-1}$ for $k=1,\ldots,\Upsilon_T$ as the controllable detection delay and further define 
\[%begin{flalign}\label{AT}
\mathcal{A}=\left\{\widehat\Upsilon_T=\Upsilon_T \text{ and for all } k\in\{1,\ldots,\Upsilon_T\}, \widehat{\tau}_k\in [\tau_k,\tau_k+d_k]\right\},
\]%end{flalign}
as the good event where all change-points are detected within the desirable detection delay. We first show that $\mathbb{P}(\mathcal{A}^c)$ is bounded by $3T^{-1}$, which serves as the foundation for the regret analysis.

Define $\mathcal{A}_{k}=\{\text{for all } j \in\{1,\ldots,k\}, \, \widehat{\tau}_j\in [\tau_j, \tau_j+d_j] \}$ as the event where the first $k$ change-points have been detected within controllable detection delay. Recall $\widehat{\tau}_0=\tau_0=0$ and let $\mathcal{A}_0=\varnothing$, we have that
\begin{flalign*}
\mathbb{P}(\mathcal{A}^c)\leq \sum_{k=1}^{\Upsilon_T+1}\mathbb{P}(\widehat{\tau}_k< \tau_k|\mathcal{A}_{k-1})+\sum_{k=1}^{\Upsilon_T}\mathbb{P}(\widehat{\tau}_k>\tau_k+d_k|\mathcal{A}_{k-1}),
\end{flalign*}
where the first term corresponds to false alarm and the second term corresponds to large delay.

By \Cref{ass_spacing} with $c_{\mathrm{snr}}^*>2c_{\mathrm{snr}}/c_m$, we have $(\tau_k-\tau_{k-1})>2(d_{k-1}\vee d_k)$ with $d_0 = d_{\Upsilon_{T+1}} = 0$. Thus, conditional on $\mathcal{A}_{k-1}$, we have $(\tau_k-\tau_{k-1}-d_{k-1})>d_k$ and $(\tau_{k+1}-\tau_k)>2d_k.$ Invoking \Cref{prop:cp}, we can show that $\mathbb{P}(\widehat{\tau}_k< \tau_k|\mathcal{A}_{k-1})<3T^{-2}$ and $\mathbb{P}(\widehat{\tau}_k>\tau_k+d_k|\mathcal{A}_{k-1})<3T^{-4}.$ Together, we have that $\mathbb{P}(\mathcal{A}^c)\leq 3T^{-1}$.

\medskip
\noindent \textbf{Step 2: }\textbf{Decomposition of the regret into regrets from different sources.}  The regret of CPDP can then be decomposed into four parts, due to the following sources: (I) price experimentation, (II) controllable delay in change-point detection, (III) parameter estimation error, and (IV) failed change-point detection, respectively. Let $\mathcal{M}=\bigcup_{k=0}^{\widehat{\Upsilon}_T}\mathcal{M}_k$ be the set of price experiments. In particular, we have that
\begin{flalign*}
R_T(\bm \theta_T)=&\mathbb{E}\left[\sum_{t=1}^Tr(p_t^*,\theta_t,z_t)-r(p_t,\theta_t,z_t)\right]
\leq C_r\mathbb{E}\left[\sum_{t=1}^T (\varphi(\theta_t, z_t)-p_t)^2\right]
\\=&C_r\mathbb{E}\left[\sum_{t=1}^T (\varphi(\theta_t, z_t)-p_t)^2\mathbb{I}(t\in\mathcal{M},\mathcal{A})\right]+C_r\sum_{k=1}^{\Upsilon_T}\mathbb{E}\left[ \sum_{t={\tau}_k+1}^{\widehat{\tau}_k
}(\varphi(\theta_t, z_t)-p_t)^2\mathbb{I}(t\in\mathcal{M}^c,\mathcal{A})\right]\\&+
C_r\sum_{k=0}^{\Upsilon_T}\mathbb{E}\left[ \sum_{t=\widehat{\tau}_{k}+1}^{{\tau}_{k+1}}(\varphi(\theta_t, z_t)-p_t)^2 %\varphi(\widehat{\theta}_t,x_t)
\mathbb{I}(t\in\mathcal{M}^c,\mathcal{A})\right]+  C_r\mathbb{E}\left[\sum_{t=1}^T(\varphi(\theta_t, z_t)-p_t)^2\mathbb{I}(\mathcal{A}^c)\right]
\\=&  R_{T,\mathrm{I}}+ R_{T,\mathrm{II}}+ R_{T,\mathrm{III}}+ R_{T,\mathrm{IV}},
\end{flalign*}
where $R_{T,i}$, $i\in\{\mathrm{I,II,III,IV}\}$ corresponds to regret due to (I)-(IV) respectively. Here, we denote $\varphi(\theta_t,z_t)=\varphi(z_t^\top \alpha_t,\beta_t)$, i.e.,\ the optimal price function defined in \eqref{eq-p-t-star-def}. For notational simplicity, we omit the superscript $\pi$ in the regret. The first inequality follows from the fact that the revenue is a smooth function of price with a bounded second order derivative~(see \Cref{sec:prop_revenue_func} of the Appendix for details).

%In view of \eqref{bound_price} and recall that $\theta_t$ is the true model parameter at time $t$, we can upper bound the revenue loss by the square of the difference between optimal price and the charged price. In particular,

\medskip
\noindent \textbf{Step 3: }\textbf{Upper-bounding regrets from different sources separately.} We then provide upper bounds for each of the four sources of regret. An important observation is that by \Cref{assum_model}, the difference $\varphi(\theta_t, z_t)-p_t=p_t^*-p_t$ is upper bounded by an absolute constant. Thus, to bound the regret $R_{T,\mathrm{I}}$ and $R_{T,\mathrm{II}}$, we can bound the number of periods in price experiments and in the controllable detection delay, respectively.

\noindent\textsc{(I). Regret due to experiments}: Conditional on $\mathcal{A}$, the shortest cycle length is $l_{\Upsilon_T}$ and thus we can bound $R_{T,\mathrm{I}}$ with
\begin{align*}
    R_{T,\mathrm{I}}\leq C_{R,\mathrm{I}}'  m\frac{T}{l_{\Upsilon_T}} \leq C_{R,\mathrm{I}} s\log(Td) \sqrt{\Upsilon_T T},
\end{align*}
where $C_{R,\mathrm{I}}>0$ is an absolute constant.

\noindent\textsc{(II). Regret due to detection delay}: Conditional on $\mathcal{A}$, the detection delay for all change-points are within $d_k$ and thus we can bound $R_{T,\mathrm{II}}$ with
\begin{align*}
    R_{T,\mathrm{II}}\leq C_{R,\mathrm{II}}'\sum_{k=1}^{\Upsilon_T} \frac{d_k}{l_{k-1}} n_{k-1}\leq C_{R,\mathrm{II}}\sum_{k=1}^{\Upsilon_T} \kappa_k^{-2} n_{k-1} \leq C_{R,\mathrm{II}}\kappa_{\min}^{-2}\sum_{k=1}^{\Upsilon_T}  \sqrt{T/k}\leq  2C_{R,\mathrm{II}}\kappa_{\min}^{-2}  \sqrt{\Upsilon_T T},
\end{align*}
where $C_{R,\mathrm{II}}>0$ is an absolute constant. Importantly, note that thanks to the accelerated exploration mechanism, the length of price exploitation $n_k=\sqrt{T/k}$ decreases with $k.$ This makes it possible to bound $R_{T,\mathrm{II}}$ with $\sqrt{\Upsilon_T T}$ instead of $\Upsilon_T\sqrt{T}$. 

\noindent\textsc{(III). Regret due to estimation error}: The analysis for bounding $R_{T,\mathrm{III}}$ is more involved. Conditional on $\mathcal{A}$, the time periods within $\{\widehat\tau_k+1, \ldots, \tau_{k+1}\}$ are stationary for $k=0,\ldots,\Upsilon_T$. By the Lipschitz condition in \eqref{C_phi} and additional technical arguments, the expected price difference can be upper bounded by the expected estimation error $\|\theta_{\tau_{k+1}}-\widehat{\theta}_t\|_2$, where $\widehat\theta_t$ is the Lasso estimator based on the accumulated price experiments.

By the design of the CPDP algorithm, for $k=0,1,\cdots,\Upsilon_T$, we can partition the price exploitation periods in $\{\widehat\tau_k+1, \ldots, \tau_{k+1}\}\cap \mathcal{M}_k^c$ into groups of size $n_{k}$. The total number of groups $J_k$ is bounded by $(\tau_{k+1}-\widehat\tau_k)/l_{k}<\tau_{k+1}-\tau_k.$ Importantly, for $j=1,2,\cdots,J_k$, the $n_k$ periods in the $j$th group share the same Lasso estimator, denoted by $\widehat\theta^j(k)$, which is estimated based on the first $j$ cycles of price experiments after $\widehat\tau_k$. Thus, we have
\begin{align*}
    R_{T,\mathrm{III}}&\leq C_{R,\mathrm{III}}'\sum_{k=0}^{\Upsilon_T}n_k\sum_{j=1}^{J_k} \|\theta_{\tau_{k+1}}-\widehat\theta^j(k)\|^2 \leq C_{R,\mathrm{III}}''\sum_{k=0}^{\Upsilon_T}n_k\sum_{j=1}^{\tau_{k+1}-\tau_k} \frac{s\log(Td)}{mj} \\
    &= C_{R,\mathrm{III}}''\sum_{k=0}^{\Upsilon_T}n_k\sum_{j=1}^{\tau_{k+1}-\tau_k} \frac{s\log(Td)}{jc_ms\log(Td)} \leq C_{R,\mathrm{III}}\sum_{k=0}^{\Upsilon_T}n_k \log(\tau_{k+1}-\tau_k) \leq 2C_{R,\mathrm{III}}\sqrt{\Upsilon_T T} \log T,
\end{align*}
where $C_{R,\mathrm{III}}>0$ is an absolute constant. The second inequality follows from the error bound for the Lasso estimator $\|\theta_{\tau_{k+1}}-\widehat\theta^j(k)\|^2$ in \Cref{lem:est} of the Appendix. Same as the case for $R_{T,\mathrm{II}}$, the accelerated exploration mechanism makes $R_{T,\mathrm{III}}$ scale with $\sqrt{\Upsilon_TT}$ instead of $\Upsilon_T\sqrt{T}.$

\noindent\textsc{(IV). Regret due to failed change-point detection}: We have $R_{T,\mathrm{IV}}\leq C_{R,\mathrm{IV}}T \mathbb P(\mathcal{A}^c) \leq 3C_{R,\mathrm{IV}}$ for an absolute constant $C_{R,\mathrm{IV}}>0$. 

Thus, combine all the bounds, we have $R_T(\bm \theta_T)\leq C \sqrt{\Upsilon_TT}\left(\kappa_{\min}^{-2}\vee {s\log (Td)}\right),$ which proves \Cref{thm:regret}. To prove \Cref{cor_regret}, it is clear that we need to sharpen the regret bound for $R_{T,\mathrm{II}}$, i.e.,\ the regret due to controllable detection delay. In particular, a bound based on the number of periods is too loose. We instead provide a sharper bound by directly analyzing the regret of all price exploitation periods in the detection delay set  $\{\tau_k+1,\ldots,\widehat\tau_k\}$. 

Using the same arguments as in $R_{T,\mathrm{III}}$, for $k=1,\ldots,\Upsilon_T$, we can partition the price exploitation periods  $\{\tau_k+1, \ldots, \widehat\tau_{k}\}\cap\mathcal{M}_{k-1}^c$ into groups of size $n_{k-1}$. Conditional on $\mathcal{A}$, the total number of groups $J_{k}^*$ is bounded by $C_J/\kappa_{k}^2$ for some absolute constant $C_J>0$. For the $n_{k-1}$ periods in the $j$th group, they share the same Lasso estimator, denoted by $\widehat\theta^j(k)$. To achieve a sharper bound, we need to analyze the estimation error $\|\theta_{\tau_{k+1}}-\widehat\theta^j(k)\|^2$. However, due to the detection delay, note that $\widehat\theta^j(k)$ is in fact estimated based on a mixture of two high-dimensional GLMs with parameters $\theta_{\tau_k}$ and $\theta_{\tau_{k+1}}.$ Thus, $\|\theta_{\tau_{k+1}}-\widehat\theta^j(k)\|^2$ consists of both bias and estimation error.

For $k=1,\cdots,\Upsilon_T+1$, denote $S_k$ as the support for $\theta_{\tau_k}$ and denote $S_k^c$ as its complement set. Given a set $S$, denote $\theta(S)$ as the sub-vector of $\theta$ on $S.$ By \Cref{assum:mixture}, for any mixture proportion $\alpha\in(0,1)$ between the two GLMs with parameters $\theta_{\tau_k}$ and $\theta_{\tau_{k+1}}$, the pseudo true parameter $\theta_\alpha^{(k)}$ satisfies that
$\big\|\theta_\alpha^{(k)}(S_k^c\cap S_{k+1}^c)\big\|_1\leq C_M\|\theta_{\tau_k}-\theta_{\tau_{k+1}}\|_1 \leq C_M\sqrt{2s}\|\theta_{\tau_k}-\theta_{\tau_{k+1}}\|_2=C_M\sqrt{2s}\kappa_k.$ In other words, though it may not be exactly sparse, the pseudo true parameter $\theta_\alpha^{(k)}$ is approximately sparse as its $\ell_1$-norm outside of $S_k\cup S_{k+1}$ is upper bounded by $O(\sqrt{s})$. Based on this, \Cref{lem_mixture} of the Appendix establishes a Lasso-type estimation error bound for $\theta_\alpha^{(k)}$. \Cref{assum:mixture} further implies $\big\|\theta^{(k)}_\alpha-\theta_{\tau_{k+1}}\big\|_2\leq \big\|\theta^{(k)}_\alpha-\theta_{\tau_{k+1}}\big\|_1 \leq C_M\|\theta_{\tau_k}-\theta_{\tau_{k+1}}\|_1\leq C_M\sqrt{2s}\kappa_k$, which controls the bias of $\theta_\alpha^{(k)}$.

 Thus, denote $\theta^j(k)$ as the pseudo true parameter for $\widehat\theta^j(k)$, we have that
 \begin{align*}
    R_{T,\mathrm{II}}&\leq C_{R,\mathrm{II}}'\sum_{k=1}^{\Upsilon_T}n_{k-1}\sum_{j=1}^{J_k^*} \|\theta_{\tau_{k+1}}-\widehat\theta^j(k)\|^2 \leq C_{R,\mathrm{II}}'\sum_{k=1}^{\Upsilon_T}n_{k-1}\sum_{j=1}^{J_k^*} \|\theta_{\tau_{k+1}}-\theta^j(k)\|^2 + \|\theta^j(k)-\widehat\theta^j(k)\|^2.
 \end{align*}
The first term is the bias term and is bounded by $C_{R,\mathrm{II}}''s\sqrt{\Upsilon_T T}$, and the second term is the estimation error term and is further bounded by $C_{R,\mathrm{II}}'''\sqrt{\Upsilon_T T}\log T$ using \Cref{lem_mixture} of the Appendix. This completes the proof.

\subsection{A Minimax Lower Bound On the Regret} \label{sec-cpdp-lower}

To further examine the role of $\Upsilon_T$ and demonstrate the optimality of the regret upper bound derived in \Cref{thm:regret} for the CPDP algorithm, in this subsection, we provide a minimax lower bound on the regret by any pricing policy in the set $\Pi$, defined in \Cref{sec-pricing-policies}.

\begin{thm}\label{thm:lowerbound}
Let $\mathcal{P}_T$ be the collection of distributions satisfying Assumptions \ref{assum_model}, \ref{assum_moment} and \ref{ass_spacing}, with the corresponding $\bm\theta_T$ possessing~$\Upsilon_T$ number of change-points, we have that
\[
    \inf_{\pi \in \Pi} \sup_{P(\bm\theta_T) \in \mathcal{P}_T} R_T^{\pi}(\bm\theta_T) \geq C\sqrt{T\Upsilon_T}/\log(\Upsilon_T),
\]
where $C>0$ is an absolute constant.
%Under Assumptions \ref{assum_model} and \ref{assum_moment}, for any pricing policy $\pi \in \Pi$, there exists a $\bm\theta_T=\{\theta_t\}_{t=1}^T$ that has exactly $\Upsilon_T$ number of change-points and satisfies \Cref{ass_spacing}, such that
%\begin{align*}
%    R_T^{\pi}(\bm\theta_T) \geq C\sqrt{T\Upsilon_T}/\log\Upsilon_T,
%\end{align*}
%where $C>0$ is an absolute constant.
\end{thm}

In terms of $T$, \Cref{thm:lowerbound} states that $R_T^{\pi}$ is lower bounded by $\sqrt{T}$.  This is consistent with the result in \cite{broder2012dynamic} and \cite{ban2021personalized} for a stationary dynamic pricing problem with non-informative prices. For dynamic pricing under non-stationarity, the lower bound $\sqrt{T}$ is derived in \cite{besbes2011minimax} under a single change-point setting.

Our lower bound in \Cref{thm:lowerbound} provides a further characterization of the problem difficulty in terms of the number of change-points $\Upsilon_T$, which is new in the dynamic pricing literature. Such a lower bound $\sqrt{\Upsilon_T}$ has appeared previously in the switching bandit literature \citep[e.g.,][]{auer2002nonstochastic, seznec2020single}. We remark that the lower bound in \Cref{thm:lowerbound} will increase to $C\sqrt{T\Upsilon_T}$, i.e.,\ without the logarithmic term, if we only require $\bm\theta_T$ to have $\Upsilon_T$ change-points without imposing the SNR condition in \Cref{ass_spacing}. %\textcolor{blue}{This suggests that the SNR condition of $\bm\theta_T$ imposed by \Cref{ass_spacing} does not make the problem easier.} \textcolor{red}{(I suggest we delete this sentence - this may be due to an artifact of the proof..)}

The lower bound is established by constructing a sequence of linear demand models that possess $\Upsilon_T$ change-points and by an application of Le Cam's lemma~\citep[e.g.,][]{tsybakov2009}. In particular, to reveal the problem difficulty in terms of the number of change-points $\Upsilon_T$, new technical arguments are developed in the proof, which may be of independent interest, as we need to handle the case of multiple change-points with $\Upsilon_T$ allowed to grow unbounded. We acknowledge that the lower bound in \Cref{thm:lowerbound} does not involve the sparsity $s$ and we conjecture that a sharper lower bound should be $\Omega(s\sqrt{T\Upsilon_T})$, as suggested by the lower bound $\Omega(s\sqrt{T})$ for dynamic pricing in a stationary GLM~\citep{ban2021personalized}. We leave this for future work.

%Note that the lower bound in \Cref{thm:lowerbound} does not involve the sparsity $s$, as for technical reasons, the lower bound instances are constructed based on a linear demand model without any covariates. However, as discussed before, for dynamic pricing under a stationary GLM, \cite{ban2021personalized} gives a lower bound of order $s\sqrt{T}$. Thus for $T$ time periods with $\Upsilon_T$ equally spaced change-points, intuitively the lower bound is of order $\Upsilon_T\cdot s\sqrt{T/\Upsilon_T}=s\sqrt{\Upsilon_T T}.$ We leave this for future research.  \textcolor{red}{(I suggest we delete this paragraph - this makes our result sound a bit weak...)}

%Our proof is inspired by the arguments used in \cite{broder2012dynamic} and \cite{keskin2017chasing} \textcolor{blue}{and is also based on Le Cam's lemma~\citep{tsybakov2009}} \textcolor{red}{(I suggest we delete this..)}.  \textcolor{blue}{ However, we develop new technical arguments and the instances constructed are completely different }  \textcolor{red}{(I find this sounds a bit weak...)} as we need to further handle non-stationarity with a growing number of change-points $\Upsilon_T$.  

%\textcolor{red}{(I suggest we rewrite the above two paragraphs as follows.)}

\section{Numerical Experiments}\label{sec:num}

In this section, we conduct extensive numerical experiments on both synthetic and real-world data, to investigate the performance of the proposed CPDP algorithm for GLM-based dynamic pricing under non-stationarity. Sections \ref{subsec:generalsetting} and \ref{sec-5.2} study the efficiency and robustness of CPDP under various simulation settings and compare its performance with popular algorithms in the literature. \Cref{subsec:autoloan} presents a real data application on online auto loan pricing to further showcase the practical utility of CPDP and the importance of handling non-stationarity in dynamic pricing.

\subsection{General Simulation Settings}\label{subsec:generalsetting}

\textbf{Choice of tuning parameters $(\lambda, m, \gamma)$.}  There are three key tuning parameters of the CPDP algorithm: (i).\ the Lasso tuning parameter $\lambda=c_{\lambda}\sqrt{\log (Td)}$, (ii).\ the price experiment length $m=c_m s\log (Td)$, and (iii).\ the change-point detection threshold $\gamma=c_{\gamma} s\log (Td).$ Note that all theoretical results in \Cref{thm:regret} hold for sufficiently large $c_{\lambda},c_m,c_{\gamma}$. In practice, we recommend setting $m=\left\lceil \{\log(Td)\}^{1.1}\right\rceil$ and $\gamma=\left\lceil \{\log(Td)\}^{1.1}\right\rceil$, which avoids the tuning for $c_m$ and $c_{\gamma}$. For any sufficiently large $Td$ and an upper bounded sparsity $s$, this provides valid theoretical results and only inflates the regret by a factor of $\{\log(Td)\}^{0.1}$.

For the choice of $\lambda$, we recommend selecting $c_{\lambda}$ via cross-validation. Specifically, before starting the CPDP algorithm, the seller can first conduct $\sqrt{T}$ pilot price experiments, where it uniformly samples price $p_t$ from the price experiment set $\widetilde{\mathcal P}$ and observe $\mathcal{M}_{\mathrm{cv}}=\left\{y_t,z_t,p_t\right\}_{t=-\sqrt{T}}^0$. Based on $\mathcal{M}_{\mathrm{cv}}$, a leave-one-out cross-validation is used to select the best $\lambda_{\mathrm{cv}}.$ Note that this pilot sample does not inflate the regret order. Numerical experiments in the following suggest that $\lambda_{\mathrm{cv}}$ provides similar finite-sample performance compared to a finely tuned $c_{\lambda}$.

\medskip
\noindent \textbf{Competing methods.} We consider several competing methods to illustrate the importance of monitoring non-stationarity and showcase the efficiency of CPDP. First, we consider four variants of CPDP. We refer to \Cref{algorithm:cpd2} for detailed definition of Step A, B and C of CPDP.
\begin{itemize}
    \itemsep-0.2em 
    \item Naive-DP: CPDP without change-point detection (Step B, thus $k\equiv 0$).
    \item Sliding window DP (SW-DP): CPDP without change-point detection (Step B, thus $k\equiv 0$). In Step (A2), only keep the most recent $\eta_{\mathrm{sw}}$ cycles of price experiments in $\mathcal{M}_0.$
    \item Discounted factor DP (DF-DP): CPDP without change-point detection (Step B, thus $k\equiv 0$). In Step (C1), estimate $\widehat{\theta}(0)$ based on $\mathcal{M}_0$ with weighted Lasso. Specifically, each data point in $\mathcal{M}_0$ is given weight $\rho_{\mathrm{df}}^n$ where $n$ is the time elapsed since that data point was observed and $\rho_{\mathrm{df}}\in(0,1)$ is the discount factor.
    \item Optimal DP (OPT-DP): CPDP with knowledge of true change-point locations. In particular, OPT-DP iterates between Step A (exploration) and Step C (exploitation). It exits Step A or C, and runs Step B (record the change and restart) once a true change-point occurs.
\end{itemize}

Compared to CPDP, which is an active adaptive strategy, SW-DP and DF-DP are two passive adaptive strategies that are popular in the switching bandit literature \citep[e.g.,][]{garivier2011upper}. They are also used in dynamic pricing for linear demand models without covariates in \cite{keskin2017chasing} to counter smooth changes. SW-DP and DF-DP adapt to potential non-stationarity by removing/discounting data points observed far in the past. Note that for $\eta_{\mathrm{sw}}=\infty$ and $\rho_{\mathrm{df}}=1$, SW-DP and DF-DP reduce to Naive-DP.

In addition, we also implement the dynamic pricing algorithm proposed in \cite{ban2021personalized}, hereafter BK-DP, which is designed for stationary high-dimensional GLM. Similar to Naive-DP, BK-DP only iterates between exploration and exploitation without an active change-point detection component, the main difference is where price experiments are conducted. In particular, BK-DP conducts price experiments on $\{L^2,L^2+1:L=1,2,\ldots\}\cap \{1, \ldots, T\}$ \citep[see eq.~(7) in][]{ban2021personalized}, while Naive-DP conducts price experiments on intervals of length $m$ that are evenly separated by $\sqrt{T}$ time periods. For a fair comparison, we implement BK-DP with price experiments on $\{L^2,L^2+1,\ldots,L^2+m-1:L=1,2,\ldots\}\cap \{1, \ldots, T\}$, which ensures that both Naive-DP and BK-DP have $O(m\sqrt{T})$ price experiments. Indeed, we find this version of BK-DP performs better as it provides more stable and accurate Lasso estimation (especially in early time periods) due to more experiments. By design, the frequency of price experiments is decreasing for BK-DP across $\{1,\ldots,T\}$ while stays the same for Naive-DP. As will be seen, this has implications on their performance under stationary and non-stationary environments. 

\textbf{Data generating process.} We consider a high-dimensional logistic regression setting, where given $(z_t,p_t)$, the consumer demand $y_t$ is a Bernoulli random variable such that
\begin{align*}
    \mathbb E(y_t|z_t,p_t)=\psi'(\alpha_t^\top z_t+\beta_t p_t),
\end{align*}
where $\psi'(\cdot)$ is the logistic function. We set dimension $d=50$ and $z_t=(1,z_{t,1},\ldots,z_{t,48})^{\top}$, where $\{z_{t,i}\}_{i = 1}^{48}$ are i.i.d.\ uniform(0,1) random variables. We set $\alpha_t=(\alpha_{t}^{*\top}, \mathbf 0_{44}^{\top})^{\top}$, where $0 \neq \alpha_t^*\in \mathbb R^5$. The sparsity is therefore $s=6$. Denote $\theta_t=(\alpha_t,\beta_t) \in \mathbb R^{50}$, in the following, we design different non-stationarity schemes by varying $\{\theta_t\}_{t=1}^T.$ 

\textbf{Implementation details.} For a fair comparison, in all experiments, we use the same Lasso tuning parameter $\lambda$ and price experiment set $\widetilde{\mathcal{P}}=[\widetilde{p}_1,\widetilde{p}_2]$ for all algorithms. In addition, we do not impose any explicit bound $C_\theta$ on $\|\theta\|_1$ in the Lasso estimation. For SW-DP and DF-DP, we experiment with $\eta_{\mathrm{sw}} \in \{4, 8, 16\}$ and $\rho_{\mathrm{df}} \in \{0.98, 0.99, 0.999\}$, and present the result for $\eta_{\mathrm{sw}}=8$ and $\rho_{\mathrm{df}}=0.99$, as they provide the best overall performance for SW-DP and DF-DP under non-stationarity. We refer to \Cref{subsec:swdf} for a sensitivity analysis of $\eta_{\mathrm{sw}}$ and $\rho_{\mathrm{df}}$.

\subsection{Experiments on Synthetic Data} \label{sec-5.2}
\subsubsection{Value of Active Change-Point Detection:}\label{subsec:valueCPD}
Set $\theta^{(1)}=(0,1,1,2,2,\mathbf{0}_{44}^{\top},-0.25)^{\top}$ and $\theta^{(2)}=(0,1,1,1,1,\mathbf{0}_{44}^{\top},-0.5)^{\top}$. Note that $\theta^{(1)}$ denotes a high-demand period with low price sensitivity $0.25$ and high attractiveness $\mathbb E(\alpha_t^\top z_t)=3$, while $\theta^{(2)}$ denotes a low-demand season with high price sensitivity $0.5$ and low attractiveness $\mathbb E(\alpha_t^\top z_t)=2.$

We first consider three scenarios with no change-point, one change-point and three change-points. In particular, $\text{(S1)}:  \theta_t\equiv\theta^{(1)}$, 
\begin{align*}
	\text{(S2)}:  \theta_t=\begin{cases} 
		\theta^{(1)}, & t\in [1,T/2],\\
		\theta^{(2)}, & t\in (T/2,T],\\
	\end{cases}\quad \mbox{and} \quad
	\text{(S3)}:  \theta_t=\begin{cases} 
		\theta^{(1)}, & t\in [1,T/4]\cup (T/2,3T/4],\\
		\theta^{(2)}, & t\in (T/4,T/2]\cup (3T/4,T].\\
	\end{cases}\quad
\end{align*}
Note that (S2) and (S3) give the scenario where the product being sold alternates between the high-demand and low-demand seasons.

We set $T \in \{5000, 10000, \ldots, 90000, 100000\}$. For the Lasso tuning parameter $\lambda$, we set $\lambda_{\mathrm{fix}}=0.2\sqrt{\log(Td)}$. We set the price experiment set as $[\widetilde{p}_l, \widetilde{p}_u]=[1,15]$ and set the price bound as $[p_l,p_u]=[0,50].$ We note that the simulation result is robust to the choice of $[\widetilde{p}_l, \widetilde{p}_u]$. In \Cref{subsec:robust_lambda}, we further report the result based on the CV-selected $\lambda_{\mathrm{cv}}$ as described in \Cref{subsec:generalsetting}. Given a horizon $T$, for each algorithm, we conduct 100 independent experiments and record their realizations of regret $\{R_T^{(i)}\}_{i=1}^{100}$. 

\begin{figure}[ht]
\vspace{-5mm}
\hspace*{-6mm} 
	\begin{subfigure}{0.32\textwidth}
		\includegraphics[angle=270, width=1.2\textwidth]{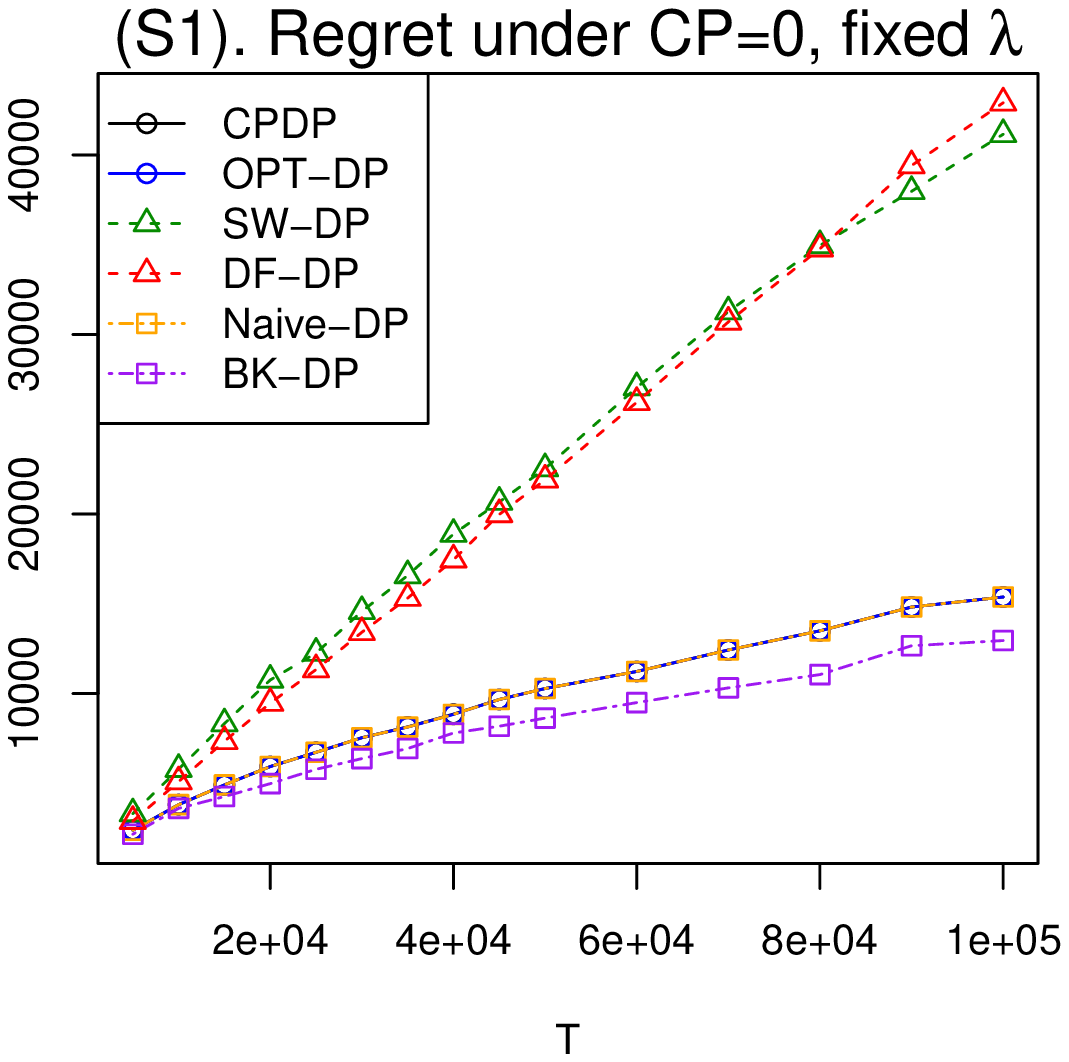}
		\vspace{-0.5cm}
	\end{subfigure}
	~
	\begin{subfigure}{0.32\textwidth}
		\includegraphics[angle=270, width=1.2\textwidth]{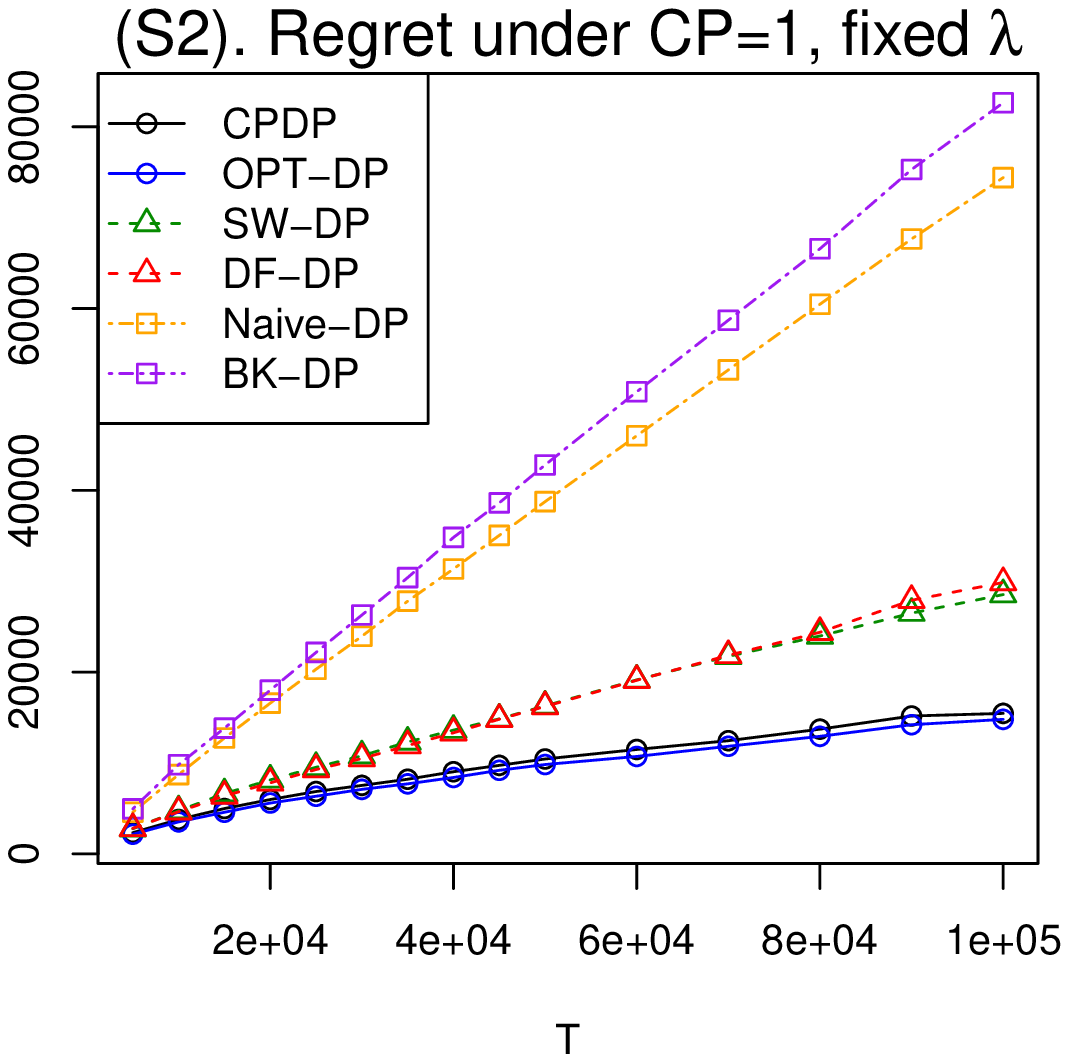}
		\vspace{-0.5cm}
	\end{subfigure}
	~
	\begin{subfigure}{0.32\textwidth}
		\includegraphics[angle=270, width=1.2\textwidth]{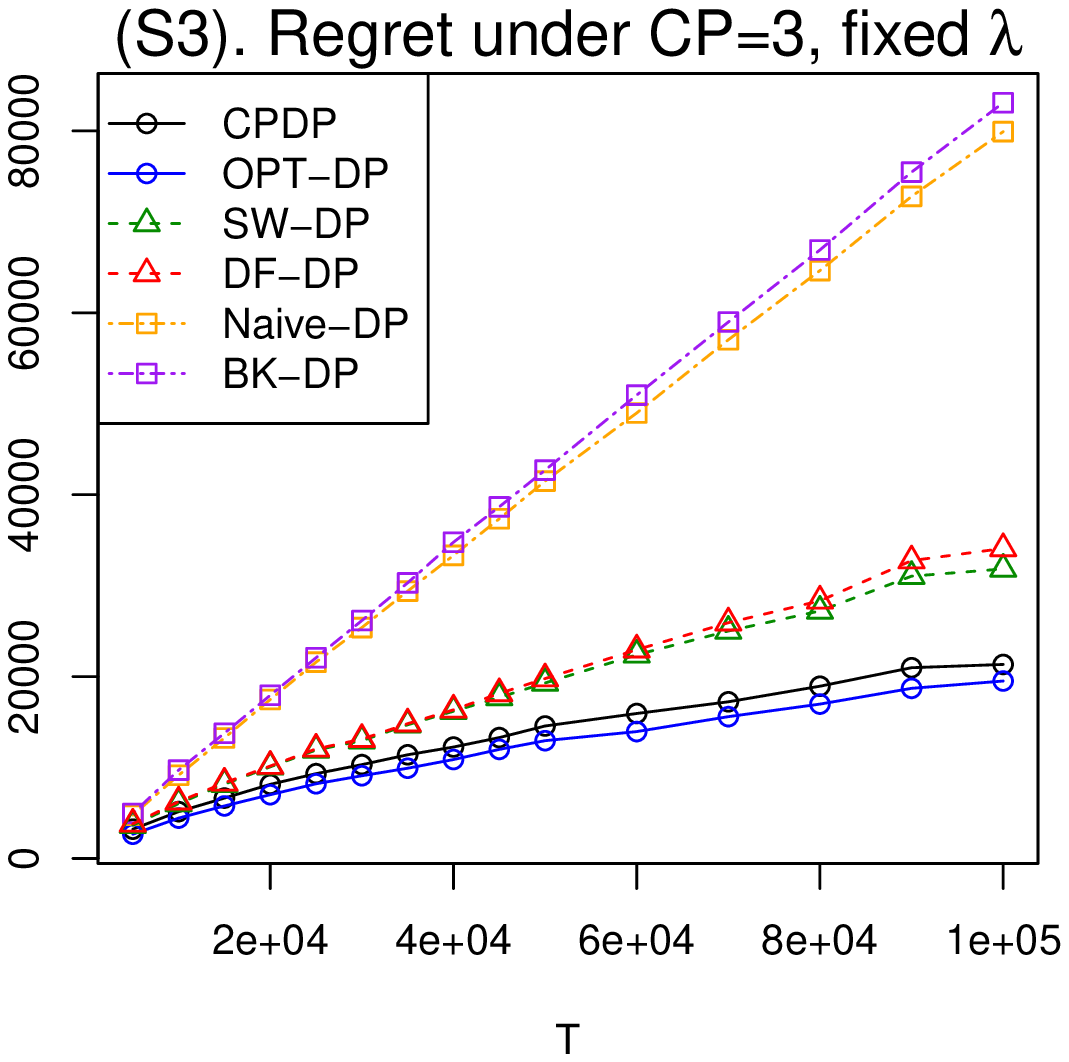}
		\vspace{-0.5cm}
	\end{subfigure}
 	\caption{ Mean regret vs.\ $T$ under (S1) [left], (S2) [middle], (S3) [right] with $\lambda_{\mathrm{fix}}=0.2\sqrt{\log(Td)}$.}
	\label{fig:fixedlambda}
\end{figure}

\Cref{fig:fixedlambda} reports the mean regret $\overline{R}_T=\sum_{i=1}^{100}R_T^{(i)}/100$ of each algorithm at different horizon~$T$ under simulation scenarios (S1), (S2) and (S3). Several comments are in order. First, under (S1) where the process is stationary, BK-DP provides the best performance, followed by Naive-DP, OPT-DP and CPDP. Indeed, OPT-DP coincides with Naive-DP as there is no change-point. CPDP gives almost identical regret as OPT-DP, suggesting that the change-point detection procedure is robust to false alarm. As discussed in \Cref{subsec:generalsetting}, compared to Naive-DP, BK-DP places more price experiments at the beginning of the learning process and thus provides more accurate parameter estimation, explaining the gap between BK-DP and Naive-DP. SW-DP and DF-DP give worst performance as useful information is discarded by design.

Under (S2) and (S3) where there is non-stationarity, OPT-DP and CPDP are the clear winners, followed by SW-DP and DF-DP. The performance gap between OPT-DP and CPDP is reasonably small, suggesting that the change-point detection procedure of CPDP is efficient (see \Cref{subsec:changesize} for more details). As expected, BK-DP and Naive-DP do not perform well due to the non-stationarity. Note that BK-DP is worse than Naive-DP, especially in (S2), as its estimation $\widehat{\theta}$ is more tilted towards $\theta^{(1)}$ due to the same reason that it places more price experiments at the beginning of the learning process.

A linear regression between $\log(\overline{R}_T)$ and $\log(T)$ for CPDP gives a slope of 0.5093, 0.5148, and 0.5269 under (S1), (S2) and (S3), providing further numerical evidence for \Cref{thm:regret}.

\begin{figure}[ht]
\vspace{-5mm}
\hspace*{-6mm}                                                           
    \begin{subfigure}{0.32\textwidth}
	\includegraphics[angle=270, width=1.2\textwidth]{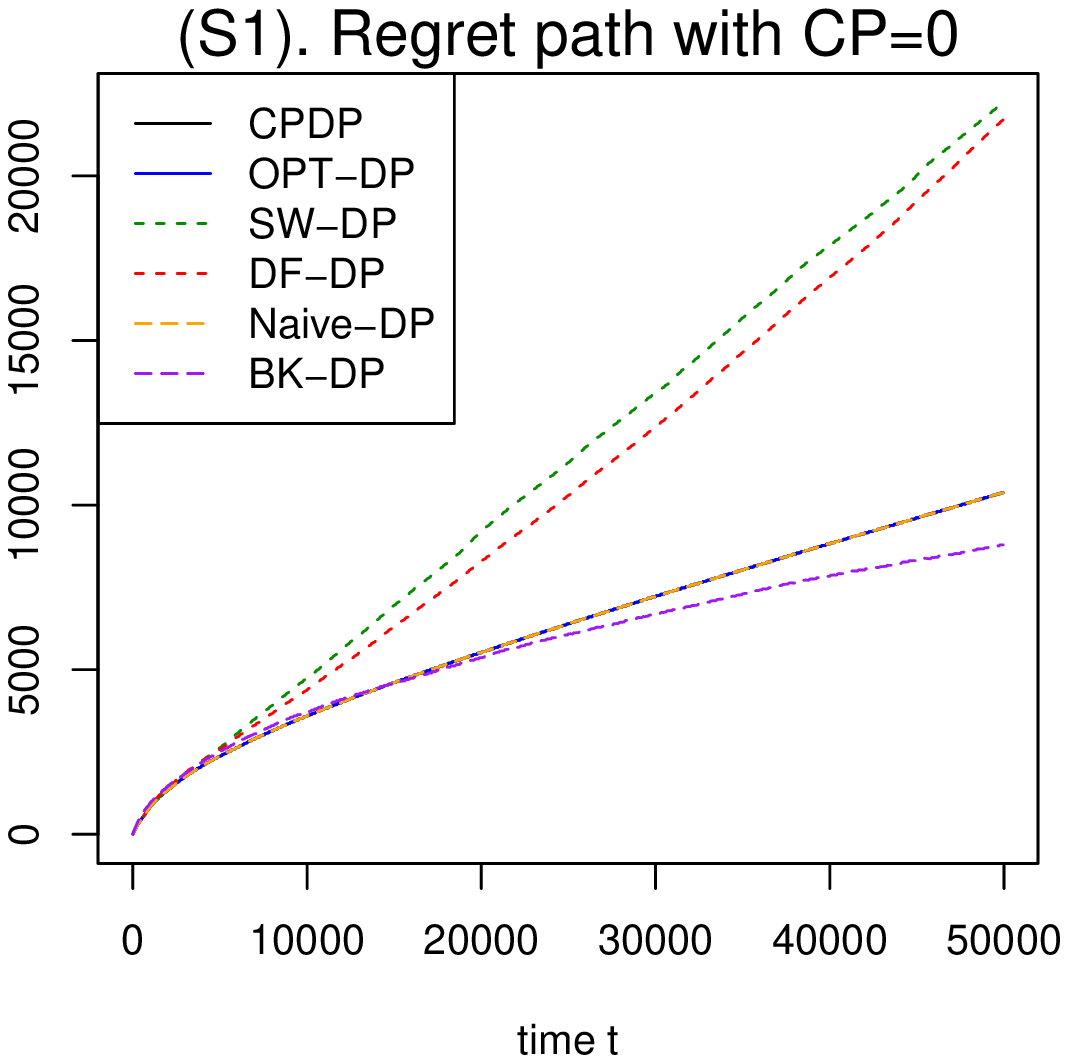}
	\vspace{-0.5cm}
    \end{subfigure}
    ~
    \begin{subfigure}{0.32\textwidth}
	\includegraphics[angle=270, width=1.2\textwidth]{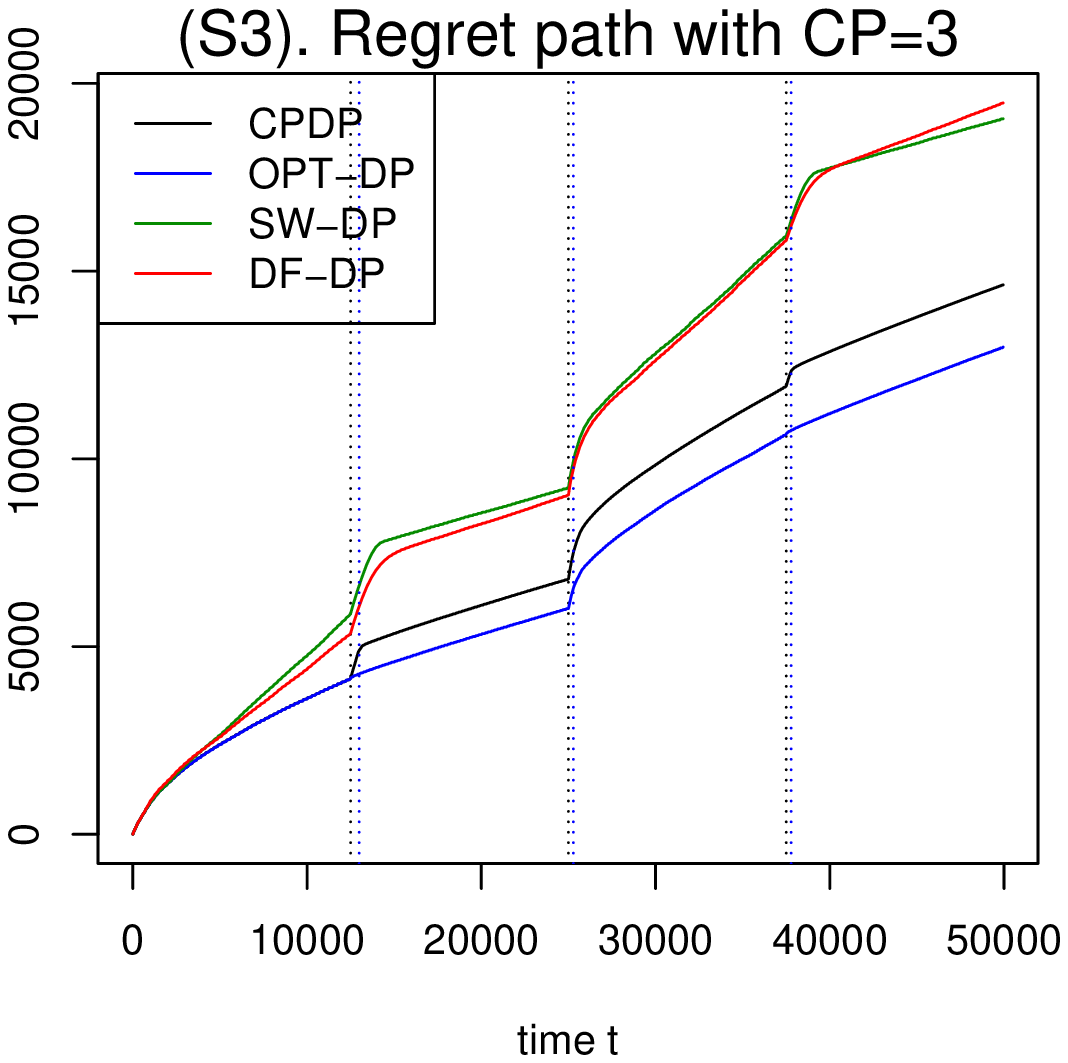}
	\vspace{-0.5cm}
    \end{subfigure}
    ~
    \begin{subfigure}{0.32\textwidth}
	\includegraphics[angle=270, width=1.2\textwidth]{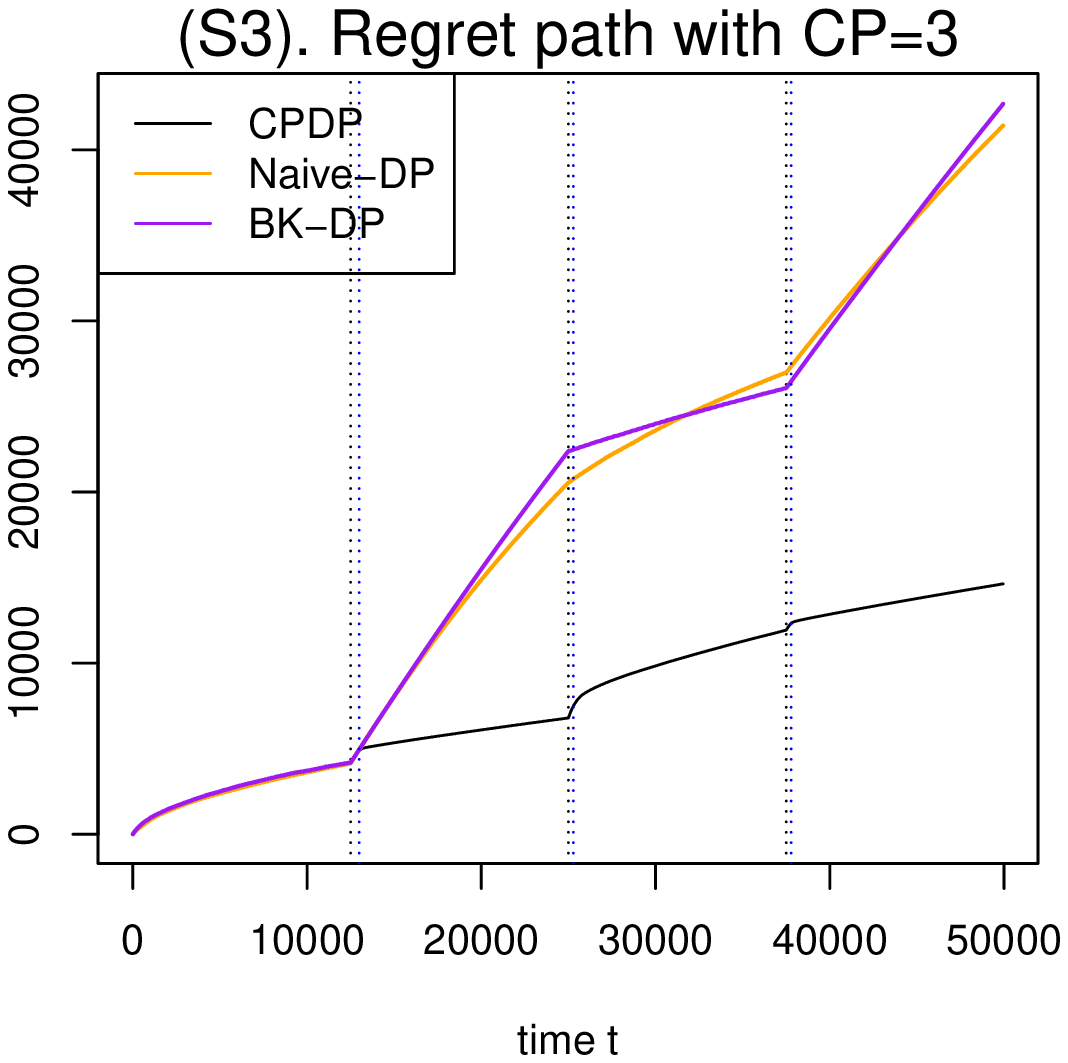}
	\vspace{-0.5cm}
    \end{subfigure}
    \caption{ Mean regret path at horizon $T=50000$ under (S1) [left] and (S3) [middle, right] with $\lambda_{\mathrm{fix}}=0.2\sqrt{\log(Td)}$. Two plots are provided under (S3) for better visualization. The vertical dotted lines (black and blue) mark the (true and estimated) change-point locations.}
    \label{fig:path}
\end{figure}

For more intuition, we further plot the regret path for each algorithm at $T=50000$. Specifically, for each algorithm, we record 100 realizations of the regret path $\{R_{T,t}^{(i)},\}_{i, t=1}^{100, T}$ at horizon $T=50000$. \Cref{fig:path} plots the mean regret path $\{\sum_{i=1}^{100}R_{T,t}^{(i)}/100\}_{t=1}^T$ by each algorithm under (S1) and (S3). The result under (S2) is similar and thus omitted to conserve space. We focus on the result for (S3). As can be seen, when a change occurs, all algorithms experience a jump on regret. For OPT-DP, the jump is purely caused by the initial estimation error after the restart. Comparing its performance on segments 1, 3 and 2, 4, it is clear that the stationary process with $\theta^{(1)}$ is more difficult to learn (i.e.,\ higher regret) than $\theta^{(2)}$. The gap between OPT-DP and CPDP is mainly caused by detection delay. SW-DP and DF-DP counter change-point by removing/discounting past observations, which is effective. However, compared to CPDP, regret accumulates at a faster rate within each stationary segment, as estimation error is larger due to less information used.

\subsubsection[]{Sensitivity Analysis of the Tuning Parameter $\lambda$:}\label{subsec:robust_lambda}
\begin{figure}[ht]
\vspace{-5mm}
\hspace*{-6mm}
    \begin{subfigure}{0.32\textwidth}
	\includegraphics[angle=270, width=1.2\textwidth]{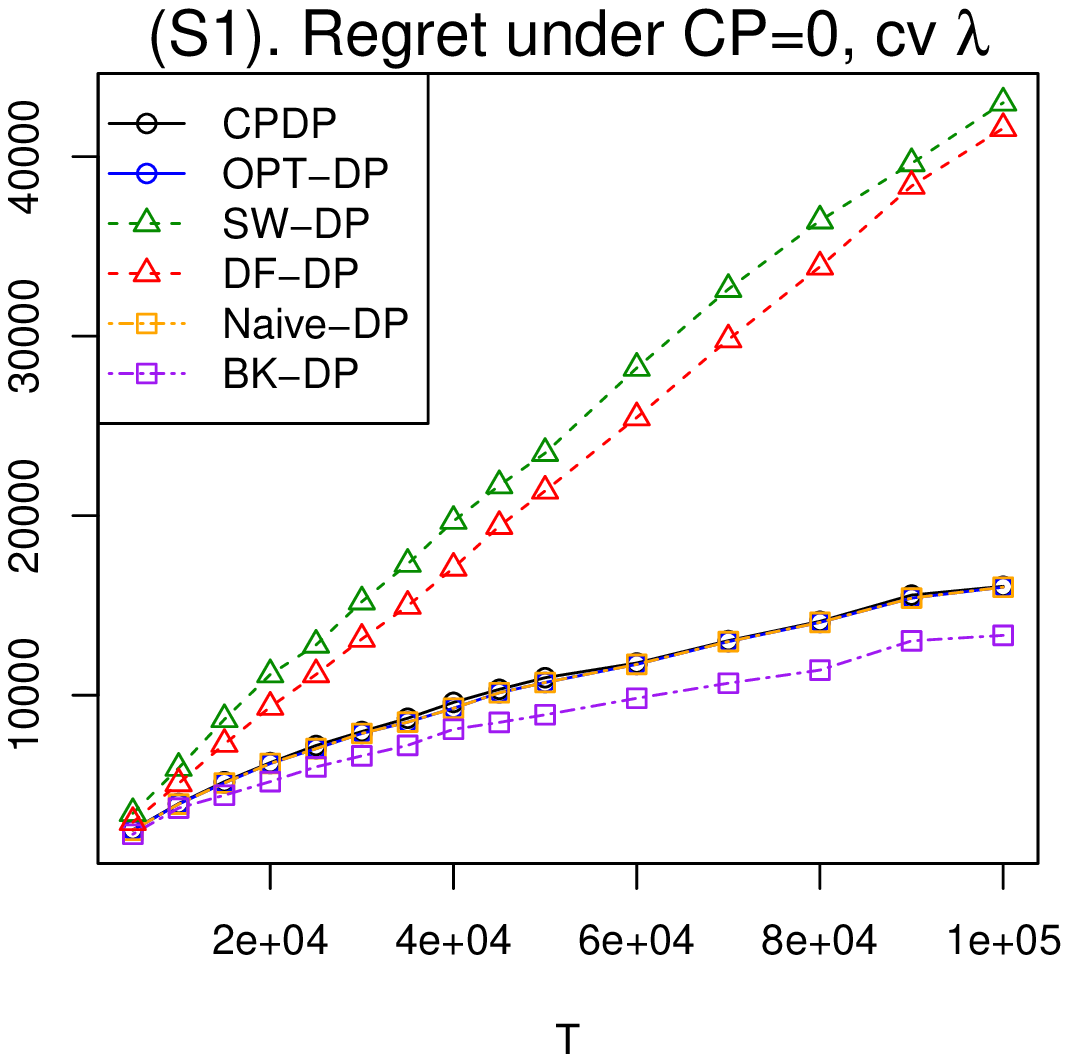}
	\vspace{-0.5cm}
    \end{subfigure}
    ~
    \begin{subfigure}{0.32\textwidth}
	\includegraphics[angle=270, width=1.2\textwidth]{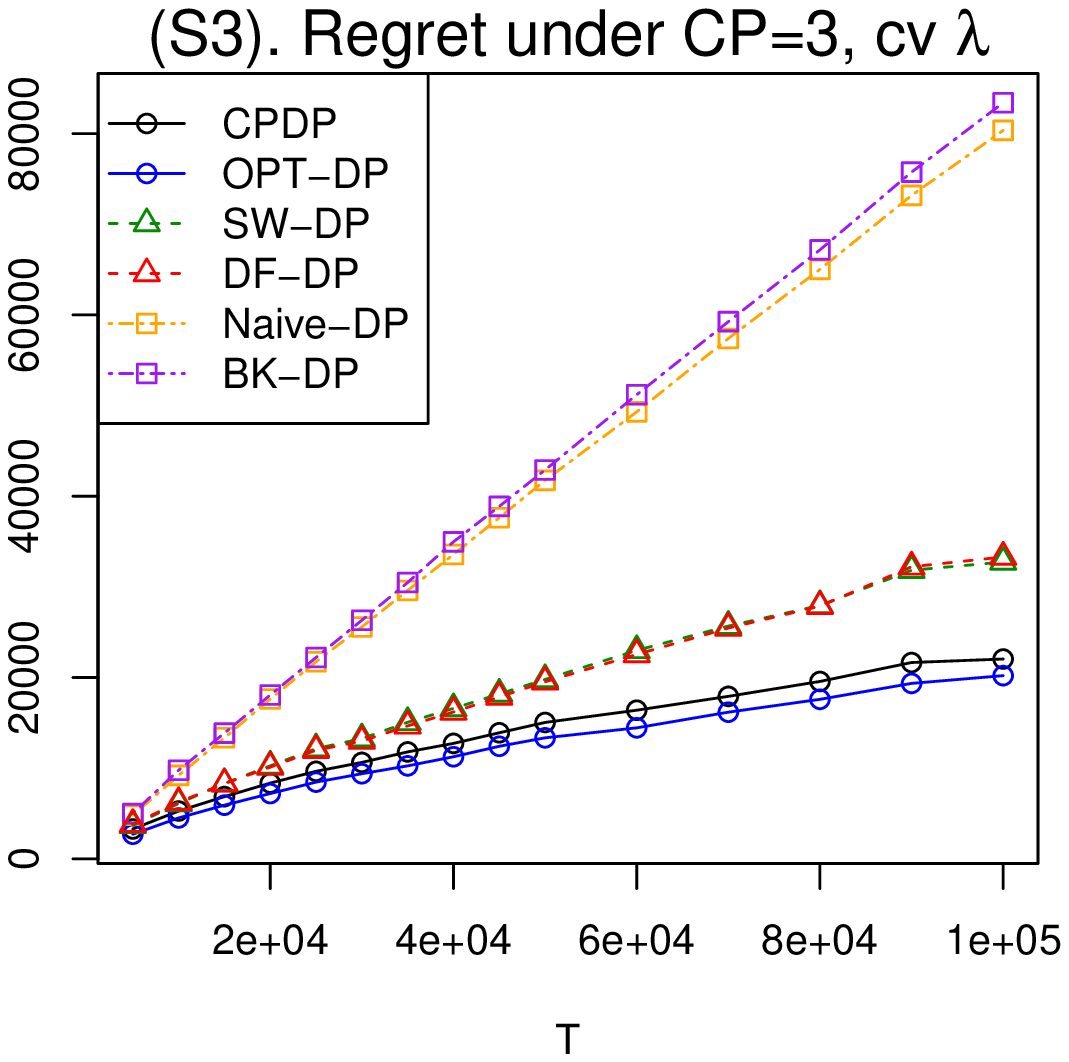}
	\vspace{-0.5cm}
    \end{subfigure}
    ~
    \begin{subfigure}{0.32\textwidth}
	\includegraphics[angle=270, width=1.2\textwidth]{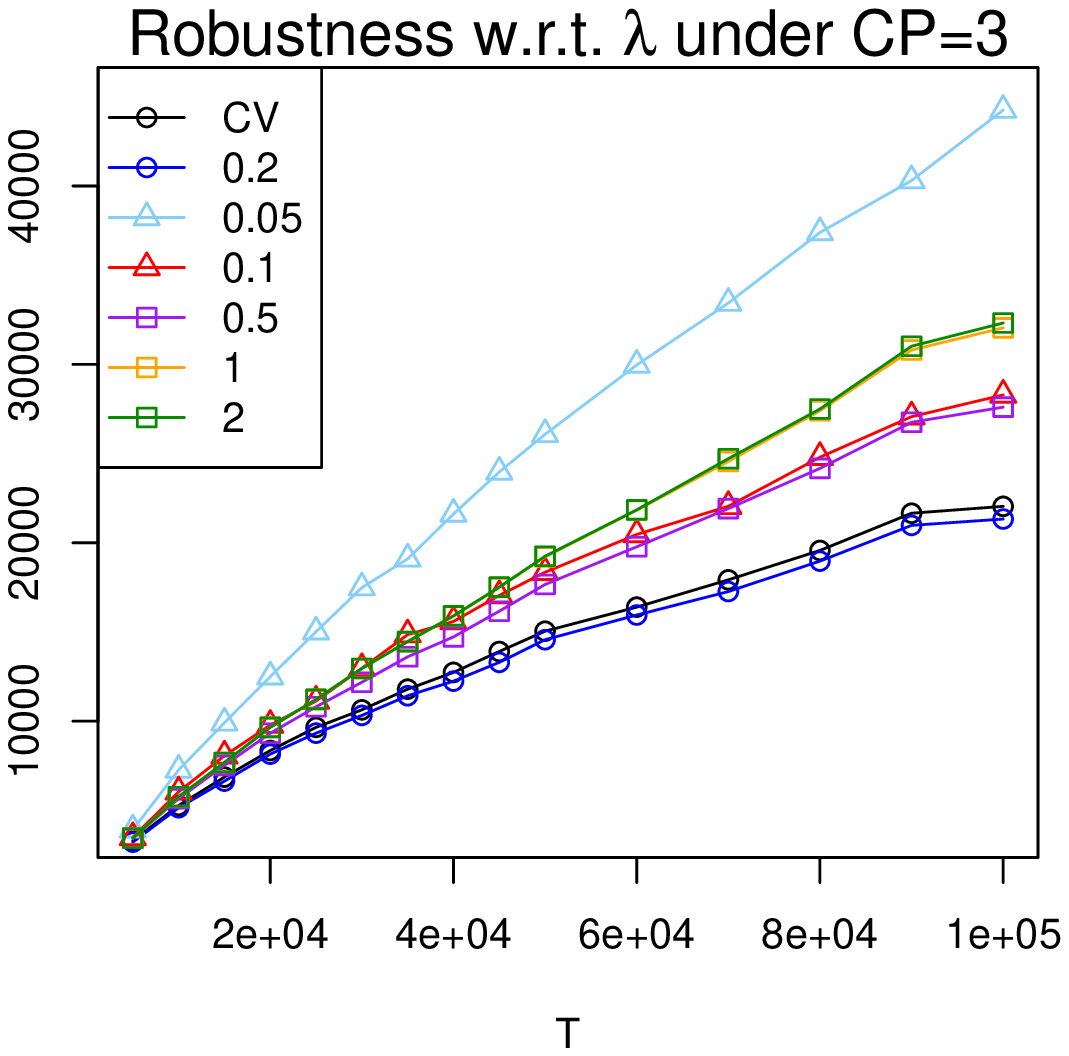}
	\vspace{-0.5cm}
    \end{subfigure}
    \caption{ [Left and middle]:  Mean regret vs.\ $T$ under (S1) and (S3) with $\lambda_{\mathrm{cv}}$. [Right]: Mean regret vs.\ $T$ by CPDP under (S3) with $\lambda_{\mathrm{cv}}$ and $\lambda_{\mathrm{fix}}=c_{\lambda}\sqrt{\log(Td)}$ at $c_{\lambda}=0.05,0.1,0.2,0.5,1,2$.}
    \label{fig:cvlambda}
\end{figure}

In this section, we further investigate the impact of Lasso tuning parameter $\lambda$ on the performance of each algorithm. \Cref{fig:cvlambda} (left and middle) reports the mean regret $\overline{R}_T=\sum_{i=1}^{100}R_T^{(i)}/100$ of each algorithm at different horizon $T$ under simulation scenarios (S1) and (S3), where $\lambda_{\mathrm{cv}}$ is used in the Lasso penalty. The result under (S2) is similar and thus omitted. Compared with \Cref{fig:fixedlambda}, the performance of each algorithm is rather stable. OPT-DP and CPDP are still the overall winners. For illustration, at $T=10000$, the minimum, 25\%, 50\%, 75\% and maximum quantiles of $\{\lambda_{\mathrm{cv}}^{(i)}\}_{i=1}^{100}$ are $0.483, 0.766, 0.848, 0.960, 1.265$, respectively, while $\lambda_{\mathrm{fix}}=0.2\sqrt{\log(Td)}=0.724.$

\Cref{fig:cvlambda} (right) further reports the mean regret of CPDP under (S3) with $\lambda_{\mathrm{cv}}$ and $\lambda_{\mathrm{fix}}=c_{\lambda}\sqrt{\log(Td)}$ at $c_{\lambda}=0.05,0.1,0.2,0.5,1,2.$ As can be seen, the performance of CPDP with $\lambda_{\mathrm{cv}}$ is comparable to the best-performing $\lambda_{\mathrm{fix}}.$ In addition, CPDP is reasonably robust to the tuning parameter $c_{\lambda}$, except for $c_{\lambda}=0.05$. In particular, an overly small $c_{\lambda}$ leads to high variance in the Lasso estimation, which in turn leads to false positives in change-point detection. For illustration, at $T=10000$, the means of the number of change-points detected at $c_{\lambda}=0.05,0.1,0.2,0.5,1,2$ are 10.79, 5.32, 3.01, 3.00, 3.00, 3.00, respectively.

\subsubsection{Performance with Varying Change Sizes:}\label{subsec:changesize}
In this section, we investigate the impact of change size on the performance of each algorithm. In particular, we modify the simulation scenarios (S2) and (S3) in \Cref{subsec:valueCPD} with $\theta^{(1)}=(0,1,1,2,2,\mathbf{0}_{44}^{\top},-0.25)^{\top}$ unchanged but replace $\theta^{(2)}$ with $\theta^{(2)*}=(0,1.5,1.5,2,2,\mathbf{0}_{44}^{\top},-0.5)^{\top}$. Denote the new scenarios as (S2)$^*$ and (S3)$^*$.

Compared to $\theta^{(2)}$, $\theta^{(2)*}$ has the same price sensitivity $0.5$ but much higher attractiveness $\mathbb E(\alpha_t^\top z_t)=3.5$ instead of 2. Thus, when switching from $\theta^{(1)}$ to $\theta^{(2)*}$, though price sensitivity is higher, due to the increased attractiveness, the optimal price may not change as much, which intuitively implies less value for change-point detection. Indeed, for $\{p_t\} \stackrel{\mbox{i.i.d.}}{\sim} \text{uniform}[\widetilde{p}_l,\widetilde{p}_u]$, the KL divergence between $\theta^{(1)}$ and $\theta^{(2)}$, $\theta^{(2)*}$ are
\begin{align*}
    \text{KL}(\theta^{(1)},\theta^{(2)})=\mathbb E_{\theta^{(1)}} \left( \log \frac{P_{\theta^{(1)}}(y_t|z_t, p_t)}{P_{\theta^{(2)}}(y_t|z_t, p_t)}\right)=0.814, \text{ and } \text{KL}(\theta^{(1)},\theta^{(2)*})=0.282.
\end{align*}
Thus, compared to $\theta^{(2)}$, it is more difficult (and also less valuable) to differentiate $\theta^{(1)}$ and $\theta^{(2)*}$.

\begin{figure}[ht]
\vspace{-5mm}
\hspace*{-6mm}
    \begin{subfigure}{0.32\textwidth}
	\includegraphics[angle=270, width=1.2\textwidth]{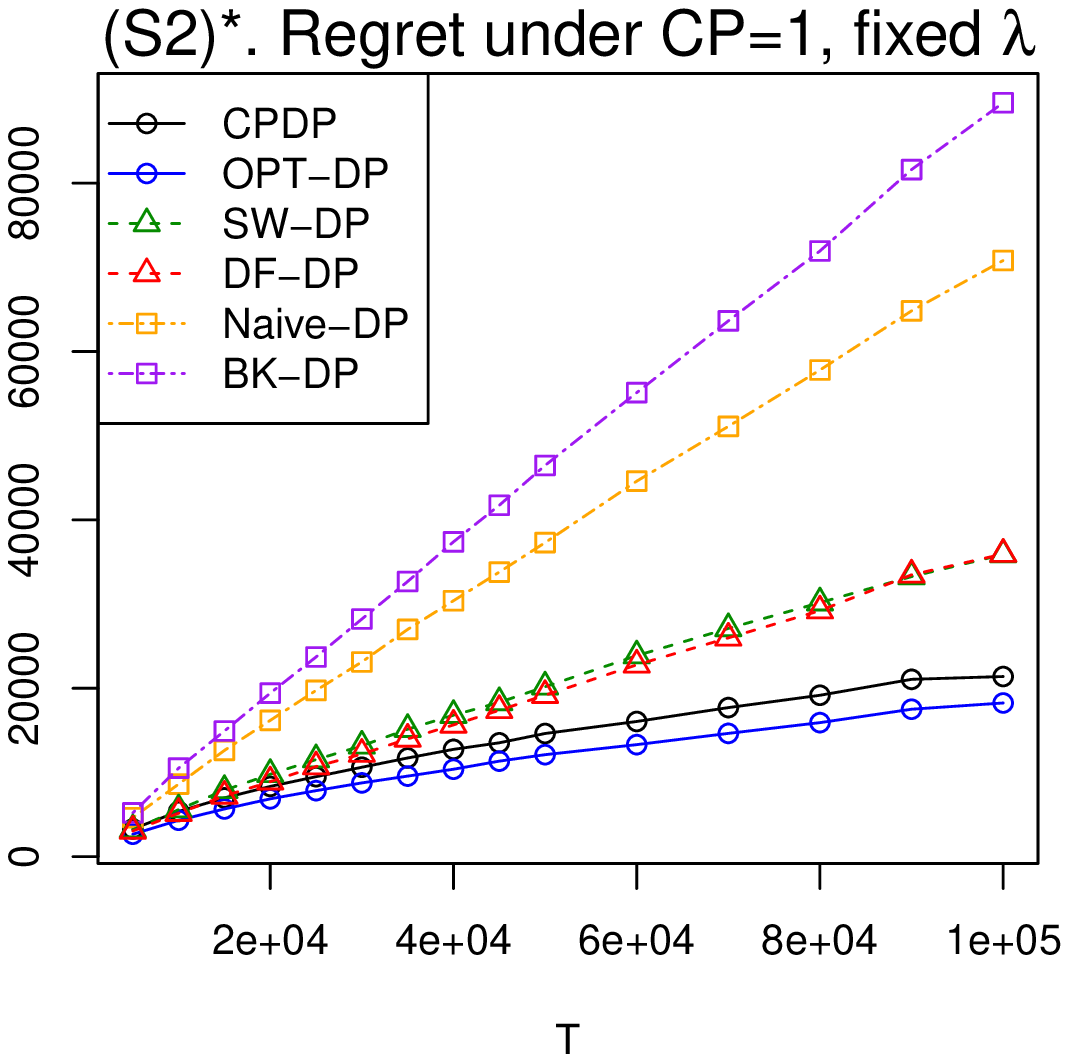}
	\vspace{-0.5cm}
    \end{subfigure}
    ~
    \begin{subfigure}{0.32\textwidth}
	\includegraphics[angle=270, width=1.2\textwidth]{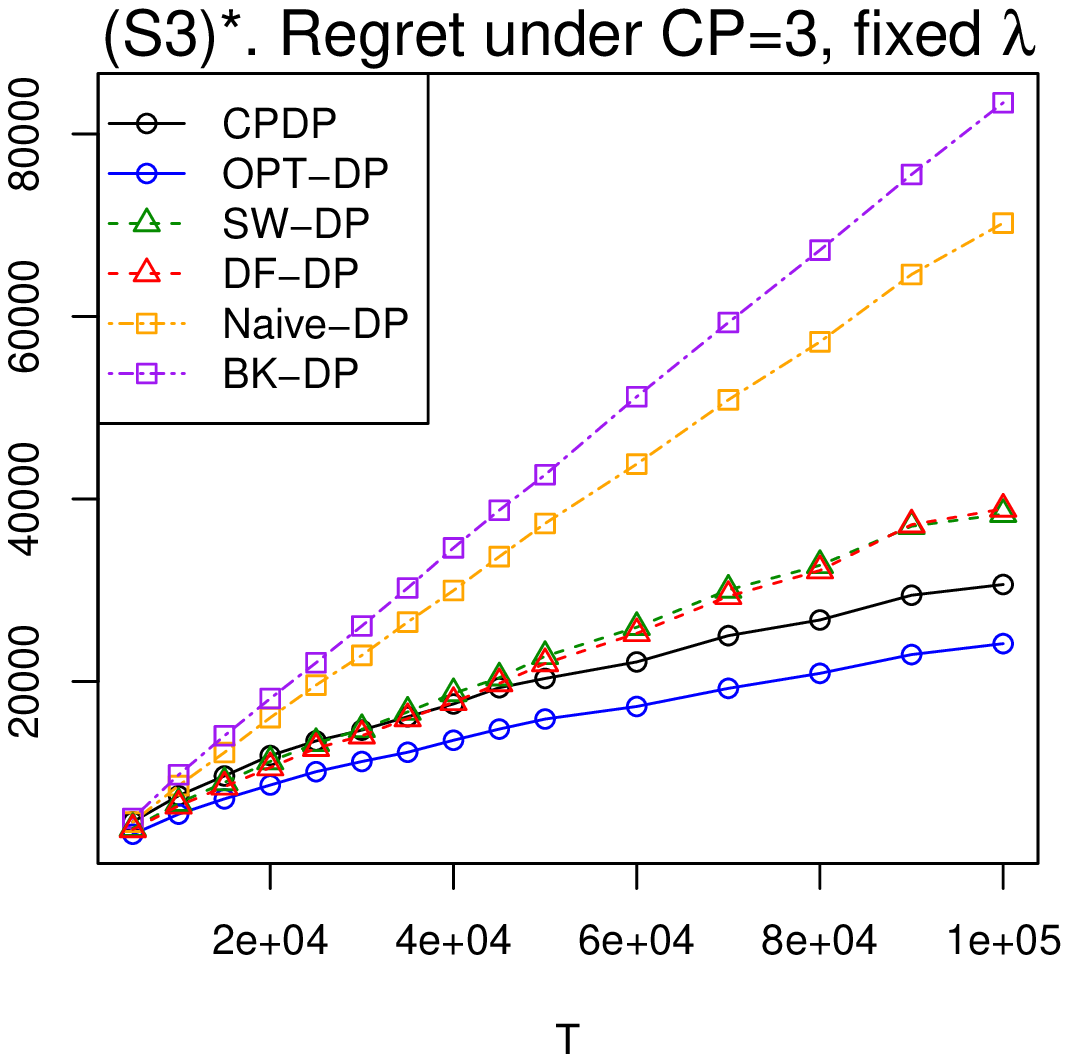}
	\vspace{-0.5cm}
    \end{subfigure}
    ~
    \begin{subfigure}{0.32\textwidth}
	\includegraphics[angle=270, width=1.2\textwidth]{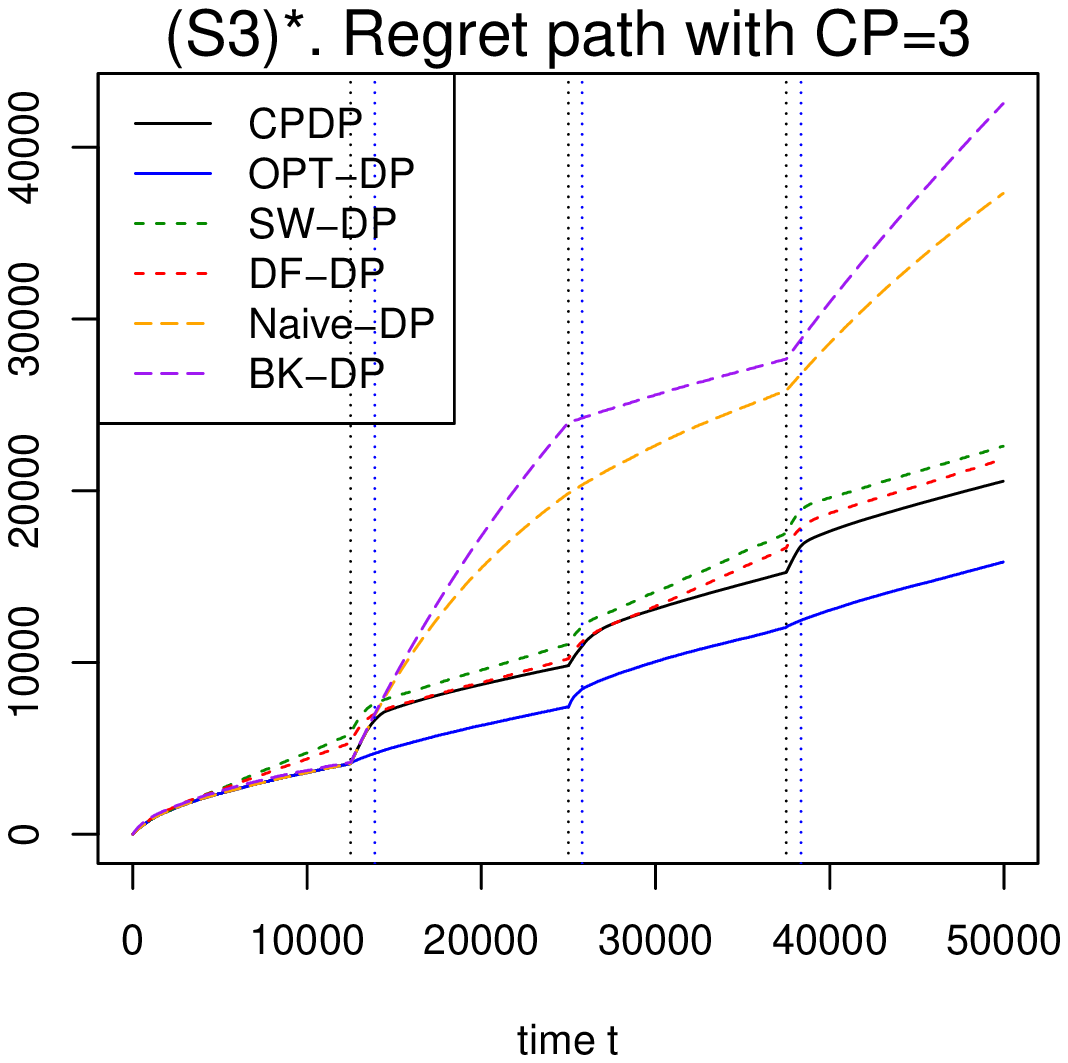}
	\vspace{-0.5cm}
    \end{subfigure}
    \caption{ [Left and middle]:  Mean regret vs.\ $T$ under (S2)$^*$ and (S3)$^*$ with $\lambda_{\mathrm{fix}}=0.2\sqrt{\log(Td)}$. [Right]: Mean regret path at horizon $T=50000$ under (S3)$^*$ with $\lambda_{\mathrm{fix}}=0.2\sqrt{\log(Td)}$. The vertical dotted lines (black and blue) mark the (true and estimated) change-point locations.}
    \label{fig:lesschange}
\end{figure}

\Cref{fig:lesschange} (left and middle) reports the mean regret $\overline{R}_T=\sum_{i=1}^{100}R_T^{(i)}/100$ of each algorithm at different horizon $T$ under simulation scenarios (S2)$^*$ and (S3)$^*$. Compared to \Cref{fig:fixedlambda}, the notable difference is that the performance gaps between CPDP and competitors are smaller. For example, at $T=50000$, for (S3), the ratio of regret by SW-DP, DF-DP, and Naive-DP with respect to CPDP are 1.32, 1.36 and 2.85, while for (S3)$^*$, the corresponding ratios are 1.12, 1.09 and 1.83. In addition, due to larger detection delay, the gaps between CPDP and OPT-DP are wider.

In particular, \Cref{tab:cpd} reports the average numbers of detected change-points across 100 experiments by CPDP under (S3) and (S3)$^*$ for different horizon $T$. As can be seen, CPDP successfully detects all change-points for (S3) but experiences some under-detection for (S3)$^*$ when $T$ is small. Denote $\widehat{\tau}_1-\tau_1$ as the detection delay for the first change-point $\tau_1=T/4$. \Cref{tab:cpd} further reports the ratio between average detection delay by CPDP under (S3)$^*$ and (S3). Compared to (S3), CPDP has around three times longer detection delay for $\tau_1$ under (S3)$^*$. This can be further seen by comparing \Cref{fig:path} (middle) and \Cref{fig:lesschange} (right), which give the mean regret path of CPDP at horizon $T=50000$ under (S3) and (S3)$^*$, respectively.

\begin{table}[ht]
\centering
\caption{Average number of detected change-points under (S3) and (S3)$^*$. ``Ratio" gives the ratio between average detection delay for the first change-point ($\tau_1$) under (S3)$^*$ and (S3).}
\label{tab:cpd}
\begin{tabular}{lccccccccccc}
\hline \hline
$T$ & 5000 & 10000 & 20000 & 30000 & 40000 & 50000 & 60000 & 70000 & 80000 & 90000 & 100000\\\hline
(S3) & 3.00 & 3.00 & 3.00 & 3.00 & 3.01 & 3.00 & 3.00 & 3.00 & 3.00 & 3.00 & 3.00 \\ 
(S3)$^*$ & 2.58 & 2.99 & 3.00 & 3.00 & 3.00 & 3.00 & 3.00 & 3.00 & 3.00 & 3.00 & 3.00 \\ 
Ratio & 2.79 & 3.02 & 2.95 & 3.64 & 3.36 & 2.70 & 3.28 & 3.00 & 3.72 & 2.67 & 3.55 \\ 
\hline \hline
\end{tabular}
\end{table}

In \Cref{subsec:halfchange}, we further modify scenarios (S1), (S2) and (S3) by replacing $\theta^{(1)}$ with $\theta^{(3)}=\theta^{(1)}/2$ and replacing $\theta^{(2)}$ with $\theta^{(4)}=\theta^{(2)}/2$, such that the change size is halved. The observation is the same: CPDP still provides the best overall performance albeit at a smaller scale of improvement.

\subsubsection[]{Scaling with respect to the Numbers of Change-Points $\Upsilon_T$:}
In this section, we investigate the performance of CPDP with respect to the number of change-points $\Upsilon_T.$ Denote $\mathbb I=10000$. We design (S4) such that 
\begin{align*}
	\text{(S4)}:  \theta_t=\begin{cases} 
		\theta^{(1)}, & t\in [k\mathbb I+1, (k+1)\mathbb I], k=0,2,4, \ldots, ~k\leq \Upsilon_T;\\
		\theta^{(2)}, & t\in [k\mathbb I+1, (k+1)\mathbb I], k=1,3,5,\ldots, ~k\leq \Upsilon_T.\\
	\end{cases}\quad
\end{align*}
By design, $T=(\Upsilon_T+1)\mathbb I$ and the process alternates between $\theta^{(1)}$ and $\theta^{(2)}$ with $\Upsilon_T$ number of change-points. We set $\lambda_{\mathrm{fix}}=0.2\sqrt{\log(Td)}$.

\begin{figure}[ht]
\vspace{-5mm}
\hspace*{16mm}
    \begin{subfigure}{0.32\textwidth}
	\includegraphics[angle=270, width=1.2\textwidth]{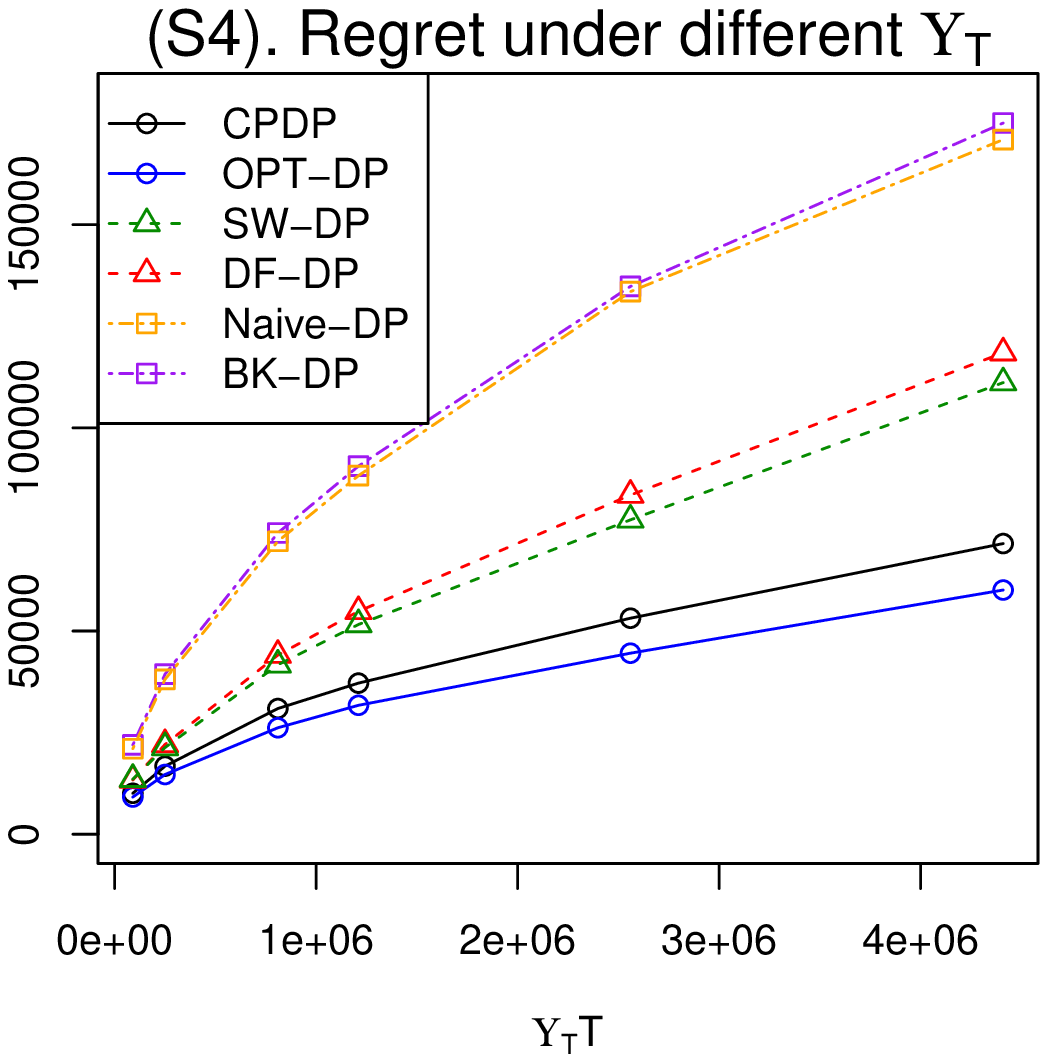}
	\vspace{-0.5cm}
    \end{subfigure}
    ~
    \begin{subfigure}{0.32\textwidth}
	\includegraphics[angle=270, width=1.2\textwidth]{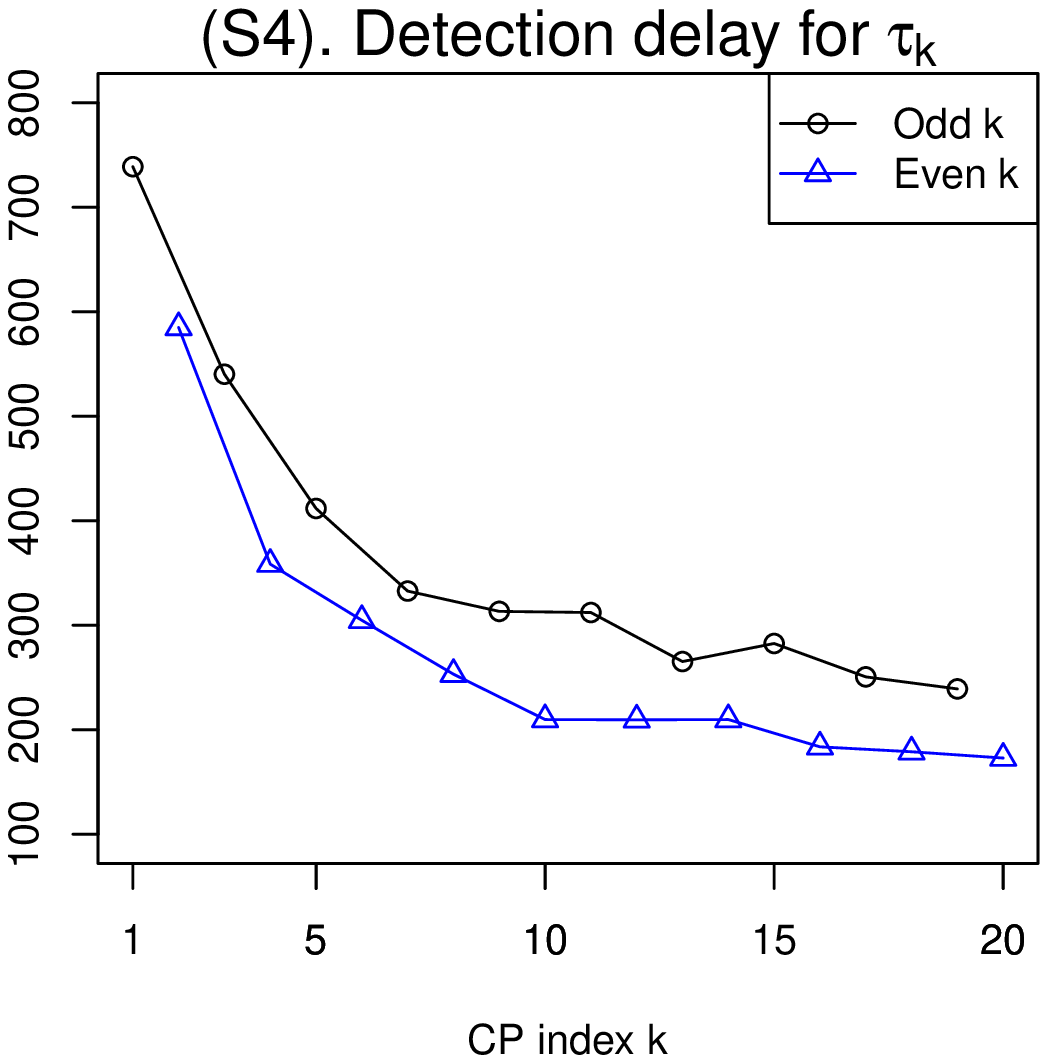}
	\vspace{-0.5cm}
    \end{subfigure}
    \caption{ [Left]:  Mean regret vs.\ $\Upsilon_TT$ for $\Upsilon_T=2,4,8,10,15,20$. [Right]: Average detection delay of $\tau_k$ for $k=1,2,\ldots, \Upsilon_T$ at $\Upsilon_T=20$.}
    \label{fig:upsilon}
\end{figure}

\Cref{fig:upsilon} (left) reports the mean regret $\overline{R}_T$ with respect to $\Upsilon_TT$ of each algorithm for $\Upsilon_T=2,4,8,10,15,20$, where the regret by CPDP resembles the shape of a square-root function. A linear regression between $\log(\overline{R}_T)$ and $\log(\Upsilon_TT)$ for CPDP gives a slope of 0.5016, providing numerical evidence that the regret by CPDP scales at the rate of $\sqrt{\Upsilon_T T}$ as in \Cref{thm:lowerbound}. \Cref{fig:upsilon} (right) plots the detection delay of $\tau_k$ (averaged across 100 experiments) for $k=1, \ldots, \Upsilon_T$ when $\Upsilon_T=20$. It is clear that the detection delay decreases with $k$, indicating the effectiveness of the accelerated exploration mechanism in \Cref{algorithm:cpd2}. Note that we separate $\tau_k$ by even and odd $k$ as the difficulty of detecting a switch from $\theta^{(1)}$ to $\theta^{(2)}$ is different than a switch from $\theta^{(2)}$ to $\theta^{(1)}$.

\subsection{Real Data Applications}\label{subsec:autoloan}
In this section, we explore practical utility of CPDP on a real-world auto loan dataset provided by the Center for Pricing and Revenue Management at Columbia University. This dataset was first studied by \cite{phillips2015effectiveness} and further used by \cite{bastani2022meta}, \cite{ban2021personalized}, \cite{luo2021distribution} to evaluate various dynamic pricing algorithms under different problem settings.

The dataset records all auto loan applications received by a major online lender in the U.S.\ from July 2002 to November 2004. We use the first 50000 loan applications for the case study. For each application, we observe information about the loan~(e.g.,\ term and amount), the prospective consumer~(e.g.,\ FICO score), and the economic environment~(e.g.,\ prime rate). We refer to \Cref{subsec:addrealdata} for detailed description of the dataset. We also observe the monthly payment required for the approved loan, which can be viewed as the pricing decision by the company, and whether the offer was accepted -- a binary purchasing decision by a consumer.

We follow the feature selection result in \cite{luo2021distribution} and \cite{bastani2022meta}, and consider the loan amount approved, term, prime rate, the competitor's rate and FICO score as covariates with non-zero coefficients, and treat the rest nine features in the dataset as zero/null covariates. The price $p$ of a loan is computed as the net present value of future payment minus the loan amount, i.e.,\ $p=\text{Monthly Payment} \times \sum_{i=1}^{\text{Term}}(1+\text{Rate})^{-i} - \text{Loan Amount}.$ We set Rate as $0.12\%$, an average of the monthly London interbank offered rate for the studied time period. Denote the $t$th consumer decision as $y_t$, non-zero covariates as $z_t^*$, all covariates as $z_t$, and loan price as $p_t.$ For convenience, we use one thousand dollars as the basic unit for loan price and loan amount. %We have $z_t^*\in \mathbb R^{5}, z_t\in\mathbb R^{14}$. 

Note that it is impossible to obtain consumers' real online responses to any dynamic pricing strategy unless it was used in the system when the data were collected. Thus, following the literature, we first estimate the demand model based on the entire 50000 observations and use it as the ground truth to generate consumer responses. Specifically, we run a standard model selection based offline change-point detection algorithm~\citep[e.g.,][]{davis2006structural, bai2003computation} on $\{(y_t,z_t^*,p_t)\}_{t=1}^{50000}$, which returns three change-points at $t=12916, 24986, 37054.$ We refer to \Cref{subsec:addrealdata} for a more detailed description of the offline change-point detection algorithm. \Cref{tab:loan} gives the estimated logistic regression based on each stationary segment, where notable changes (such as sensitivity to price and competitor's rate) can be observed. As discussed in the \Cref{sec-intro}, this confirms the informal change-point analysis in \cite{besbes2011minimax}.

\begin{table}[ht]
\centering
\caption{Estimated logistic regression on each stationary segments. Coefficients for FICO and Term are in the scale of $\times 10^{-3}$ and $\times 10^{-2}$.}
\label{tab:loan}
\begin{tabular}{lrrrrrrr}
  \hline\hline
Segment & Intercept & Price & FICO & Competitor Rate & Amount & Prime Rate & Term \\ 
  \hline
$[1,12916]$ & -7.63 & -0.42 & 3.20 & 1.83 & -0.11 & -2.00 & 1.47 \\ 
$[12917,24986]$ & -12.29 & -0.60 & 0.35 & 2.69 & -0.07 & -0.63 & 1.43 \\ 
$[24986,37054]$ & -12.81 & -1.27 & -7.24 & 4.77 & -0.02 & -1.92 & 2.89 \\ 
$[37054,50000]$ & -7.66 & -0.88 & -2.37 & 2.77 & -0.09 & -1.02 & 0.57 \\ 
   \hline\hline
\end{tabular}
\end{table}

We run all algorithms based on the ground truth model in \Cref{tab:loan}. For the Lasso tuning parameter $\lambda$, we use both $\lambda_{\mathrm{fix}}=0.02\sqrt{\log(Td)}$ and $\lambda_{\mathrm{cv}}$. We set the price experiment set $\widetilde{\mathcal{P}}$ as $[\widetilde{p}_l, \widetilde{p}_u]=[0.2,6]$ and set the price bound as $[p_l,p_u]=[0,14].$ For reference, the min, max, median and mean prices given by the company are $0.2, 13.5, 2.6$ and $2.9$ in the studied period. We remark that the result is robust to the choice of $[\widetilde{p}_l, \widetilde{p}_u]$. For each algorithm, we conduct 100 experiments and record regret $\{R_T^{(i)}\}_{i=1}^{100}$. Note that for all experiments, we keep the sample path of the covariates $\{z_t\}_{t=1}^{50000}$ fixed as that in the dataset.

\begin{figure}[ht]
\vspace{-5mm}
\hspace*{-6mm}
    \begin{subfigure}{0.32\textwidth}
	\includegraphics[angle=270, width=1.2\textwidth]{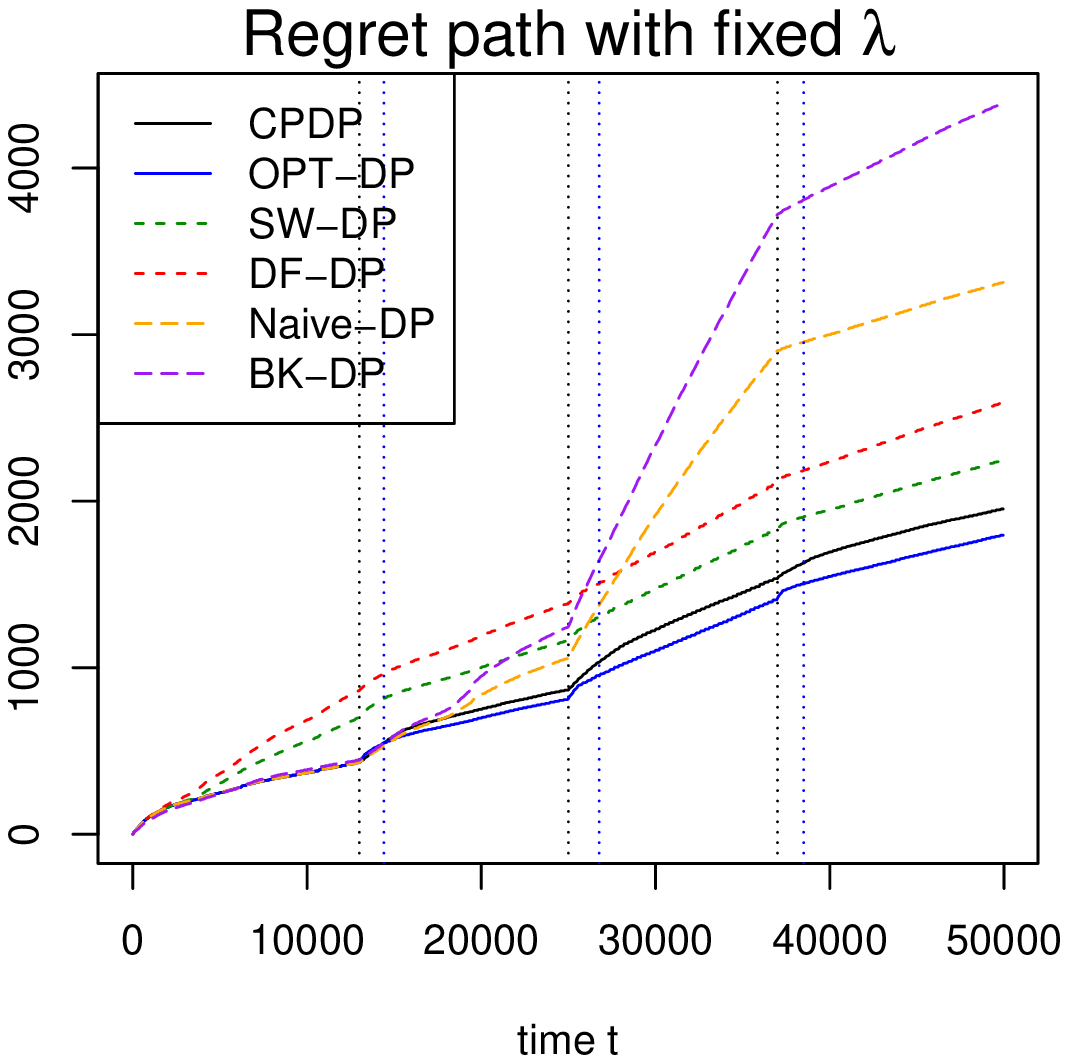}
	\vspace{-0.5cm}
    \end{subfigure}
    ~
    \begin{subfigure}{0.32\textwidth}
	\includegraphics[angle=270, width=1.2\textwidth]{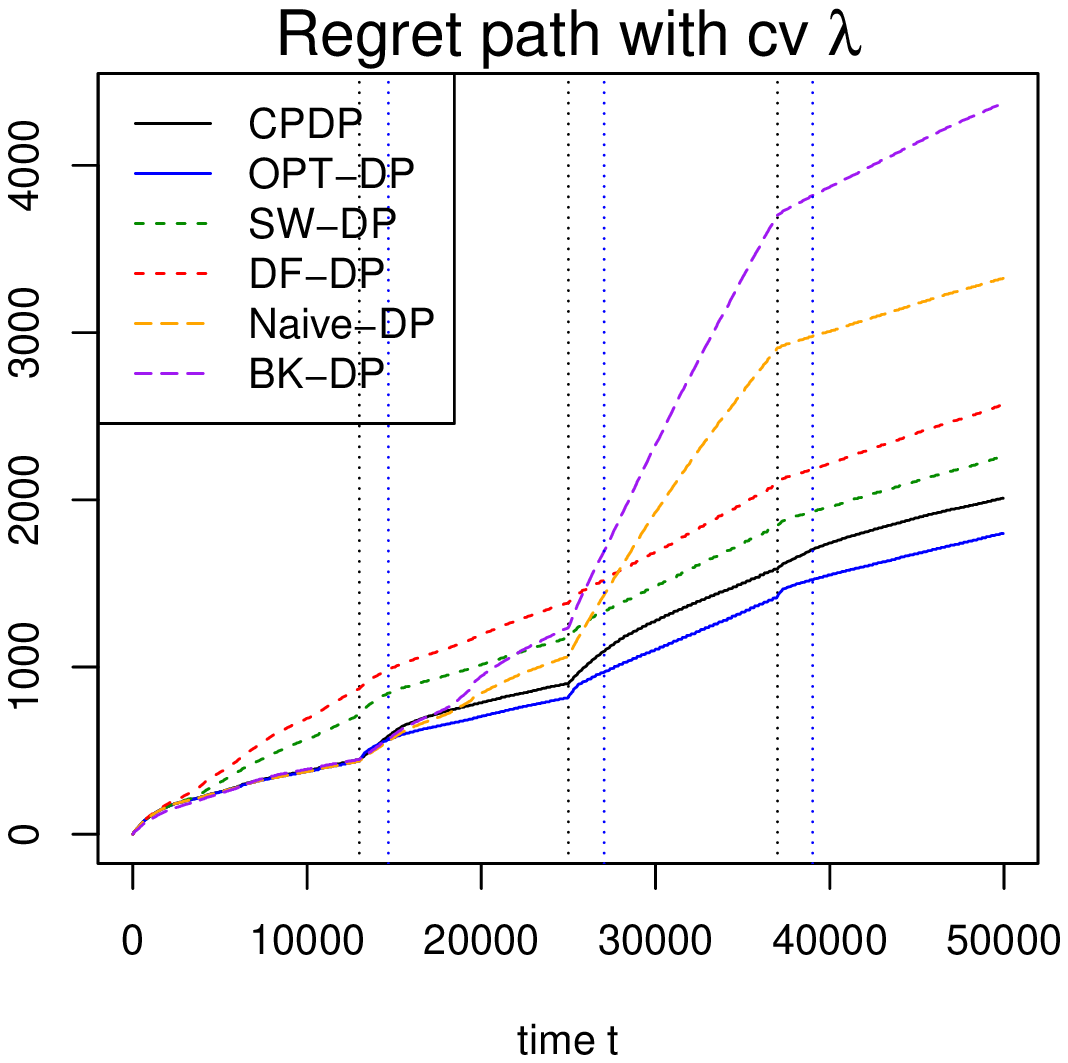}
	\vspace{-0.5cm}
    \end{subfigure}
    ~
    \begin{subfigure}{0.32\textwidth}
	\includegraphics[angle=270, width=1.2\textwidth]{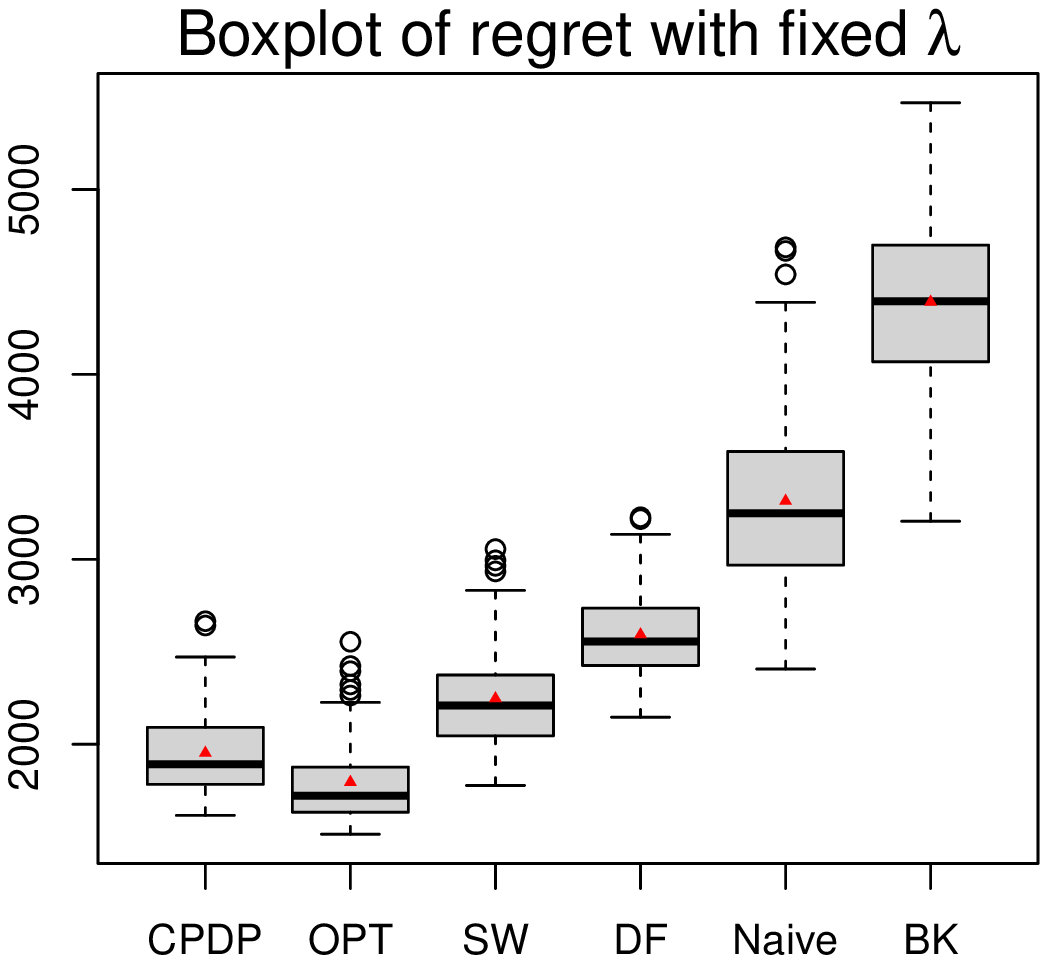}
	\vspace{-0.5cm}
    \end{subfigure}
    \caption{ [Left and middle]: Mean regret path with $\lambda_{\mathrm{fix}}=0.02\sqrt{\log(Td)}$ and $\lambda_{\mathrm{cv}}$. The vertical dotted lines (black and blue) mark the (true and estimated) change-point locations. [Right]: Boxplot of regret based on 100 experiments with $\lambda_{\mathrm{fix}}=0.02\sqrt{\log(Td)}$. Red dots indicate mean.}
    \label{fig:autoloan}
\end{figure}

\Cref{fig:autoloan} (left and middle) give the mean regret path of each algorithm based on $\lambda_{\mathrm{fix}}$ and $\lambda_{\mathrm{cv}}$, where the results are almost identical, indicating the robustness of the finding. For reference, the minimum, 25\%, 50\%, 75\% and maximum quantiles of $\{\lambda_{\mathrm{cv}}^{(i)}\}_{i=1}^{100}$ are $0.0119, 0.106, 0.164, 0.221$ and $0.525$, respectively, while $\lambda_{\mathrm{fix}}=0.02\sqrt{\log(Td)}=0.0735.$ 

\Cref{fig:autoloan} (right) further gives the boxplot of $\{R_T^{(i)}\}_{i=1}^{100}$ for each algorithm, where CPDP and OPT-DP are the clear overall winners. In particular, the mean regret (in thousands) by CPDP, OPT-DP, SW-DP, DF-DP, Naive-DP and BK-DP are 1953, 1795, 2247, 2593, 3315 and 4392. For reference, the regret incurred by the company pricing policy is 7206. Thus, compared to Naive-DP and BK-DP, CPDP reduces revenue loss by 1.3$\sim$2.3 million dollars, indicating the importance of handling non-stationarity in dynamic pricing. The expected revenue by optimal pricing is 25078. It is easy to see that CPDP lifts revenue by 29\%, 11.4\% and 6.1\% when compared with the company policy, BK-DP and Naive-DP.

\section{Conclusion and Discussion}\label{sec:conclusion}
%\textcolor{blue}{temporal dependence, robust, (contamination, non sub Gaussian...), quantile model, multiple datasets. other methodology (CPT in MNL), applications}

In this paper, we study a dynamic pricing problem under non-stationarity, where the demand model is a GLM with high-dimensional contextual information. We propose the CPDP algorithm, which achieves a near-optimal regret of order $O(s\sqrt{\Upsilon_T T}\log(Td)).$ To our best knowledge, CPDP is the first algorithm in the dynamic pricing literature that handles non-stationary demand models with covariates. Compared to existing works, CPDP does not require the knowledge of a lower bound on the change size, and its regret scales optimally with the number of change-points $\Upsilon_T$, owing to a novel accelerated exploration mechanism. As an important byproduct, we develop an online change-point detection algorithm for high-dimensional GLMs, which on its own, is optimal and novel in the online change-point detection literature.  

In this paper, we assume   full access to  consumer-level transaction data, which may include sensitive information such as consumer demographics. With the increasing awareness of privacy, it would be interesting and timely to develop a dynamic pricing policy that can handle non-stationarity subject to privacy constraints. We leave this for future research.

\setlength{\bibsep}{0.2pt plus 1ex}
\vspace{0.2cm}
\begin{spacing}{1.14}
\bibliographystyle{informs2014}
\bibliography{reference}
\end{spacing}

\clearpage

\setcounter{page}{1}

\setcounter{equation}{0}
\renewcommand\theequation{EC.\arabic{equation}}

\setcounter{section}{0}
\renewcommand\thesection{\Alph{section}}

\renewcommand\thelemma{EC.\arabic{lemma}}

 \begin{center}
{\Large Online Appendix\\
\vspace{.1in}
\large High-Dimensional Dynamic Pricing under Non-Stationarity: Learning and Earning with Change-Point Detection}
\end{center}
\vspace{.1in}

The Appendix is organized as follows.  \Cref{sec:prop_revenue_func} derives useful properties of the revenue function, which are used throughout the proof. \Cref{sec:GLM} and \Cref{sec:GLM_mix} provide estimation error bounds of the Lasso estimator for a single GLM and a mixture of two GLMs, respectively. \Cref{sec:proof_prop} proves the theoretical guarantee of the change-point detector given in \Cref{prop:cp}. \Cref{sec:proof_regret} proves the regret upper bound of CPDP in \Cref{thm:regret} and \Cref{cor_regret}. \Cref{sec:proo_lower} proves the regret lower bound in \Cref{thm:lowerbound}. \Cref{sec:add_simu} gives additional numerical results. % Note we assume $p_u\leq C_b$ throughout the appendix. \textcolor{red}{move the last sentence somewhere else.}

The estimation error bounds derived in \Cref{sec:GLM} are used in \Cref{sec:proof_prop} for establishing theoretical guarantees (i.e.,\ false alarm, detection delay) of the online change-point detection algorithm. The estimation error bounds in \Cref{sec:GLM}, \Cref{sec:GLM_mix} and the theoretical guarantees of the online change-point detection algorithm in \Cref{sec:proof_prop} are further used in \Cref{sec:proof_regret} for establishing the regret upper bound of the CPDP algorithm.

\section{Properties of the Revenue Function}\label{sec:prop_revenue_func}
This section derives two useful properties of the revenue function. The first property shows  $\varphi$ is Lipschitz as promised in \eqref{C_phi}, and the second property suggests the revenue loss (instant regret) can be bounded above by  the squared price differences. 
\subsection[]{Lipschitz property of $\varphi$}
\begin{lemma}\label{lem_punique}
Under \Cref{assum_model}, there exists a continuous function ${\varphi}(u,v)$ such that
$$p_t^*={\varphi}(\alpha_t^{\top}z_t,\beta_t),$$
for all $t\in\{1,2,\cdots,T\}.$
   %For all $\theta_t,z_t$ that satisfy Assumption \ref{assum_model}, we have $$p_t^*=\widetilde{\varphi}(\theta_t,z_t)={\varphi}(\alpha_t^{\top}z_t,\beta_t),$$ where ${\varphi}(u,v):\mathbb{U}\times \mathbb{V}\mapsto [p_l,p_u]$ is a continuous function with $\alpha_t^{\top}z_t\in \mathbb{U}\subset \mathbb{R}$, and $\beta_t\in \mathbb{V}\subset \mathbb{R}$.
\end{lemma}

By Lemma \ref{lem_punique},  $p_t^*$ is a continuous function of $\alpha_t^{\top}z_t$ and $\beta_t$. Under Assumption \ref{assum_model}, we further have $|z_t^\top \alpha_t| \leq \|z_t\|_\infty \|\alpha_t\|_1 \leq C_b C_\theta$ and $|\beta_t|\leq C_\theta$. Together with \Cref{lem_punique}, this implies the Lipschitz continuity in \eqref{C_phi}, as a continuous function is Lipschitz continuous on a compact set. Throughout the Appendix, with a slight abuse of notation, we further denote $p_t^*={\varphi}(\alpha_t^{\top}z_t,\beta_t)={\varphi}(\theta_t,z_t)$. 
\medskip

%By Assumption \ref{assum_model}(iii), the revenue function $r(p_t,\theta_t,x_t)$ is continuously differentiable w.r.t.\ $p_t$ .  In particular, given $z_t$ and $\theta_t$, the optimal price $p_t^*$ solves the following first order condition,
%\begin{flalign*} 
%\psi'(\alpha_t^{\top}z_t+\beta_tp_t^*)+\beta_tp_t^*\psi''(\alpha_t^{\top}z_t+\beta_tp_t^*)=0.  %\end{flalign*}

\noindent\textbf{Proof of \Cref{lem_punique}}:
    Recall that  $r(p_t, \theta_t, z_t)= p_t \mathbb E[y_t|p_t, z_t]= p_t\psi'(z_t^\top \alpha_t+\beta_tp_t).$ By definition,
    \begin{align*}
        p_t^*=\argmax_{p\in[p_l,p_u]} p\cdot \psi'(\alpha_t^\top z_t +\beta_t p).
    \end{align*}

Thus, $p_t^*$ depends on $\theta_t$ and $z_t$ via $\alpha_t^{\top} z_t$ and $\beta_t$. By Assumption \ref{assum_model}(iv), for any $\theta_t$ and $z_t$, $p_t^*$ is unique. Thus, $p_t^*$ is in fact a function of $\alpha_t^{\top} z_t$ and $\beta_t$ and can be written as $p_t^*={\varphi}(\alpha_t^{\top}z_t,\beta_t)$. We next prove that $ {\varphi}(u,v)$ is continuous. 

To apply Berge's Maximum Theorem \citep{berge1957two}, we first define $\Gamma(u,v)=[p_l,p_u]$ for any $u,v$. Note that $\Gamma(u,v)$ is a set-valued function and is continuous at any $(u,v)$. By the definition of $\varphi(u,v)$, we have that
    \begin{align*}
        \varphi(u,v)=\argmax_{p\in[p_l,p_u]} p\cdot \psi'(u +v p)=\argmax_{p\in\Gamma(u,v)} p\cdot \psi'(u +v p).
    \end{align*}

By \Cref{assum_model}(iii), $p\cdot \psi'(u +v p)$ is a continuous function of $(p,u,v)$. Together with the fact that $\Gamma(u,v)$ is continuous at any $(u,v)$, by Berge's Maximum Theorem, we have that $\varphi(u,v)$ (as a set-valued function) is upper hemicontinuous. However, by \Cref{assum_model}(iv), $\varphi(u,v)$ is indeed a function. Thus, by the definition of upper hemicontinuity, we have that $\varphi(u,v)$ is continuous. \qed

    %For all $\theta_t,z_t$, let $u=\alpha_t^{\top}z_t$, $v=\beta_t$, then for  $(u,v)\in \mathbb{U}\times \mathbb{V}$, we have  $r(p,\theta_t,z_t)=\widetilde{r}(p,u,v)$.   Define a correspondence (set-valued function) $\Gamma$ from $\mathbb{U}\times \mathbb{V}$ to $[p_l,p_u]$ such that for any $(u,v)$, $\Gamma(u,v)=[p_l,p_u]$. Clearly, $\Gamma$  is continuous, compact and non-empty for all $(u,v)\in \mathbb{U}\times \mathbb{V}$. 
      
   %Define $h(u,v)=\max_{p\in\Gamma(u,v)} r(p,u,v)$, and hence $\widetilde{\varphi}(u,v)=\{p\in\Gamma(u,v): h(u,v)=r(p,u,v)\}$. Then, by Berge's Maximum Theorem \citep{berge1957two}, we know that ${\varphi}(u,v)$ is upper hemi-continuous. In addition, by Assumption \ref{assum_model}(iv),  we know that ${\varphi}(u,v)$ is indeed a function, hence it is continuous. 

\subsection{Bound instant regret by square of price difference}
As another consequence of Assumption \ref{assum_model},  given $\theta_t$ and $z_t$, for any fixed price $p_t$, 
we have 
\begin{align*}
r(p_t, \theta_t, z_t)=r(p_t^*, \theta_t, z_t)+\underbrace{\frac{\partial r(p_t^*, \theta_t, x_t)}{\partial p} }_{=0}(p_t-p_t^*) +\frac{1}{2} \frac{\partial^2 r(\widetilde{p}, \theta_t, z_t)}{\partial p^2} (p_t-p_t^*)^2,
\end{align*}
for some $\widetilde{p}$ lies between $p_t$ and $p_t^*$ with $p_t^*={\varphi}(\theta_t, z_t)$. This implies the instant regret is quadratic in terms of the price difference between $p_t$ and optimal price $p_t^*$.  

Furthermore,  \Cref{assum_model} also implies that  
\begin{equation}\label{C_r}
   \sup_{\|z\|_{\infty}\leq C_b, p\in [p_l, p_u],\|\theta\|_1\leq C_\theta}\left|\frac{1}{2}\frac{\partial^2 r(p,\theta,z)}{\partial p^2}\right|= C_r<\infty.
\end{equation}
Therefore, 
\begin{flalign}\label{bound_price}
    r(p_t^*,\theta_t,z_t)-r(p_t,\theta_t,z_t)\leq C_r({\varphi}(\theta_t, z_t)-p_t)^2.
\end{flalign}

% A direct consequence  is that 
% \begin{equation}\label{C_phi1}
%     |\varphi(\theta^{(1)},z)-\varphi(\theta^{(2)},z)|\leq C_{\varphi}\left\{|(\alpha^{(1)}-\alpha^{(1)})^{\top}z|+|\beta^{(1)}-\beta^{(2)}|\right\}.
% \end{equation}

\section{GLM Estimation under Stationarity}\label{sec:GLM}
%\zifeng{ToDo: 1. Track the constants}
This section provides results of the Lasso estimator for a single GLM. In particular, 
\Cref{lem:DB} provides a deviation bound for the first order derivatives of the log-likelihood function and \Cref{lem:RE} derives the restricted strong convexity. Based on \Cref{lem:DB} and \Cref{lem:RE}, the estimation error bounds of the Lasso estimator is established in \Cref{lem:est}.

Recall the parameter space $\Theta=\{\theta\in\mathbb{R}^d:\|\theta\|_1\leq C_{\theta}\}$ for some $C_{\theta}>0.$ Without loss of generality, we assume $C_{\theta}\geq 1$, and $p_u\leq C_b$.  Note that $\|\theta\|_1\leq C_{\theta}$ implies $\|\theta\|_2\leq C_{\theta}$ and $\langle x_t, \theta \rangle \leq \|x_t\|_\infty \|\theta\|_1 \leq C_b C_{\theta}$, where  $x_t=(z_t^{\top},p_t)^{\top}$. We denote  $\psi''_u=\sup_{|x|\leq (C_{\theta}+1)C_b} \psi''(x) $, and $\psi''_l=\inf_{|x|\leq C_bC_{\theta}} \psi''(x)$ such that $0<\psi''_l\leq \psi''_u<\infty$. For any $I\subset \{1,2,\ldots,T\}$, define $\nabla L(\theta, I)=\sum_{t \in I}\{\psi'(x^{\top}_t\theta) - y_t\}x_t$ as the first order derivative of $L(\theta, I) = \sum_{t \in I} \{\psi(x^{\top}_t\theta) - y_t x^{\top}_t\theta\}$ w.r.t.\ $\theta.$ Note that in this section, we assume the true model parameter is constant over $I$, i.e., $\theta_t\equiv\theta^*$, for all $ t\in I$. 

\subsection{Deviation bound}

%\textcolor{red}{($\theta^*$ needs to be a function of $t$.  need to specify $\Theta$)}
% \begin{ass}\label{assum1}
%     (i) The covariate $x_t\in \mathbb{R}^d$ are i.i.d.\ random vectors with $\|x_t\|_\infty < b.$ (ii) The true parameter $\theta^* \in \Theta$ and has support $S\subseteq \mathbb{R}^d$ with cardinality $|S|=s.$
% \end{ass}
% and $\lambda_{max}(\Sigma)=\sigma_u<\infty.$

\begin{lemma}\label{lem:DB}
Under Assumptions \ref{assum_model} and \ref{assum_moment}, we have that 
$$\mathbb{P}\left( \|\nabla L(\theta^*, I)\|_\infty \geq  2C_b\sqrt{2\psi''_u|I|\log(Td)} \right)\leq 2d^{-3}T^{-4},$$
for all $|I|\geq {8\log(Td)}/({\psi''_uC_b^2}).$
\end{lemma}

\noindent\textbf{Proof of \Cref{lem:DB}}:
For $j=1,2,\cdots, d$, denote $v_{tj}=\{\psi'(x^{\top}_t\theta^*) - y_t\}x_{tj}$. Conditioning on $\{z_t,t\in I\}$, $y_t$ is drawn from model \eqref{model_GLM} with parameter $\theta^*$ and uniformly and independently sampled $p_t\in[\widetilde{p}_l,\widetilde{p}_u]$. In the following, we use Chernoff's method, see e.g.,  Chapter 2 in \ \cite{wainwright2019high}, to bound $\sum_{t\in I} v_{tj}.$

For any $u\in \mathbb{R}$, by the property of the exponential family \citep{agresti2015foundations}, the cumulant function takes the form 
\begin{align}\label{eq_cumu}
\begin{split}
    \log \mathbb{E} [\exp(uv_{tj}) |x_t]& = \log \left\{\mathbb{E} [\exp(-uy_tx_{tj}) |x_t] \exp [\psi'(x^{\top}_t\theta^*)ux_{tj}] \right\} \\
    &=\psi(x_t^\top \theta^*-ux_{tj})-\psi(x_t^\top \theta^*)+\psi'(x^{\top}_t\theta^*)ux_{tj}\\
    &=\frac{1}{2}\psi''(x_t^\top \theta^*-a_tux_{tj}) (ux_{tj})^2,
\end{split}
\end{align}
for some $a_t \in (0,1)$, where the last equality follows from Taylor's expansion.

In the following, consider $|u|< 1.$ Thus, we have $|x_t^\top \theta^*-a_tux_{tj}|<C_b(C_{\theta}+1).$ Recall $\psi''_u=\sup_{|z|< C_b(C_{\theta}+1)} \psi''(z) $, we have that
\begin{align*}
    \frac{1}{|I|}\sum_{t\in I} \log \mathbb{E} [\exp(uv_{tj}) |x_t] \leq \frac{1}{2} \psi''_u C_b^2u^2, \text{~~for all } |u|<1.
\end{align*}

Thus, by Chernoff's bound,  we have that, for any $\delta>0$,
\begin{align*}
    \mathbb{P}\left[\left|\frac{1}{|I|}\sum_{t\in I}v_{tj}\right|\geq \delta\right] \leq 2\exp\left\{ |I|\left(\frac{1}{2} \psi''_u C_b^2u^2-u\delta \right)\right\} \text{~~for all } |u|<1.
\end{align*}
Thus, for any $\delta \in [0, \psi''_uC_b^2)$, we can set $u=\delta/(\psi''_uC_b^2)<1$ so that we have
\begin{align*}
    \mathbb{P}\left[\left|\frac{1}{|I|}\sum_{t\in I}v_{tj}\right|\geq \delta\right] \leq 2\exp \left(-\frac{|I|\delta^2}{2\psi''_uC_b^2} \right).
\end{align*}
Thus, by union bound, we have
\begin{align*}
    \mathbb{P}\left( \|\nabla L(\theta^*, I)\|_\infty \geq|I|\delta \right) \leq 2\exp\left(-\frac{|I|\delta^2}{2\psi''_uC_b^2} +\log d \right) \text{~~for all } \delta \in [0, \psi''_uC_b^2).
\end{align*}
Set $\delta=2C_b\sqrt{\frac{2\psi''_u\log(Td)}{|I|}}$. Note that we have $\delta < \psi''_uC_b^2$ as long as $$|I|\geq \frac{8\log(Td)}{\psi''_uC_b^2}.$$
Thus, we have that
\begin{align*}
    &\mathbb{P}\left( \|\nabla L(\theta^*, I)\|_\infty \geq|I|\delta \right)
    =\mathbb{P}\left( \|\nabla L(\theta^*, I)\|_\infty \geq  2C_b\sqrt{2\psi''_u|I|\log(Td)} \right)
    \\
    \leq& 2\exp\left(-\frac{|I|\delta^2}{2\psi''_uC_b^2} +\log d \right) =2\exp\left(-4\log(T) -3\log d \right).
\end{align*}
This completes the proof. \qed

\subsection{Restricted strong convexity}
%\textcolor{red}{would it be more readable to denote it as $\mathrm{d}L$?}  

% \textcolor{red}{For $\Delta \in \mathbb{R}^d$, $I$ ... $\theta^*$ also should depend on $t$?} \zifeng{\Cref{lem:RE} works for any $\theta^*$ satisfies \Cref{assum1} and \Cref{assum:moment}. Thus I think it is okay to skip $t$ but emphasize that $\theta^*$ can be anything in $\Theta$.}

For any $\theta\in\Theta$, let $\Delta=\theta-\theta^*\in\mathbb{R}^d$, we define
\begin{equation}\label{eq:dlike}
    \mathrm{d} L(\Delta,\theta^*,I)= \left\{L(\theta^*+\Delta,I)-L(\theta^*,I)-\langle \nabla L(\theta^*, I), \Delta \rangle\right\}/|I|, 
\end{equation}which is the key quantity used in the definition of restricted strong convexity.

% \textcolor{red}{in the range of inf, shouldn't it be $b(W+1)$?} \zifeng{$C_bC_{\theta}$ is sufficient. See third line of proof of \Cref{lem:RE}.}

%\zifeng{Check, also check moment condition here about mean of $x_t.$}

% \begin{ass}\label{assum:moment}
%  The second-order moment matrix $\mathbb E(x_tx_t^\top)=\Sigma$ satisfies that $\lambda_{\min}(\Sigma)=\sigma_l>0$. In addition, there exists $\delta >0$ such that $\sup_{\|\Delta\|_2=1} \mathbb{E}\{(\Delta^\top x_t)^{2+\delta}\} =\sigma_u <\infty$. %{\color{blue} multiple $\delta$ defined}
% \end{ass}
% \Cref{assum:moment} is a mild moment condition of the covariate $x_t$ and further implies that the maximum eigenvalue of $\mathbb E(x_tx_t^\top)=\Sigma$ is upper bounded by $\sigma_u^{2/(2+\delta)}.$ Add an explanation with $\delta=0$ will inflate by a $\log^3$ factor, see Section 4 in \cite{rudelson2012reconstruction}. Remark that this also implies that the maximum eigenvalue of $E(z_iz_i^{\top})$ is also bounded using the property of submatrix of a positive definite matrix.

\begin{lemma}\label{lem:RE}
Under Assumptions \ref{assum_model} and \ref{assum_moment},  there exist absolute positive constants $c_0$, $\omega_1$ and $\omega_2$, depending only on constants in Assumptions \ref{assum_model} and \ref{assum_moment} %($\psi''_l$, $C_b$, $\sigma_l$, $\sigma_u,$ $\delta$), 
such that
\begin{align}\label{eq:RE}
   \mathrm{d} L(\Delta,\theta^*,I)\geq \omega_1 \|\Delta\|_2\left\{\|\Delta\|_2-\omega_2\sqrt{\frac{\log d}{|I|}}\|\Delta\|_1 \right\} \text{ for all } \|\Delta\|_2\leq 2C_{\theta},
\end{align}
with probability at least $1-2\exp(-c_0|I|)$. In particular, we can set $\omega_1=\psi''_l \sigma_l /4$, $\omega_2=64\sqrt{2}\ell C_b/\sigma_l$, and $c_0=\sigma_l^2/(8\ell^4),$ with $\ell=2(2\sigma_u/\sigma_l)^{1/\delta}$.
\end{lemma}

\noindent\textbf{Proof of \Cref{lem:RE}}:
By a second order Taylor expansion, we have that
\begin{align*}
    \mathrm{d} L(\Delta,\theta^*,I) =\frac{1}{|I|} \sum_{t\in I} \psi''(\langle \theta^*, x_t \rangle +a_t\langle \Delta, x_t \rangle) (\Delta^\top x_t)^2, \text{ for } a_t \in (0,1). 
\end{align*}
Note that $|\langle \theta^*, x_t \rangle +a_t\langle \Delta, x_t \rangle|\leq C_bC_{\theta}$, as $\theta^*+a_t\Delta \in \Theta$ using the compactness of $\Theta$ by Assumption \ref{assum_model} (ii). Thus, we have that
\begin{align}\label{eq:lowerbound_dlike}
    \mathrm{d} L(\Delta,\theta^*,I) \geq \psi''_l\frac{1}{|I|}\sum_{t\in I} (\Delta^\top x_t)^2.
\end{align}

Thus, to establish \eqref{eq:RE}, we only need to show
\begin{align}\label{eq:bound_quadratic}
    \psi''_l\frac{1}{|I|}\sum_{t\in I} (\Delta^\top x_t)^2\geq \omega_1 \|\Delta\|_2\left\{\|\Delta\|_2-\omega_2\sqrt{\frac{\log d}{|I|}}\|\Delta\|_1 \right\} \text{ for all } \|\Delta\|_2\leq 2C_{\theta},
\end{align}
with high probability. 
%\textcolor{red}{This part later is incorrect as we cannot further assume after rescaling, $\Delta$ still has $\|\Delta\|_1\leq 2W.$ We might further need the sub-Gaussianity assumption. Otherwise, we know in the cone, $\|\Delta\|_1\leq 4\sqrt{s}\|\Delta\|_2=4\sqrt{s},$ instead of $2W.$ But we lose an $s^2$ instead of $s$ term. 2. Maybe use the RE in van de Geer and Bulhamn as in Javanmard? Check van de Geer's result.}
In addition, note that by scaling the above equation with $\|\Delta\|_2$, we only need to show that
\begin{align}\label{eq:RE_scaled}
    \psi''_l\frac{1}{|I|}\sum_{t\in I} (\Delta^\top x_t)^2\geq \omega_1 \left\{1-\omega_2\sqrt{\frac{\log d}{|I|}}\|\Delta\|_1 \right\} \text{ for all } \|\Delta\|_2=1,
\end{align}
with high probability. On the other hand, note that after scaling with $\|\Delta\|_2$, the $L_1$ norm $\|\Delta\|_1$ is no longer bounded by $2C_{\theta}$. Thus, $|\Delta^\top x_t|$ may not be bounded by $2C_bC_{\theta}$, which complicates the analysis and requires additional technical arguments.

Specifically, we adapt the truncation strategy in \cite{negahban2012unified}. Set the truncation level $\ell=2(2\sigma_u/\sigma_l)^{1/\delta}$, define the function
\begin{align*}
    \phi_\ell(u)=\begin{cases}
        u^2 &  \text{ if } |u|\leq \ell/2,\\
        (\ell-u)^2 & \text{ if } \ell/2\leq |u| \leq \ell,\\
        0 & \text{ otherwise.}
    \end{cases}
\end{align*}
By definition, $\phi_\ell(u)\leq u^2$ and $\phi_\ell(u)$ is a Lipschitz function with parameter $\ell.$ Thus, to establish \eqref{eq:RE_scaled}, we only need to show  
\begin{align*}
        \psi''_l\frac{1}{|I|}\sum_{t\in I} \phi_\ell\left(\Delta^\top x_t\right)\geq \omega_1 \left\{1-\omega_2\sqrt{\frac{\log d}{|I|}}\|\Delta\|_1 \right\} \text{ for all } \|\Delta\|_2=1,
\end{align*}

%Define the event
%\begin{align*}
%    \mathcal{E}(r)=\left\{\psi''_l\frac{1}{|I|}\sum_{t\in I} (\Delta^\top x_t)^2 < \omega_1\left[1-\omega_2\sqrt{\frac{\log d}{|I|}}r \right], \text{ for some } \Delta \in \mathbb{B}_1(r)\cap \mathbb{S}_2(1) \right\}.
%\end{align*}
%We first show that, for any $r>0,$ the probability of $\mathcal{E}(r)$ is small for suitably chosen $\omega_1$ and $\omega_2.$

Denote $\mathbb{S}_2(1)=\{\Delta\in \mathbb{R}^d:\|\Delta\|_2=1\}$ and $\mathbb{B}_1(r)=\{\Delta\in \mathbb{R}^d:\|\Delta\|_1 \leq r\}$ for any $r>0.$ First, we show that $\mathbb{E}\{\phi_\ell(\Delta^\top x_t)\}\geq \sigma_l/2$ for $\Delta \in \mathbb{B}_1(r)\cap \mathbb{S}_2(1).$ Specifically, we have
\begin{align}\label{ineq_phi}
\begin{split}
    &\mathbb{E}\{\phi_\ell(\Delta^\top x_t)\}\geq \mathbb{E}\{(\Delta^\top x_t)^2\}- \mathbb{E}\{(\Delta^\top x_t)^2\mathbb{I}(|\Delta^\top x_t|>\ell/2)\}\\
    \geq & \sigma_l - \frac{2^\delta\mathbb{E}\{(\Delta^\top x_t)^{2+\delta}\}}{\ell^\delta}=\sigma_l - (2/\ell)^\delta \sigma_u = \sigma_l/2,
\end{split}
\end{align}
where the second inequality holds by Assumption \ref{assum_moment} and Markov's inequality.

Second, we provide a high probability bound on the concentration of $1/{|I|}\sum_{t\in I} \phi_\ell\left(\Delta^\top x_t\right)$ around its expectation $\mathbb{E}\{\phi_\ell(\Delta^\top x_t)\}.$ Without loss of generality, we can assume $I=\{1,2,\cdots, |I|\}$, and define the random variable
\begin{align*}
    Z(r)=f(x_1,x_2,\cdots x_{|I|})=\sup_{\Delta \in \mathbb{B}_1(r)\cap \mathbb{S}_2(1)} \left| \frac{1}{|I|}\sum_{t\in I} \phi_{\ell}(\Delta^\top x_t)- \mathbb{E}\{\phi_\ell(\Delta^\top x_t)\}\right|.
\end{align*}
We provide a high probability bound for $Z(r)$. By triangle inequality, we have that
\begin{align*}
    &\mathrm{D}_k f(x_1,\cdots,x_{|I|})\\ 
    :=&\sup_{y\in \mathcal{X}} f(x_1,x_2,\cdots,x_{k-1},y,x_{k+1},\cdots,x_{|I|})-\inf_{y\in \mathcal{X}} f(x_1,x_2,\cdots,x_{k-1},y,x_{k+1},\cdots,x_{|I|})\\
    \leq & \left(\sup_{\Delta \in \mathbb{B}_1(r)\cap \mathbb{S}_2(1)} \left| \frac{1}{|I|}\sum_{t\in I, i\neq k} \phi_{\ell}(\Delta^\top x_t) - \mathbb{E}\{\phi_\ell(\Delta^\top x_t)\}\right|+ \frac{1}{4|I|}\ell^2\right)\\
    &-\left(\sup_{\Delta \in \mathbb{B}_1(r)\cap \mathbb{S}_2(1)} \left| \frac{1}{|I|}\sum_{t\in I, i\neq k} \phi_{\ell}(\Delta^\top x_t) - \mathbb{E}\{\phi_\ell(\Delta^\top x_t)\} \right|- \frac{1}{4|I|}\ell^2\right)\\
    = &\frac{1}{2|I|}\ell^2,
\end{align*}
where $\mathcal{X}$ denotes the support of $x_t$.

Thus, by McDiarmid’s inequality, we have that for any $z>0$,
\begin{align}\label{eq:McDiarmid}
    \mathbb{P}\left[Z(r)\geq \mathbb{E}(Z(r))+ z \right]\leq \exp \left( -\frac{2z^2}{\sum_{t\in I}\|\mathrm{D}_t f(x_1,\cdots,x_{|I|})\|_\infty^2}\right) \leq \exp \left( -\frac{8|I|z^2}{\ell^4}\right).
\end{align}

We now bound $\mathbb{E}(Z(r)).$ Let $\{\varepsilon_t\}_{t=1}^{|I|}$ be an i.i.d.\ sequence of Rademacher variables, a standard symmetrization argument gives that (see e.g., Proposition 4.11 in \cite{wainwright2019high})
\begin{align*}
    \mathbb{E}(Z(r))\leq 2\mathbb{E}\left[ \sup_{\Delta \in \mathbb{B}_1(r)\cap \mathbb{S}_2(1)} \left| \frac{1}{|I|}\sum_{t\in I} \varepsilon_t\phi_{\ell}(\Delta^\top x_t)\right|\right].
\end{align*}
Since the function $\phi_{\ell}(u)$ is Lipschitz with parameter $\ell$, by the Ledoux-Talagrand contraction inequality, we have that
\begin{align*}
    \mathbb{E}(Z(r))\leq& 2\times 2\ell\mathbb{E}\left[ \sup_{\Delta \in \mathbb{B}_1(r)\cap \mathbb{S}_2(1)} \left| \frac{1}{|I|}\sum_{t\in I} \varepsilon_t\Delta^\top x_t\right|\right]\\
    =& 4\ell \mathbb{E}\left[ \sup_{\Delta \in \mathbb{B}_1(r)\cap \mathbb{S}_2(1)} \left| \Delta^\top\frac{1}{|I|}\sum_{t\in I} \varepsilon_t x_t\right|\right]\\
    \leq & 4\ell  r \mathbb{E}\left[ \left\|\frac{1}{|I|}\sum_{t\in I} \varepsilon_t x_t\right\|_\infty \right].
\end{align*}

For $j=1,2,\cdots,d$, note that $\varepsilon_t x_{tj}$ is a sequence of i.i.d.\ mean-zero bounded random variable. By Hoeffding's lemma, we have $\mathbb{E}(\exp(u \frac{1}{|I|}\sum_{t\in I}\varepsilon_t x_{tj}))\leq \exp(u^2C_b^2/(2|I|)).$ Thus, by standard bounds on expectations of maxima (see e.g.,\ (2.66) in \cite{wainwright2019high}), we have that 
$$\mathbb{E}\left[ \left\|\frac{1}{|I|}\sum_{t\in I} \varepsilon_t x_t\right\|_\infty \right]\leq C_b\sqrt{\frac{2 \log d}{|I|}},$$
which implies that $\mathbb{E}(Z(r))\leq 4\ell r C_b \sqrt{{2 \log d}/{|I|}}.$

Set $z=\sigma_l/8+4\ell r C_b\sqrt{2\log d/|I|}$. Define $g(r)=\sigma_l/8+8\ell r C_b\sqrt{2\log d/|I|}$ and $c=2/\ell^4.$ In addition, define $f(\Delta;I)=\left| \frac{1}{|I|}\sum_{t\in I} \phi_{\ell}(\Delta^\top x_t)- \mathbb{E}\{\phi_\ell(\Delta^\top x_t)\}\right|$. Then \eqref{eq:McDiarmid} implies that
\begin{align*}
    &\mathbb{P}\left[ Z(r)\geq \sigma_l/8+8 \ell r C_b\sqrt{2\log d/|I|} \right]\\
    =&\mathbb{P}\left[ \sup_{\Delta \in \mathbb{B}_1(r)\cap \mathbb{S}_2(1)} f(\Delta;I)\geq g(r) \right]
    \leq \exp(-c|I|g(r)^2).
\end{align*}
Thus, by the peeling device in Lemma 3 of \cite{raskutti2010restricted}, we have that
\begin{align}\label{ineq_peeling}
    \mathbb{P}\left\{ \exists \Delta \in \mathbb{S}_2(1) \text{ such that } f(\Delta,I)\geq 2g(\|\Delta\|_1) \right\} \leq 2\exp(-4c|I|(\sigma_l/8)^2)= 2\exp\left(-\frac{|I|\sigma_l^2}{8\ell^4}\right).
\end{align}
Thus, with probability at least $1-2\exp\left(-|I|\sigma_l^2/(8\ell^4)\right),$ we have that for all $\Delta \in \mathbb{S}_2(1)$,
\begin{align*}
    &\frac{1}{|I|}\sum_{t\in I} (\Delta^\top x_t)^2 \geq \frac{1}{|I|}\sum_{t\in I} \phi_{\ell}(\Delta^\top x_t) \geq \mathbb{E}\{\phi_\ell(\Delta^\top x_t)\} - f(\Delta;I)\\
    \geq & \sigma_l/2 - 2g(\|\Delta\|_1)=\frac{\sigma_l}{4}\left(1-\frac{64 \ell C_b}{\sigma_l}\sqrt{\frac{2\log d}{|I|}}\|\Delta\|_1\right),
\end{align*}
where the first inequality holds by the definition of $\phi_{\ell}(z)$, the second by the definition of $f(\Delta;I)$, and the third  by \eqref{ineq_phi} and  \eqref{ineq_peeling}.

Thus, in view of \eqref{eq:RE_scaled}, the restricted strong convexity in \eqref{eq:RE} holds with $\omega_1=\psi''_l \sigma_l /4$, $\omega_2=64\sqrt{2}\ell C_b/\sigma_l$, and $c_0=\sigma_l^2/(8\ell^4),$ where recall $\ell=2(2\sigma_u/\sigma_l)^{1/\delta}$.  \qed

\subsection{Estimation error bounds}
Recall the Lasso estimator on $I$ takes the form
\begin{align*}
    \widehat \theta_I = \arg\min_{\theta \in \Theta} L(\theta, I)+ \lambda\sqrt{|I|}\|\theta\|_1= \arg\min_{\theta \in \Theta} \sum_{t\in I} \{\psi(x^{\top}_t\theta) - y_t x^{\top}_t\theta\}+ \lambda\sqrt{|I|}\|\theta\|_1.
\end{align*}
In what follows, for a generic parameter vector $\theta\in\Theta$, we denote $\theta(S)$ as the parameter vector by setting all coordinates of $\theta$ in $S^c$ as 0, and $\theta(S^c)=\theta-\theta(S)$.  

% \zifeng{$\lambda$ can be set as any $c_\lambda\sqrt{\log(Td)}$ with $c_\lambda\geq 4b\sqrt{2\psi''_u}.$ Make the error bound be in terms of $\lambda$, instead of $\sqrt{\log(Td)}$, this makes the later proof more transparent. The general strategy: keep $\lambda$ in the proof as much as possible instead of writing it as $c_\lambda \sqrt{\log(Td)}.$ Same to all the results later.}

\begin{lemma}\label{lem:est}
Suppose Assumptions \ref{assum_model} and \ref{assum_moment} hold. Set $\lambda =c_{\lambda}\sqrt{\log(Td)}$, with $c_{\lambda}\geq  4C_b\sqrt{2\psi''_u}$. There exist absolute positive constants $c_1, c_2, c_3$ that only depend on constants in Assumptions \ref{assum_model} and \ref{assum_moment}, such that for any $|I| > c_1 s \log(Td)$, we have that
    \begin{align*}
     &\|\widehat{\theta}_{I}-\theta^*\|_2\leq c_2 \lambda \sqrt{|I|^{-1}s},\\
     &\|\widehat{\theta}_{I}-\theta^*\|_1 \leq 4c_2 \lambda s\sqrt{|I|^{-1}},\\
     &  \frac{1}{|I|}\sum_{t\in I}(x_t^\top (\widehat{\theta}_I-\theta^*))^2 \leq c_3\lambda^2s|I|^{-1},
    \end{align*}
with probability at least $1-T^{-4}$. In particular, we can set $ c_1=\max\{{8}/{(\psi''_uC_b^2)}, 64\omega_2^2, 4/c_0\}, $ $c_2=3\omega_1^{-1}$ and $c_3=9/(2\psi''_l\omega_1)$, with $\omega_1,\omega_2,c_0$ given in \Cref{lem:RE}.
\end{lemma}

\noindent\textbf{Proof of \Cref{lem:est}}:
In the following, assume we are in the good event where both \Cref{lem:DB} and \Cref{lem:RE} hold. By union bound, this holds with probability at least
$$1- 2\exp\left(-4\log(Td) +\log d \right) - 2 \exp(-c_0|I|),$$
for any $|I|> [{8 {(\psi''_uC_b^2)}^{-1} \vee  4 c_0^{-1}] \log(Td)}$ and $d>1$.

For  $\lambda\geq 4C_b\sqrt{2\psi''_u\log(Td)}$, when \Cref{lem:DB} holds, we have \begin{equation}\label{eq:bound_dlike}
    2\| \nabla L(\theta^*, I)\|_\infty \leq \lambda\sqrt{|I|}.
\end{equation}  By the definition of $\widehat{\theta}_I$, we have that
\begin{align*}
   L(\widehat{\theta}_I, I) + \lambda\sqrt{|I|}\|\widehat{\theta}_I\|_1\leq L(\theta^*,I)+\lambda \sqrt{|I|}\|\theta^*\|_1.
\end{align*}

Define $\widehat\Delta_I=\widehat{\theta}_I-\theta^*$. Recall  the definition of $\mathrm{d} L(\Delta,\theta^*,I)$  in \eqref{eq:dlike}, by a second order Taylor expansion and \eqref{eq:bound_dlike}, we have that
\begin{align*}
    &\mathrm{d} L(\widehat \Delta_I,\theta^*,I)|I|+\lambda\sqrt{|I|}\|\widehat{\theta}_I\|_1\leq |\langle \nabla L(\theta^*, I), \widehat\Delta_I \rangle| + \lambda \sqrt{|I|}\|\theta^*\|_1 \nonumber\\
    \leq & \| \nabla L(\theta^*, I)\|_\infty \| \widehat\Delta_I\|_1 + \lambda \sqrt{|I|}\|\theta^*(S)\|_1\leq \lambda/2 \sqrt{|I|} \| \widehat\Delta_I\|_1 + \lambda \sqrt{|I|}\|\theta^*(S)\|_1.
\end{align*}

This implies that
\begin{align*}
    &\mathrm{d} L(\widehat \Delta_I,\theta^*,I)|I|+\lambda\sqrt{|I|}\|\widehat{\theta}_I(S^c)\|_1=\mathrm{d} L(\widehat \Delta_I,\theta^*,I)|I|+\lambda\sqrt{|I|}\|\widehat{\Delta}_I(S^c)\|_1\\
    \leq &  \lambda/2 \sqrt{|I|}\| \widehat\Delta_I\|_1 + \lambda \sqrt{|I|}\|\theta^*(S)\|_1-\lambda\sqrt{|I|}\|\widehat{\theta}_I(S)\|_1\leq  \lambda/2 \sqrt{|I|}\| \widehat\Delta_I\|_1 +\lambda \sqrt{|I|}\| \widehat\Delta_I(S)\|_1 \\
    =&\lambda/2 \sqrt{|I|}\| \widehat\Delta_I(S^c)\|_1 +3\lambda /2\sqrt{|I|}\| \widehat\Delta_I(S)\|_1.
\end{align*}
where the second inequality is due to triangle inequality. Thus, we have that
\begin{align}\label{eq:MLE_property}
\mathrm{d} L(\widehat \Delta_I,\theta^*,I)|I|+\lambda/2\sqrt{|I|}\|\widehat{\Delta}_I(S^c)\|_1 \leq 3\lambda /2\sqrt{|I|}\| \widehat\Delta_I(S)\|_1.
\end{align}

Recall by \eqref{eq:lowerbound_dlike}, $\mathrm{d} L(\widehat \Delta_I,\theta^*,I)>0$,  then \eqref{eq:MLE_property} implies that $\|\widehat{\Delta}_I(S^c)\|_1\leq 3\|\widehat{\Delta}_I(S)\|_1$, and hence $\|\widehat{\Delta}_I\|_1\leq 4\|\widehat{\Delta}_I(S)\|_1$. Together with the restricted strong convexity in \Cref{lem:RE}, and that $\|\widehat{\Delta}_I(S)\|_1\leq \sqrt{s}\|\widehat{\Delta}_I\|_2$, we have that
\begin{align*}
    \mathrm{d} L(\widehat \Delta_I,\theta^*,I) \geq \omega_1\left(1-4\omega_2\sqrt{\frac{s\log d}{|I|}}\right)\|\widehat \Delta_I\|_2^2\geq \frac{\omega_1}{2} \|\widehat \Delta_I\|_2^2,
\end{align*}
for any $|I|>64\omega_2^2s \log(d).$

Thus, \eqref{eq:MLE_property} further implies that
\begin{align*}
    \frac{\omega_1}{2}|I|\|\widehat\Delta_I\|_2^2 \leq \mathrm{d} L(\widehat \Delta_I,\theta^*,I)|I| \leq 3\lambda/2\sqrt{|I|}\|\widehat\Delta_I(S)\|_1\leq 3\lambda /2\sqrt{s|I|}\|\widehat\Delta_I\|_2.
\end{align*}
Thus, set $|I|=c_1 s\log(Td)$ with $c_1=\max\{{8}/{(\psi''_uC_b^2)}, 64\omega_2^2, 4/c_0\}$, we have with probability at least $1-T^{-4}$ such that
\begin{align*}
    \|\widehat\Delta_I\|_2\leq \frac{3\lambda}{\omega_1}\sqrt{\frac{s}{|I|}}.
    %= c_2\sqrt{\frac{s\log(Td)}{|I|}},
\end{align*}
%where $c_2=12b\sqrt{2\psi''_u}/\omega_1.$

In addition, recall \eqref{eq:lowerbound_dlike}, we have
\begin{align*}
    \frac{1}{|I|}\sum_{t\in I}(x_t^\top \widehat{\Delta}_I)^2 \leq \frac{1}{\psi''_l}\mathrm{d} L(\widehat{\Delta}_I,\theta^*,I)\leq  \frac{1}{\psi''_l} 3\lambda /2\sqrt{s/|I|}\|\widehat\Delta_I\|_2\leq \frac{9\lambda^2}{2\psi''_l\omega_1}\frac{s}{|I|}.
    %=c_3\frac{s\log(Td)}{|I|},
\end{align*}
% where the first inequality follows from the proof of \Cref{lem:RE} and $c_3=144b^2\psi''_u/(\psi''_l\omega_1).$

Finally, we have that
\begin{align*}
    \|\widehat\Delta_I\|_1\leq 4\|\widehat\Delta_I(S)\|_1 \leq 4\sqrt{s}\|\widehat\Delta_I\|_2\leq  \frac{12\lambda}{\omega_1}\frac{s}{\sqrt{|I|}},
\end{align*}
which completes the proof.

In particular, recall $\omega_1$, $\omega_2$ and $c_0$ in Lemma \ref{lem:RE}, we set  here 
\begin{equation}\label{bound_c1}
   c_1=\max\{{8}/{(\psi''_uC_b^2)}, 64\omega_2^2, 4/c_0\}, 
\end{equation}$c_2=3\omega_1^{-1}$ and $c_3=9/(2\psi''_l\omega_1)$.  \qed

\section{GLM Estimation under Mixture}\label{sec:GLM_mix}
This section provides the estimation error bounds of the Lasso estimator for a mixture of two high-dimensional GLMs with the additional Assumption \ref{assum:mixture}, which is later used in the proof of regret upper bound in \Cref{cor_regret}.

%We only consider the mixture of two high-dimensional GLMs, which is sufficient for our regret analysis.  
% Denote $\alpha\in (0,1)$ as the mixture proportion. Define the pseudo true parameter $\theta^*_\alpha$ as
% \begin{align*}
%     \theta^*_\alpha=\arg\min_\theta \mathbb L_\alpha(\theta)=\arg\min_\theta \mathbb E\left[ \alpha\{\psi(x^{\top}_1\theta) - y_1 x^{\top}_1\theta\} + (1-\alpha)\{\psi(x^{\top}_2\theta) - y_2 x^{\top}_2\theta\} \right],
% \end{align*}
% where $(y_k|p_k,\theta_k^*,z_k$) with i.i.d. $x_k=(z_k^{\top},p_k)^{\top}$, $k=1,2$  is generated from GLM model \eqref{model_GLM} with model parameter $\theta_k^*$, $k=1,2$, respectively. 

% \begin{ass}\label{assum:mixture}
%     For any $\alpha\in (0,1)$, the pseudo true parameter $\theta^*_\alpha$ is in the interior of $\Theta$ and $\max\{\|\theta^*_\alpha-\theta_1^*\|_1, \|\theta^*_\alpha-\theta_2^*\|_1\}\leq C_M\|\theta_1^*-\theta_2^*\|_1$ for some absolute constant $C_M.$
% \end{ass}

Let $I_1,I_2 \subset \{1,\cdots,T\}$ such that  $I_1\cap I_2=\varnothing$ and  $I=I_1\cup I_2$.  Suppose for $t\in I_k$, $\theta_t=\theta_k^*$ with support $S_k$, $k=1,2$, with  $\|\theta_1^* - \theta_2^*\|_2 = \kappa > 0$, and  that $\{x_t,y_t\}_{t\in I_k} \sim $ GLM($\theta_k^*$) for $k=1,2.$ Define the Lasso estimator as
\begin{align*}
    \widehat \theta_I = \arg\min_{\theta \in \Theta} L(\theta, I)+ \lambda\sqrt{|I|}\|\theta\|_1= \arg\min_{\theta \in \Theta} \sum_{t\in I} \{\psi(x^{\top}_t\theta) - y_t x^{\top}_t\theta\}+ \lambda\sqrt{|I|}\|\theta\|_1.
\end{align*}
Define $\psi'_u=\sup_{|v|\leq C_bC_{\theta}}|\psi'(v)|$. Furthermore, define
 $\alpha=|I_1|/(|I_1|+|I_2|) \in (0,1)$ as the mixture proportion and define the pseudo true parameter $\theta^*_\alpha$ as
\begin{align*}
    \theta^*_\alpha=\arg\min_{\theta\in\Theta} \mathbb L_\alpha(\theta)=\arg\min_{\theta\in\Theta} \mathbb E\left[ \alpha\{\psi(x^{\top}_{t_1}\theta) - y_{t_1} x^{\top}_{t_1}\theta\} + (1-\alpha)\{\psi(x^{\top}_{t_2}\theta) - y_{t_2} x^{\top}_{t_2}\theta\} \right],
\end{align*}
for any $t_1\in I_1$ and $t_2\in I_2$. In the following, we skip the $\alpha$ subscript in $\theta^*_\alpha$ for notational simplicity.

\begin{lemma}\label{lem_mixture}
  Suppose  Assumptions \ref{assum_model}, \ref{assum_moment} and \ref{assum:mixture} hold. Set $\lambda=c_{\lambda}\sqrt{\log(Td)}$, with $c_{\lambda}\geq  8C_b(\sqrt{2\psi''_u}+\psi'_u) $. There exist absolute positive constants $c_1',c_2',c_3'$ that only depend on constants in Assumptions \ref{assum_model}, \ref{assum_moment} and \ref{assum:mixture}, such that for any intervals $I_1, I_2$ with $\min\{|I_1|,|I_2|\} > c_1' s \log(Td)$, we have that
    \begin{align*}
     &\|\widehat{\theta}_{I}-\theta^*\|_2^2\leq c_2' {\frac{s\lambda^2}{|I|}}+c_3'\kappa^2,
    \end{align*}
with probability at least $1-T^{-4}$. In particular, we can set $c_1'=\max\{8/(\psi''_uC_b^2),128\omega_2^2,4/c_0\}$,  $c_2'={144}/{\omega_1^2}+{8}/{\omega_1}$ and $ c_3'=\left(8+{8}/{\omega_1}\right)C_M^2$, with $\omega_1,\omega_2,c_0$ given in \Cref{lem:RE}.

\end{lemma}

\noindent\textbf{Proof of \Cref{lem_mixture}}:
The proof is based on the same technique as that used in the proof of \Cref{lem:est} with some modification. 

Let $S=S_1\cup S_2$, by \Cref{assum:mixture}, we have that \begin{equation}\label{eq:mixture_sparse}
    \|\theta^*(S^c)\|_1=\|\theta^*(S^c)-\theta_1^*(S^c)\|_1\leq C_M\|\theta_1^*-\theta_2^*\|_1 \leq C_M\sqrt{2s} \|\theta_1^*-\theta_2^*\|_2=C_M\sqrt{2s}\kappa,
\end{equation}
which essentially implies that the pseudo true parameter $\theta^*$ is approximately sparse.

%  Furthermore, denote 
%  $$\mathbb L_\alpha(\theta)= \mathbb E\left[ \alpha\{\psi(x^{\top}_1\theta) - y_1 x^{\top}_1\theta\} + (1-\alpha)\{\psi(x^{\top}_2\theta) - y_2 x^{\top}_2\theta\} \right],$$ which is clearly differentiable, so that 
% \begin{align}\label{eq:pseudo_score}
%     \frac{\partial}{\partial \theta} \mathbb L_\alpha(\theta^*_\alpha)&=\mathbb E\left[ \alpha\{\psi'(x^{\top}_1\theta^*_\alpha) - y_1\}x_1 + (1-\alpha)\{\psi'(x^{\top}_2\theta^*_\alpha) - y_2\}x_2\right]\nonumber\\
%     &=\mathbb E\left[ \alpha\{\psi'(x^{\top}_1\theta^*_\alpha) - \psi'(x^{\top}_1\theta^*_1)\}x_1 + (1-\alpha)\{\psi'(x^{\top}_2\theta^*_\alpha) - \psi'(x^{\top}_2\theta^*_2)\}x_2 \right].
% \end{align}

\noindent \textbf{Step 1.} [Deviation Bound]
Define $\nabla L(\theta^*, I)= \sum_{t\in I} \{\psi'(x^{\top}_t\theta^*) - y_t\}x_t$. We have that
\begin{align*}
    \nabla L(\theta^*, I)=\sum_{k=1}^2 \sum_{t\in I_k} \{\psi'(x^{\top}_t\theta^*_k) - y_t\}x_t + \sum_{k=1}^2 \sum_{t\in I_k} \{\psi'(x^{\top}_t\theta^*) - \psi'(x^{\top}_t\theta^*_k)\}x_t = M_1+M_2.
\end{align*}
By \Cref{lem:DB}, for $\min\{|I_1|,|I_2|\}\geq 8\log(Td)/(\psi''_uC_b^2)$, we have that with probability larger than $1-4d^{-3}T^{-4}$, $\|M_1 \|_\infty \leq 2C_b\sqrt{2\psi''_u\log(Td)}\sum_{k=1}^2 \sqrt{|I_k|} \leq 4C_b\sqrt{\psi''_u\log(Td)}\sqrt{|I|}$.

For $M_2$, by \Cref{assum:mixture}, we have that $\mathbb E M_2=0_d$ and that $\max\{|x^{\top}_t\theta^*_1|, |x^{\top}_t\theta^*_2|, |x^{\top}_t\theta^*|\} \leq C_bC_{\theta}$. For $j=1,2,\cdots, d$, define
$v_{tj}=(\psi'(x^{\top}_t\theta^*) - \psi'(x^{\top}_t\theta^*_k))x_{tj}$. For all $j$, we have that $|v_{tj}|\leq 2C_b\psi'_u$ and $\mathbb E (\sum_{k=1}^2 \sum_{t\in I_k} v_{tj})=0$. Thus, by Hoeffding's inequality, we have that
\begin{align*}
    &\mathbb P\left( \left\|M_2\right\|_\infty \geq 4\sqrt{2}C_b\psi'_u\sqrt{|I|\log(Td)}  \right)\\
    \leq &\sum_{j=1}^d\mathbb P\left( \left|\sum_{k=1}^2 \sum_{t\in I_k} v_{tj}\right| \geq 4\sqrt{2}C_b\psi'_u\sqrt{|I|\log(Td)}\right) \leq 2 d^{-3}T^{-4}.
\end{align*}
Thus, we have that with probability at least $1- 6d^{-3}T^{-4}$,
$$\|\nabla L(\theta^*, I)\|_\infty \leq 4C_b(\sqrt{2\psi''_u}+\psi'_u) \sqrt{\log(Td)}\sqrt{|I|}\leq \frac{\lambda}{2}\sqrt{|I|}. $$ 

\medskip

\noindent \textbf{Step 2.} [Restricted Strong Convexity].  Recall $\mathrm{d} L(\Delta,\theta^*,I)$ as defined in \eqref{eq:dlike}. Follow the same arguments as that in the proof of \Cref{lem:RE}, we can show that
\begin{align*}
       \mathrm{d} L(\Delta,\theta^*,I)\geq \omega_1 \|\Delta\|_2\left\{\|\Delta\|_2-\omega_2\sqrt{\frac{\log d}{|I|}}\|\Delta\|_1 \right\} \text{ for all } \|\Delta\|_2\leq 2C_{\theta},
\end{align*}
with probability at least $1-2\exp(-c_0|I|)$, where $\omega_1,\omega_2,c_0$ are the same as the ones given in \Cref{lem:RE}.

\medskip

\noindent \textbf{Step 3.} [Estimation Error] This part is adapted from the proof of \Cref{lem:est}. 
By the definition of $\widehat{\theta}_I$, we have that
\begin{align*}
   L(\widehat{\theta}_I, I) + \lambda\sqrt{|I|}\|\widehat{\theta}_I\|_1\leq L(\theta^*,I)+\lambda \sqrt{|I|}\|\theta^*\|_1.
\end{align*}

Define $\widehat\Delta_I=\widehat{\theta}_I-\theta^*$. By the definition of $\mathrm{d} L(\Delta,\theta^*,I)$ and a second order Taylor expansion, we have that
\begin{align*}
    &\mathrm{d} L(\widehat \Delta_I,\theta^*,I)|I|+\lambda\sqrt{|I|}\|\widehat{\theta}_I\|_1\leq |\langle \nabla L(\theta^*, I), \widehat\Delta_I \rangle| + \lambda \sqrt{|I|}\|\theta^*\|_1 \nonumber\\
    \leq & \| \nabla L(\theta^*, I)\|_\infty \| \widehat\Delta_I\|_1 + \lambda \sqrt{|I|}\|\theta^*(S)\|_1 +  \lambda \sqrt{|I|}\|\theta^*(S^c)\|_1 \\
    \leq & \lambda/2 \sqrt{|I|} \| \widehat\Delta_I\|_1 + \lambda \sqrt{|I|}\|\theta^*(S)\|_1+ \lambda \sqrt{|I|}\|\theta^*(S^c)\|_1.%\\
   % \leq & \lambda/2 \sqrt{|I|} \| \widehat\Delta_I\|_1 + \lambda \sqrt{|I|}\|\theta^*(S)\|_1+\lambda \sqrt{|I|} C_0\|\theta_1^*-\theta_2^*\|_1\\
    %\leq & \lambda/2 \sqrt{|I|} \| \widehat\Delta_I\|_1 + \lambda \sqrt{|I|}\|\theta^*(S)\|_1+\lambda \sqrt{|I|} C_0 \sqrt{s}\kappa,
\end{align*}
This implies that
\begin{align*}
    &\mathrm{d} L(\widehat \Delta_I,\theta^*,I)|I|+\lambda\sqrt{|I|}\|\widehat{\Delta}_I(S^c)\|_1\\
    \leq &\mathrm{d} L(\widehat \Delta_I,\theta^*,I)|I|+\lambda\sqrt{|I|}\|\widehat{\theta}_I(S^c)\|_1+\lambda\sqrt{|I|}\|\theta^*(S^c)\|_1\\
    \leq &  \lambda/2 \sqrt{|I|}\| \widehat\Delta_I\|_1 + \lambda \sqrt{|I|}\|\theta^*(S)\|_1+2\lambda\sqrt{|I|}\|\theta^*(S^c)\|_1-\lambda\sqrt{|I|}\|\widehat{\theta}_I(S)\|_1\\
    \leq & \lambda/2 \sqrt{|I|}\| \widehat\Delta_I\|_1 +\lambda \sqrt{|I|}\| \widehat\Delta_I(S)\|_1+2\lambda\sqrt{|I|}\|\theta^*(S^c)\|_1  \\
    \leq &\lambda/2 \sqrt{|I|}\| \widehat\Delta_I(S^c)\|_1 +3\lambda /2\sqrt{|I|}\| \widehat\Delta_I(S)\|_1+2C_M\lambda\sqrt{2s|I|}\kappa,
\end{align*}
where the first and third inequality is due to triangle inequality, and the last inequality is based on \Cref{assum:mixture} and  \eqref{eq:mixture_sparse}.

Thus, we have that
\begin{align}\label{eq:MLE_property_mixture}
\mathrm{d} L(\widehat \Delta_I,\theta^*,I)|I|+\lambda/2\sqrt{|I|}\|\widehat{\Delta}_I(S^c)\|_1 \leq 3\lambda /2\sqrt{|I|}\| \widehat\Delta_I(S)\|_1+2\sqrt{2}C_M\lambda\sqrt{s|I|}\kappa.
\end{align}

Thus, we have that $\|\widehat{\Delta}_I(S^c)\|_1\leq 3\|\widehat{\Delta}_I(S)\|_1+4\sqrt{2}C_M\sqrt{s}\kappa$. Together with the restricted strong convexity in \textbf{Step 2}, it implies that
\begin{align*}
    \mathrm{d} L(\widehat \Delta_I,\theta^*,I) \geq& \omega_1 \|\widehat \Delta_I\|_2^2-\omega_1\omega_2\sqrt{\frac{\log d}{|I|}}\|\widehat \Delta_I\|_2\|\widehat \Delta_I\|_1 \\
    \geq &
    \omega_1 \|\widehat \Delta_I\|_2^2-4\omega_1\omega_2\sqrt{\frac{\log d}{|I|}}\|\widehat \Delta_I\|_2 (\|\widehat{\Delta}_I(S)\|_1+C_M\sqrt{2s}\kappa)\\
    \geq & \omega_1 \|\widehat \Delta_I\|_2^2-4\omega_1\omega_2\sqrt{\frac{2s\log d}{|I|}}\|\widehat \Delta_I\|_2 (\|\widehat{\Delta}_I\|_2+C_M\kappa)\\
    \geq & \frac{\omega_1}{2} \|\widehat \Delta_I\|_2^2-\frac{C_M\omega_1\kappa}{2}\|\widehat{\Delta}_I\|_2,
\end{align*}
for any $|I|>128\omega_2^2s \log(d).$  Note that the events in \textbf{Step 1} and \textbf{Step 2} hold with probability larger than $1-6d^{-3}T^{-4}-2\exp(c_0|I|)$. Then, above results hold probability larger than $1-T^{-4}$ by letting $\min\{|I_1|,|I_2|\}\geq c_1's\log Td$ with $c_1'=\max\{8/(\psi''_uC_b^2),128\omega_2^2,4/c_0\}$.

Thus, \eqref{eq:MLE_property_mixture} further implies that
\begin{align*}
    &\frac{\omega_1}{2}\|\widehat\Delta_I\|_2^2 \leq \mathrm{d} L(\widehat \Delta_I,\theta^*,I) + \frac{C_M\omega_1\kappa}{2}\|\widehat{\Delta}_I\|_2\\
    \leq & \frac{3\lambda} {2\sqrt{|I|}}\|\widehat\Delta_I(S)\|_1 + 2\sqrt{2}C_M\lambda\sqrt{s/|I|}\kappa + \frac{C_M\omega_1\kappa}{2}\|\widehat{\Delta}_I\|_2\\
    \leq & \frac{3\lambda\sqrt{2s}} {2\sqrt{|I|}}\|\widehat\Delta_I\|_2+ 2C_M\lambda\sqrt{2s/|I|}\kappa + \frac{C_M\omega_1\kappa}{2}\|\widehat{\Delta}_I\|_2\\
    \leq & (\frac{3\lambda\sqrt{2s}} {2\sqrt{|I|}}+\frac{C_M\omega_1\kappa}{2})\|\widehat\Delta_I\|_2 + C_M^2\kappa^2 + 2s\lambda^2/|I|,
\end{align*}
where the last inequality holds by Cauchy-Schwartz inequality.

This implies that either $\|\widehat\Delta_I\|_2 \leq 2\omega_1^{-1} (3\lambda \sqrt{2s/|I|}+C_M\omega_1\kappa)$ or $\|\widehat\Delta_I\|_2^2\leq 4\omega_1^{-1}(C_M^2\kappa^2 + 2s\lambda^2/|I|)$ holds. Hence,
\begin{align*} 
    \|\widehat\Delta_I\|_2^2 &\leq \frac{8}{\omega_1^2}(18\lambda^2 s/|I| + C_M^2\omega_1^2\kappa^2)+ \frac{8}{\omega_1}(C_M^2\kappa^2 + s\lambda^2/|I|)\\
    &=\left(\frac{144}{\omega_1^2}+\frac{8}{\omega_1} \right)\frac{s\lambda^2}{|I|} + \left(8+\frac{8}{\omega_1}\right) C_M^2\kappa^2.
\end{align*}
In particular, we can choose  $c_2'={144}/{\omega_1^2}+{8}/{\omega_1}$ and $c_3'=\left(8+{8}/{\omega_1}\right)C_M^2$. \qed

\section[]{Proof of \Cref{prop:cp}}\label{sec:proof_prop}
This section gives the proof of \Cref{prop:cp}, which provides theoretical guarantees for the proposed online change-point detection algorithm. The proof of \Cref{prop:cp} is directly built on \Cref{lem-no-change} and \Cref{lem-one-change_alter} below, which characterizes the performance of the online change-point detection algorithm under (i) no change-point and (ii) a single change-point scenarios. In addition, \Cref{lem:large-prob-1} provides a novel deviation bound for the interaction between any vector with a single change-point and the first order derivatives of the log-likelihood function, which is used in the  proof of \Cref{lem-one-change_alter} and can be of independent interest.
\vskip 1mm

\begin{lemma}[No change-point]\label{lem-no-change}
 Let $I_1,I_2 \subset \{1,\cdots,T\}$ such that  $I_1\cap I_2=\varnothing$ and $I=I_1\cup I_2$.  Suppose Assumptions \ref{assum_model} and \ref{assum_moment} hold, and for all $t\in I$, $\theta_t=\theta^*$.   For absolute constants $(c_1,c_2,c_3)$ taken from Lemma \ref{lem:est}, let
    \[
         \lambda= c_{\lambda}\sqrt{\log(Td)}\quad \text{with   }c_{\lambda}\geq  4C_b\sqrt{2\psi_{u}''},\quad  \mbox{and  } \quad \gamma=c_\gamma s\lambda^2\quad \text{with   }c_\gamma\geq 22c_2+\psi''_u c_3 /2.
    \]
    Given that $\min\{|I_1|,|I_2|\}\geq c_1s\log(Td)$, it holds that
    \[
        L(\widehat{\theta}_{I_1}, I_1) + L(\widehat{\theta}_{I_2}, I_2) + \gamma \geq L(\widehat{\theta}_I, I)+\lambda \sqrt{|I_1|} \|\widehat{\theta}_I-\widehat{\theta}_{I_1}\|_1 + \lambda \sqrt{|I_2|} \|\widehat{\theta}_I-\widehat{\theta}_{I_2}\|_1,
    \]
    with probability at least $1-3T^{-4}$.
\end{lemma}

\begin{lemma}[One change-point] \label{lem-one-change_alter}
Let $I_1,I_2 \subset \{1,\cdots,T\}$ such that  $I_1\cap I_2=\varnothing$ and  $I=I_1\cup I_2$.  Suppose Assumptions \ref{assum_model} and \ref{assum_moment} hold, and for $t\in I_k$, $\theta_t=\theta_k^*$ with support $S_k$, $k=1,2$, and $\|\theta_1^* - \theta_2^*\|_2 = \kappa > 0$. Let  $\gamma$ be defined as in \Cref{lem-no-change}, and $$
        \lambda= c_{\lambda}\sqrt{\log(Td)}\quad \text{with   }c_{\lambda}\geq 8\omega_1\omega_2C_{\theta}\vee 4C_b\sqrt{2\psi_{u}''},$$
        where $\omega_1,\omega_2$ are taken from \Cref{lem:RE}.
   If $\min\{|I_1|,|I_2|\}\geq \{c_1s\log(Td)\}\vee \{c_5s\lambda^2/\kappa^2\}$ with $c_1$ taken from Lemma \ref{lem:est} and $c_5$ being an absolute constant~(see definition later), it holds that 
    \begin{equation}\label{eq-cp-lem-cond_alter}          L(\widehat{\theta}_{I_1}, I_1) + L(\widehat{\theta}_{I_2}, I_2) + \gamma < L(\widehat{\theta}_I, I)+\lambda \sqrt{|I_1|} \|\widehat{\theta}_I-\widehat{\theta}_{I_1}\|_1 + \lambda \sqrt{|I_2|} \|\widehat{\theta}_I-\widehat{\theta}_{I_2}\|_1,
    \end{equation}
 with probability at least $1-3T^{-4}$. In particular, we can set $$c_5=\frac{16}{\omega_1}  \left\{c_{\gamma}+\left[12c_2+\psi''_uc_3+\frac{2c_4^2}{\omega_1c^2_{\lambda}}\right]\right\},$$ with the absolute constants $c_2,c_3$ taken from \Cref{lem:est} and the absolute constant $c_4$ taken from \Cref{lem:large-prob-1}.
 \end{lemma}

\noindent\textbf{Proof of \Cref{prop:cp}}: In the following, suppose $c_m\geq c_1$, with $c_1$ being the absolute constant taken from \Cref{lem:est} and $c_\lambda, c_\gamma$ are sufficiently large absolute constants that satisfy the conditions in \Cref{lem-no-change} and \Cref{lem-one-change_alter}. In addition, suppose the SNR condition in \eqref{eq-snr-prop-3.1} is satisfied with the absolute constant $c_{\text{snr}}= c_5c_\lambda^2$, with $c_5$ being the absolute constant defined in \Cref{lem-one-change_alter}.

Thus, under scenario (i), we have $$\mathbb{P}(\mathcal{T}_n=0)\leq \sum_{t=m}^{n-m} \mathbb{P}(\mathcal{D}(t,n)>\gamma)\leq 3(n-2m+1)T^{-4}\leq 3T^{-3},$$
where the first inequality holds by union bound, and the second inequality holds by \Cref{lem-no-change}. 

Under scenario (ii), we further have 
\begin{flalign*}
    \mathbb{P}(\mathcal{T}_n=1)
    %\leq& \mathbb{P}\left(\bigcap_{t=\tau}^{\tau+c_{\mathrm{snr}}s\log(Td)/\kappa^2}\mathcal{D}(t,n)\leq \gamma\right)
    \geq &  \mathbb{P}\left(\mathcal{D}(\tau,n)> \gamma\right)\geq 1-3T^{-4},
\end{flalign*}
where the second inequality holds by \Cref{lem-one-change_alter}, where we set $I_1=\{1,2,\cdots,\tau\}$, $I_2=\{\tau+1,\cdots,n\}$ and $I=\{1,2,\cdots,n\}.$
\qed

\medskip

\subsection{Proofs of Lemmas \ref{lem-no-change}-\ref{lem:large-prob-1}}

\noindent\textbf{Proof of \Cref{lem-no-change}}:
    We prove by contradiction.  Assume that
    \[
        L(\widehat{\theta}_{I_1}, I_1) + L(\widehat{\theta}_{I_2}, I_2) + \gamma <L(\widehat{\theta}_I, I)+\lambda \sqrt{|I_1|} \|\widehat{\theta}_I-\widehat{\theta}_{I_1}\|_1 + \lambda \sqrt{|I_2|} \|\widehat{\theta}_I-\widehat{\theta}_{I_2}\|_1.
    \]
    Denote $\Delta_I=\widehat{\theta}_I-\theta^*$, by a Taylor expansion, we have with probability at least $1-T^{-4}$,
    \begin{align*}
    L(\widehat{\theta}_I, I)= & L({\theta}^*, I) + \langle \nabla L(\theta^*, I), \Delta_I \rangle +\frac{1}{2}\sum_{t\in I} \psi''(\langle \theta^*, x_t \rangle +a_t\langle \Delta_I, x_t \rangle) (\Delta_I^\top x_t)^2, \text{ for } a_t \in (0,1),\\
    \leq & L({\theta}^*, I) + \frac{\lambda\sqrt{|I|}}{2}\|\Delta_I\|_1 +\frac{1}{2}\psi''_u \sum_{t\in I}(\Delta_I^\top x_t)^2 \\
    \leq & L({\theta}^*, I) + {2c_2s\lambda^2} +\frac{1}{2}\psi''_u c_3 s\lambda^2=L({\theta}^*, I) +  s\lambda^2(2c_2+\psi''_u c_3 /2),
    \end{align*}
    where the first inequality follows from \Cref{lem:DB} and the second inequality follows from \Cref{lem:est}.

    In addition, by \Cref{lem:est} and the triangle inequality, we have that
    \begin{align*}
        \lambda \sqrt{|I_1|} \|\widehat{\theta}_I-\widehat{\theta}_{I_1}\|_1 + \lambda \sqrt{|I_2|} \|\widehat{\theta}_I-\widehat{\theta}_{I_2}\|_1 \leq 16 c_2s\lambda^2,
    \end{align*}
    with probability at least $1-T^{-4}.$
    
    Based on similar arguments, we have with probability at least $1-T^{-4}$, on $I_1$
    \begin{align*}
        L(\widehat{\theta}_{I_1}, I_1)= & L({\theta}^*, I_1) + \langle \nabla L(\theta^*, I_1), \Delta_{I_1} \rangle +\frac{1}{2}\sum_{t\in I_1} \psi''(\langle \theta^*, x_t \rangle +a_t\langle \Delta_{I_1}, x_t \rangle) (\Delta_{I_1}^\top x_t)^2, \text{ for } a_t \in (0,1),\\
    \geq & L({\theta}^*, I_1) - \frac{\lambda\sqrt{|I_1|}}{2}\|\Delta_{I_1}\|_1 
    \geq L({\theta}^*, I_1) - {2}c_2s\lambda^2,
    \end{align*}
Same holds for $I_2,$ i.e., 
$
L(\widehat{\theta}_{I_1}, I_1)\geq  L({\theta}^*, I_2) - {2}c_2s\lambda^2.
$
    
    Thus, combine all inequalities above, we have that
    \begin{align*}
        &L(\theta^*, I_1) + L(\theta^*, I_2) - 4 c_2 s\lambda^2 +\gamma \leq L(\widehat{\theta}_{I_1}, I_1) + L(\widehat{\theta}_{I_2}, I_2) + \gamma \\<&L(\widehat{\theta}_I, I)+\lambda \sqrt{|I_1|} \|\widehat{\theta}_I-\widehat{\theta}_{I_1}\|_1 + \lambda \sqrt{|I_2|} \|\widehat{\theta}_I-\widehat{\theta}_{I_2}\|_1 < L({\theta}^*, I) +(18c_2+\psi''_u c_3 /2)s\lambda^2,
    \end{align*}
    which implies that
    \[
       c_{\gamma} < 22c_2+\psi''_u c_3 /2.
    \]
    This contradicts with the choice of $c_\gamma$ and thus concludes the proof. \qed
\medskip

% \zifeng{Same here. $\gamma$ can be the same as in \Cref{lem-no-change}. $\lambda$ needs a $c_\lambda$ adjustment, see below. We fix the constants in \Cref{lem:est} and \Cref{lem_mixture} and use them as it is. Track more explicitly for the role of $\lambda$ instead of quickly making it $c_\lambda \sqrt{\log(Td)}.$}

% \textcolor{red}{Link this penalty with the penalized log-likelihood.}

\noindent\textbf{Proof of \Cref{lem-one-change_alter}}:
 We again prove by contradiction. Suppose \eqref{eq-cp-lem-cond_alter} fails, we have
    \begin{align*}
         &\sum_{t \in I_1} \{\psi(x_t^{\top} \widehat{\theta}_I) - \psi(x_t^{\top}\widehat\theta_{I_1})\} + \sum_{t \in I_2} \{\psi(x_t^{\top} \widehat{\theta}_I) - \psi(x_t^{\top}\widehat\theta_{I_2})\} - \sum_{t \in I_1} y_t x_t^{\top} (\widehat{\theta}_I-\widehat\theta_{I_1}) -\sum_{t \in I_2} y_t x_t^{\top} (\widehat{\theta}_I-\widehat\theta_{I_2}) \\
         \leq &\gamma-\lambda \sqrt{|I_1|} \|\widehat{\theta}_I-\widehat{\theta}_{I_1}\|_1 - \lambda \sqrt{|I_2|} \|\widehat{\theta}_I-\widehat{\theta}_{I_2}\|_1.
    \end{align*}
Denote 
    \[
        \Delta_t = \begin{cases}
            \Delta^{(1)}=\widehat{\theta}_{I} - \theta^*_1, & t \in I_1, \\
            \Delta^{(2)}=\widehat{\theta}_{I} - \theta^*_2, & t \in I_2.
        \end{cases} 
    \]
        By Taylor expansions w.r.t. $\widehat\theta_{I_1}$ and $\widehat\theta_{I_2}$ at $\theta_1^*$ and $\theta_2^*$ respectively, we have that
    \begin{align}\label{eq:onecp_bound}
         &\sum_{t \in I_1} \{\psi(x_t^{\top} \widehat{\theta}_I) - \psi(x_t^{\top}\theta^*_1)\} + \sum_{t \in I_2} \{\psi(x_t^{\top} \widehat{\theta}_I) - \psi(x_t^{\top}\theta^*_2)\} - \sum_{t \in I} y_t x_t^{\top} \Delta_t \nonumber\\
        \leq & \gamma + e_I-\lambda \sqrt{|I_1|} \|\widehat{\theta}_I-\widehat{\theta}_{I_1}\|_1 - \lambda \sqrt{|I_2|} \|\widehat{\theta}_I-\widehat{\theta}_{I_2}\|_1,
    \end{align}
    where
    \begin{align*}
        e_I= &\sum_{t \in I_1}(\psi'(x_t^\top \theta_1^*)-y_t)x_t^\top(\widehat{\theta}_{I_1}-\theta_1^*) + \sum_{t \in I_2}(\psi'(x_t^\top \theta_2^*)-y_t)x_t^\top(\widehat{\theta}_{I_2}-\theta_2^*) \\
        +& \frac{1}{2}\psi''_u\sum_{t \in I_1} (x_t^\top(\widehat{\theta}_{I_1}-\theta_1^*))^2 + \frac{1}{2}\psi''_u \sum_{t \in I_2}(x_t^\top(\widehat{\theta}_{I_2}-\theta_2^*))^2.
    \end{align*}
    %for some absolute constant $c>0$ due to the boundedness of the parameter space and feature space. %\zifeng{On the other hand, I wonder if the compactness of parameter space $\|\theta\|_1<C$ is a reasonable assumption or not? Javanmard2018 used it. } {\color{blue} I think we need this assumption in regret analysis}

    Note that $e_I$ can be well controlled   using the results in \Cref{lem:DB} and \Cref{lem:est}. Specifically, we have
    \begin{align*}
        &\left|\sum_{t \in I_1}(\psi'(x_t^\top \theta_1^*)-y_t)x_t^\top(\widehat{\theta}_{I_1}-\theta_1^*)\right| = \langle\nabla L(\theta^*,I_1),\widehat{\theta}_{I_1}-\theta_1^*\rangle \\
        \leq &\left\|\nabla L(\theta^*,I_1)\right\|_\infty \|\widehat{\theta}_{I_1}-\theta_1^*\|_1
        \leq 4c_2 s\lambda [2C_b\sqrt{2\psi_u''\log(Td)}]\leq 2c_2s\lambda^2,
    \end{align*}
    and
    \begin{align*}
        \frac{1}{2}\psi''_u\sum_{t \in I_1} (x_t^\top(\widehat{\theta}_{I_1}-\theta_1^*))^2 \leq \frac{1}{2}\psi''_uc_3 s \lambda^2,
    \end{align*}    
    with probability at least $1-T^{-4}$. Same holds for $I_2$.
    
    Let $S=S_1\cup S_2$, and clearly $|S|\leq 2s$. Then \eqref{eq:onecp_bound} implies that
    \begin{align}\label{eq:onecp_bound2}
       &L(\widehat{\theta}_I, I)-L(\theta_1^*,I_1)-L(\theta_2^*,I_2)\leq \gamma+ (4c_2+\psi''_uc_3) s \lambda^2-\lambda \sqrt{|I_1|} \|\widehat{\theta}_I-\widehat{\theta}_{I_1}\|_1 - \lambda \sqrt{|I_2|} \|\widehat{\theta}_I-\widehat{\theta}_{I_2}\|_1,\nonumber\\
       \leq &\gamma+ (4c_2+\psi''_uc_3) s \lambda^2-\lambda \sqrt{|I_1|} \|\widehat{\theta}_I(S^c)-\widehat{\theta}_{I_1}(S^c)\|_1 - \lambda \sqrt{|I_2|} \|\widehat{\theta}_I(S^c)-\widehat{\theta}_{I_2}(S^c)\|_1\nonumber\\
       \leq &\gamma+ (4c_2+\psi''_uc_3) s \lambda^2-\lambda \sqrt{|I_1|} \|\widehat{\theta}_I(S^c)\|_1+\lambda \sqrt{|I_1|}\|\widehat{\theta}_{I_1}(S^c)\|_1 -\lambda \sqrt{|I_2|} \|\widehat{\theta}_I(S^c)\|_1+\lambda \sqrt{|I_2|}\|\widehat{\theta}_{I_2}(S^c)\|_1\nonumber\\
       \leq &\gamma+(12c_2+\psi''_uc_3) s \lambda^2 -\lambda \sqrt{|I_1|} \|\widehat{\theta}_I(S^c)\|_1-\lambda \sqrt{|I_2|} \|\widehat{\theta}_I(S^c)\|_1
    \end{align}
    with probability at least $1-2T^{-4}$, where the last inequality holds by that $$\|\widehat{\theta}_{I_k}(S^c)\|_1=\|\widehat{\theta}_{I_k}(S^c)-\theta_k^*(S^c)\|_1\leq \|\widehat{\theta}_{I_k}-\theta_k^*\|_1\leq 4c_2s\lambda\sqrt{|I|_k^{-1}},\quad k=1,2$$ using Lemma \ref{lem:est}.  %{\color{blue} I am not sure of the last inequality, $|I_1|$ and $|I_2|$ are only lower bounded. }

    Denote $\epsilon_t=\psi'(x_t^{\top}\theta^*_1)-y_t$  for $t\in I_1$ and $\epsilon_t=\psi'(x_t^{\top}\theta^*_2)-y_t$ for $t\in I_2$. With a further Taylor expansion, we have that for some $a_t\in(0,1)$,
    \begin{align*}
        &L(\widehat{\theta}_I, I)-L(\theta_1^*,I_1)-L(\theta_2^*,I_2)\\
        = &\sum_{t \in I_1} \epsilon_t x_t^\top (\widehat{\theta}_I-\theta^*_1) + \sum_{t \in I_2} \epsilon_t x_t^\top (\widehat{\theta}_I-\theta^*_2) +\sum_{t\in I}\frac{1}{2}\psi''(x_t^\top\theta_t^*+a_tx_t^\top \widehat{\theta}_I)(\Delta_t^\top x_t)^2\\
        \geq & \sum_{j\in S}\sum_{t\in I} \Delta_{tj}x_{tj}\epsilon_t +\sum_{j\in S^c}\widehat{\theta}_{Ij}\sum_{t\in I_1} x_{tj}\epsilon_t + \sum_{j\in S^c}\widehat{\theta}_{Ij}\sum_{t\in I_2} x_{tj}\epsilon_t + \frac{\psi''_l}{2}\sum_{t\in I}(\Delta_t^\top x_t)^2\\
        \geq & -\left|\sum_{j\in S}\sum_{t\in I} \Delta_{tj}x_{tj}\epsilon_t\right| - \|\widehat{\theta}_I(S^c)\|_1\left\|\sum_{t\in I_1} x_t\epsilon_t\right\|_\infty -
        \|\widehat{\theta}_I(S^c)\|_1\left\|\sum_{t\in I_2} x_t\epsilon_t\right\|_\infty + \frac{\psi''_l}{2}\sum_{t\in I}(\Delta_t^\top x_t)^2\\
        \geq & -\left|\sum_{j\in S} \sqrt{\sum_{t\in I} \Delta_{tj}^2}\sum_{t\in I}\frac{\Delta_{tj}}{\sqrt{\sum_{t\in I} \Delta_{tj}^2}}x_{tj}\epsilon_t\right| - \|\widehat{\theta}_I(S^c)\|_1\left\|\sum_{t\in I_1} x_t\epsilon_t\right\|_\infty -
        \|\widehat{\theta}_I(S^c)\|_1\left\|\sum_{t\in I_2} x_t\epsilon_t\right\|_\infty + \frac{\psi''_l}{2}\sum_{t\in I}(\Delta_t^\top x_t)^2\\
        \geq & -c_4{ \sqrt{ 2s\sum_{t\in I}\|\Delta_t(S)\|_2^2\log(Td)}} - \|\widehat{\theta}_I(S^c)\|_1\frac{\lambda}{2}\sqrt{|I_1|}-\|\widehat{\theta}_I(S^c)\|_1\frac{\lambda}{2}\sqrt{|I_2|} + \frac{\psi''_l}{2}\sum_{t\in I}(\Delta_t^\top x_t)^2,
    \end{align*}
    with probability at least $1-2T^{-4}$, where the first inequality uses the fact that $\theta_{kj}^*=0$, $k=1,2$ for $j\in S^c$, the third inequality holds by Cauchy-Schwartz inequality, and the last inequality follows from \Cref{lem:large-prob-1} and \Cref{lem:DB}.
    
    Thus, combined with \eqref{eq:onecp_bound2}, we have that    
    \begin{align*}
        & \frac{\psi''_l}{2}\sum_{t \in I} (\Delta_t^\top x_t)^2 \leq c_4\sqrt{2 s \sum_{t\in I}\|\Delta_t(S)\|_2^2\log(Td)} - \lambda/2 (\sqrt{|I_1|} +\sqrt{|I_2|}) \|\widehat{\theta}_I({S^c})\|_1+\gamma+(12c_2+\psi''_uc_3) s \lambda^2. 
    \end{align*}    
    By \eqref{eq:bound_quadratic}, we further have that 
    \begin{align*}
        &\frac{\psi''_l}{2}\sum_{t \in I_1} (\Delta_t^\top x_t)^2\geq |I_1|\omega_1 \|\Delta^{(1)}\|_2\left\{\|\Delta^{(1)}\|_2-\omega_2\sqrt{\frac{\log d}{|I_1|}}\|\Delta^{(1)}\|_1 \right\}\\
        \geq &|I_1|\omega_1 \|\Delta^{(1)}\|_2^2-\omega_1\omega_2 2C_{\theta} \sqrt{{\log d}{|I_1|}} \|\widehat \theta_I(S^c)\|_1-\omega_1\omega_2 \|\Delta^{(1)}\|_2 \sqrt{{\log d}{|I_1|}} \| \Delta^{(1)}(S)\|_1\\
        \geq & |I_1|\omega_1 \|\Delta^{(1)}\|_2^2-\omega_1\omega_2 2C_{\theta} \sqrt{{\log d}{|I_1|}} \|\widehat \theta_I(S^c)\|_1 -\omega_1\omega_2 \|\Delta^{(1)}\|_2^2 \sqrt{2s{\log d}{|I_1|}}\\
        \geq & \frac{\omega_1}{2}|I_1| \|\Delta^{(1)}\|_2^2-\omega_1\omega_2 2C_{\theta} \sqrt{{\log d}{|I_1|}} \|\widehat \theta_I(S^c)\|_1,
    \end{align*}
    where the last inequality follows from the assumption that $|I_1|\geq c_1 s\log(d)$ and that $c_1\geq 8\omega_2^2$ in \eqref{bound_c1}. The same holds for $I_2.$ %{\color{blue} in the second inequality, $s$ should be outside square root?}
    
    Thus, for $\lambda= c_{\lambda}\sqrt{\log(Td)}$, with $c_{\lambda} \geq  8\omega_1\omega_2C_{\theta}$, together we have that, 
    \begin{align*}
        &\frac{\omega_1}{2} \sum_{t\in I} \|\Delta_t\|_2^2 =\frac{\omega_1}{2}|I_1| \|\Delta^{(1)}\|_2^2 + \frac{\omega_1}{2}|I_2| \|\Delta^{(2)}\|_2^2 \\
        \leq & c_4\sqrt{2s \sum_{t\in I}\|\Delta_t(S)\|_2^2\log(Td)} - \lambda/4 (\sqrt{|I_1|} +\sqrt{|I_2|}) \|\widehat{\theta}({S^c})\|_1 + \gamma + (12c_2+\psi''_uc_3)s\lambda^2\\
        \leq &  c_4^2(2/\omega_1)s\log(Td)+ \frac{\omega_1}{4} \sum_{t \in I} \|\Delta_t(S)\|_2^2 + \gamma + (12c_2+\psi''_uc_3)s\lambda^2,
    \end{align*}
    where the last inequality is due to Cauchy-Schwartz inequality. 
    
    %and the last inequality follows from the fact that $\Delta_t(S)$ can be viewed as a part of a $p$-dimensional vector in the cone and thus the restricted eigenvalue condition in \Cref{lem:RE} hold.

    %On the other hand, by \Cref{lem:RE} and the proof of \Cref{lem:est}~(i.e., restricted strong convexity), we have that \zifeng{This step has a gap on whether $\Delta_t$ is in the cone. But I think we can directly bound $\sum_{t \in I} \Delta_t^{\top} x_t x_t^{\top} \Delta_t$ based on $\Delta^*=\|\theta_1^*-\theta_2^*|$, which is in the cone.}
    %\[
    %    \frac{\psi''_l}{2}\sum_{t \in I} \Delta_t^{\top} x_t x_t^{\top} \Delta_t \geq \frac{\omega_1}{2} \sum_{t \in I}\|\Delta_t\|^2.
    %\]

   We therefore have that
    \[
        \frac{\omega_1}{4}\sum_{t \in I} \|\Delta_t\|_2^2 \leq \gamma + (12c_2+\psi''_uc_3+\frac{2c_4^2}{\omega_1c^2_{\lambda}})s\lambda^2.
    \]
    On the other hand, note that by triangle inequality, we must have $\|\Delta^{(1)}\|_2 + \|\Delta^{(2)}\|_2 \geq \|\theta_1^*-\theta_2^*\|_2$, which implies that $\max_{k=1,2}\{\|\Delta^{(1)}\|_2,\|\Delta^{(2)}\|_2\}\geq \kappa/2$. Therefore,     
    \begin{align*}
       \sum_{t \in I} \|\Delta_t\|_2^2  \geq \min\{|I_1|,|I_2|\} \kappa^2/4.
    \end{align*}

    We then have that 
    \[
        \min\{|I_1|,|I_2|\} \leq \frac{16}{\omega_1}  [c_{\gamma}+(12c_2+\psi''_uc_3+\frac{2c_4^2}{\omega_1c^2_{\lambda}})]\frac{  s\lambda^2}{\kappa^2}=c_5s\lambda^2/\kappa^2,
    \]
which contradicts the condition given in the lemma.  \qed

%\zifeng{An alternative strategy: To proceed, we need to further \textcolor{red}{lower bound} $ \sum_{t \in I} \Delta_t^{\top} x_t x_t^{\top} \Delta_t$ based on \Cref{lem:triangle} to
%\[
%  \sum_{t \in I} \Delta_t^{\top} x_t x_t^{\top} \Delta_t \geq \min\{|I_1|, |I_2|\} \kappa^2/2.
%\]
%This part requires some bounds on sample covariance matrix for $\|\Delta\|_2\leq 2W.$}
    
%\begin{lemma}\label{lem:triangle}
%    Given a positive semi-definite matrix $A \in \mathbb{R}^{p\times p}$ and any vectors $v_1,v_2,v_3\in \mathbb{R}^d$, the triangle inequality holds $$\sqrt{(v_1-v_2)^\top A (v_1-v_2)} \leq \sqrt{(v_1-v_3)^\top A (v_1-v_3)} + \sqrt{(v_2-v_3)^\top A (v_2-v_3)},$$ which implies $\max\{(v_1-v_3)^\top A (v_1-v_3), (v_2-v_3)^\top A (v_2-v_3) \}\geq (v_1-v_2)^\top A (v_1-v_2)/4.$
%\end{lemma}

% \zifeng{Is this $c_1$ same as the ones in \Cref{lem:est} and \Cref{lem-one-change_alter}? Or what notation should we use here?}
\begin{lemma}\label{lem:large-prob-1}
Let $I_1,I_2 \subset \{1,\cdots,T\}$ such that  $I_1\cap I_2=\varnothing$ and  $I=I_1\cup I_2$.  Suppose Assumptions \ref{assum_model} and \ref{assum_moment} hold, and for $t\in I_k$, $\theta_t=\theta_k^*$ with support $S_k$, $k=1,2$. Denote by $\mathcal{S}$ the linear subspace of all piecewise-constant vectors $v\in\mathbb{R}^{|I|}$ with the only change-point at $\tau=|I_1|$ and denote $\mathcal E_1=\{ v\in \mathbb{R}^{|I|}:\|v\|_2=1\}$ as the unit sphere in $\mathbb R^{|I|}$. If $\min\{|I_1|,|I_2|\}\geq c_1s\log(Td)$ with $c_1$ taken from Lemma \ref{lem:est}, it holds that 
\begin{align*}
    \mathbb P\left\{\max_{j = 1, \ldots, d} \sup_{v\in \mathcal{S}\cap \mathcal{E}_1} \sum_{t \in I} v_t x_{tj} \{y_t-\psi'(x_t^{\top} \theta_t)\} \geq c_4\log^{1/2}(Td) \right\}\leq T^{-4},
\end{align*}
where $c_4=4\sqrt{3\psi''_uC_b^2}\vee [(\psi''_uC_b^2+10)/\sqrt{2c_1}]$ is an absolute constant.
\end{lemma}
\medskip

\noindent\textbf{Proof of \Cref{lem:large-prob-1}}:
     Let $\mathcal{N}_{1/|I|}$ be a $1/|I|$-net of $\mathcal{S} \cap \mathcal{E}_1$. Since $\mathcal{S}$ is an affine subspace with dimension 2 and $\mathcal{E}_1$ is of diameter $1$, by Lemma 4.1 in \cite{pollard1990empirical}, $\mathcal{N}_{1/|I|}$ can be chosen such that $|\mathcal{N}_{1/|I|}| \leq (3|I|)^2=9|I|^2$.

    Denote $\epsilon_t=y_t-\psi'(x_t^{\top} \theta_t).$ It then holds for any fixed $j \in \{1, \ldots, d\}$, we have that 
    \begin{align*}
        & \mathbb{P}\left\{\sup_{v \in \mathcal{S} \cap \mathcal{E}_1} \sum_{t \in I} v_t x_{tj} \epsilon_t \geq c_4\log^{1/2}(Td)\right\} \\
        \leq & \mathbb{P}\left\{\sup_{v \in \mathcal{N}_{1/|I|}} \sum_{t \in I} v_t x_{tj} \epsilon_t + \sup_{u\in \mathcal{S} \cap \mathcal{E}_1}\inf_{v\in \mathcal{N}_{1/|I|}} |\sum_{t \in I} (u_t- v_t) x_{tj} \epsilon_t| \geq c_4\log^{1/2}(Td) \right\} \\
        \leq & \mathbb{P}\left\{\sup_{v \in \mathcal{N}_{1/|I|}} \sum_{t \in I} v_t x_{tj} \epsilon_t + \max_{t \in I}| x_{tj} \epsilon_t| \sup_{u\in \mathcal{S} \cap \mathcal{E}_1}\inf_{v\in \mathcal{N}_{1/|I|}} \|v-u\|_1 \geq c_4\log^{1/2}(Td) \right\} \\
        \leq & \mathbb{P}\left\{\sup_{v \in \mathcal{N}_{1/|I|}} \sum_{t \in I} v_t x_{tj} \epsilon_t + \max_{t \in I}| x_{tj} \epsilon_t| \sqrt{|I|}/|I| \geq c_4\log^{1/2}(Td) \right\} \\
        %\leq & \mathbb{P}\left\{\sup_{v \in \mathcal{N}_{1/|I|}} \sum_{t \in I} v_t x_{tj} \epsilon_t + |I|^{-1/2}\max_{t \in I} | x_{tj} \epsilon_t| \geq c_4\log^{1/2}(Td) \right\} \\
        \leq & \mathbb{P}\left\{\sup_{v \in \mathcal{N}_{1/|I|}} \sum_{t \in I} v_t x_{tj} \epsilon_t \geq c_4/2\log^{1/2}(Td)\right\}  +  \mathbb{P}\left\{|I|^{-1/2}\max_{t \in I} |x_{tj} \epsilon_t| \geq c_4/2\log^{1/2}(Td)\right\}\\
        \leq & 9|I|^2 \sup_{v \in \mathcal{N}_{1/|I|}} \mathbb{P}\left\{\sum_{t \in I} v_t x_{tj} \epsilon_t \geq c_4/2\log^{1/2}(Td)\right\} + \mathbb{P}\left\{|I|^{-1/2}\max_{t \in I} |x_{tj} \epsilon_t| \geq c_4/2\log^{1/2}(Td)\right\}
        %\\\leq & 144 \exp \left\{\frac{C \log(Td)}{\sum_{t \in I} v_t^2}\right\} + 2T \exp\left(-c_4\log(Td)\right) \leq (T\vee p)^{-c},
    \end{align*}
    where the third inequality follows from the definition of $\mathcal{N}_{1/|I|}$ and the fact that $\|u\|_1\leq \sqrt{|I|}\|u\|_2$ for $u\in\mathbb R^{|I|}$, and the last inequality follows as $|\mathcal{N}_{1/|I|}| \leq (3|I|)^2=9|I|^2$.

    Based on the same argument as that in the proof of \Cref{lem:DB} (i.e., Chernoff bound) and the fact that $\min\{|I_1|,|I_2|\}\geq c_1s\log(Td)$, it is easy to show that
    \begin{align*}
        \mathbb{P}\left\{\sum_{t \in I} v_t x_{tj} \epsilon_t \geq c_4/2\log^{1/2}(Td)\right\} \leq 2d^{-5}T^{-6} \text{ for all } v\in \mathcal{S} \cap \mathcal{E}_1,
    \end{align*}
    as $c_4\geq 4\sqrt{3\psi''_uC_b^2}.$

    In addition, by Markov inequality, we have that
    \begin{align*}
         & \mathbb{P}\left\{|I|^{-1/2}\max_{t \in I} |x_{tj} \epsilon_t| \geq c_4/2\log^{1/2}(Td)\right\}
        \leq |I| \mathbb{P}\left\{|I|^{-1/2} |x_{1j} \epsilon_1| \geq c_4/2\log^{1/2}(Td)\right\}\\
    \leq & |I| \frac{\mathbb E [\exp(|x_{1j}\epsilon_1|)]}{\exp (c_4\sqrt{|I|\log(Td)}/2 )} \leq 2|I| \exp\left\{\frac{1}{2}\psi''_uC_b^2-c_4\sqrt{|I|\log(Td)}/2\right\}\\
    \leq & 2T\exp\left(\frac{1}{2}\psi''_uC_b^2\right) (Td)^{-c_4\sqrt{c_1/2}} \leq 2d^{-5}T^{-4},
    \end{align*}
where the second inequality holds by noting $\mathbb E [\exp(|x_{1j}\epsilon_1|)]\leq \mathbb E [\exp(x_{1j}\epsilon_1)]+\mathbb E [\exp(-x_{1j}\epsilon_1)]$, and taking $u=1$ and $-1$ respectively in \eqref{eq_cumu}, and the third inequality
uses the fact $2c_1s\log(Td)\leq |I_1|+|I_2|= |I|\leq T$ and the last by $c_4\geq (\psi''_uC_b^2+10)/\sqrt{2c_1}.$ 
    
    Putting everything together, we conclude the proof. % we have that $\mathbb{P}\{\mathcal{A}(I)\} \leq 2d^{-5}T^{-4}.$
\qed

% {\color{blue} As discussed last time, if change-point testing failed, can we say something about the degree of change is small enough and estimation error is ignorable?}  \Yi{yes and no..}

% \zifeng{Consider an at-most-one-change-point model with high-dimensional linear regression. I believe that the change-point detection algorithm is adaptive in the sense that its regret will also be $O(\sqrt{T})$ when $\Delta^2 \tau<\sqrt{T}$. The reason is that the regret depends on $\Delta^2$ but not $\Delta$. This does not hold true for MAB in Maillard as regret in MAB depends on $\Delta$ not $\Delta^2.$}

% \zifeng{Our current analysis allows $z_t$ to change distributions across different segments.}

\section[]{Proof of regret upper bounds in \Cref{sec-cpdp-upper-bound}} \label{sec:proof_regret}

This section gives the proof of \Cref{thm:regret} and \Cref{cor_regret}, which provides the regret upper bound for the CPDP algorithm. The proof follows the basic steps given in the sketch of proofs in \Cref{sec-cpdp-upper-bound} of the main text.
\medskip

\noindent\textbf{Proof of \Cref{thm:regret}}:
In the following, suppose $c_m\geq c_1$, with $c_1$ being the absolute constant taken from \Cref{lem:est} and $c_\lambda, c_\gamma$ are sufficiently large absolute constants that satisfy the conditions in \Cref{prop:cp}~(see \Cref{lem-no-change} and \Cref{lem-one-change_alter} for more details). In addition, suppose the SNR condition in \Cref{ass_spacing} is satisfied with an absolute constant $c_{\text{snr}}^*>2c_5c_\lambda^2/c_m,$ with $c_5$ being the absolute constant defined in \Cref{lem-one-change_alter}. Recall that \Cref{prop:cp} holds for $c_{\text{snr}}=c_5c_\lambda^2$ in the SNR condition in \eqref{eq-snr-prop-3.1} (see its proof for details). Thus, we have $c_{\text{snr}}^*>2c_{\text{snr}}/c_m.$

Let $\{\widehat{\tau}_k\}_{k=1}^{\widehat{\Upsilon}_T}$ be the estimated change-points with $\widehat{\tau}_0=0$ and $\widehat{\tau}_{\widehat\Upsilon_T+1}=T$. By design, each of the cycles between $\widehat{\tau}_k$ and $\widehat{\tau}_{k+1}$ is of length $l_k=n_k+m$. Define $d_k=\lceil c_{\mathrm{snr}}/(c_m\kappa_k^2)\rceil l_{k-1}$ for $k=1,\ldots,\Upsilon_T$ as the controllable detection delay and further define 
\[%begin{flalign}\label{AT}
\mathcal{A}=\left\{\widehat\Upsilon_T=\Upsilon_T \text{ and for all } k\in\{1,\ldots,\Upsilon_T\}, \widehat{\tau}_k\in [\tau_k,\tau_k+d_k]\right\},
\]%end{flalign}
as the good event where all change-points are detected within the desirable detection delay. We show in \Cref{lem_regret_bad} that $\mathcal{A}$ is a high probability event.

Let $\mathcal{M}=\bigcup_{k=0}^{\widehat{\Upsilon}_T}\mathcal{M}_k$ be the set of price experiments. By \eqref{bound_price}, we have that
\begin{flalign*}
R_T(\bm \theta_T)=&\mathbb{E}\left[\sum_{t=1}^Tr(p_t^*,\theta_t,z_t)-r(p_t,\theta_t,z_t)\right]
\leq C_r\mathbb{E}\left[\sum_{t=1}^T (\varphi(\theta_t, z_t)-p_t)^2\right]
\\=&C_r\mathbb{E}\left[\sum_{t=1}^T (\varphi(\theta_t, z_t)-p_t)^2\mathbb{I}(t\in\mathcal{M},\mathcal{A})\right]+C_r\sum_{k=1}^{\Upsilon_T}\mathbb{E}\left[ \sum_{t={\tau}_k+1}^{\widehat{\tau}_k
}(\varphi(\theta_t, z_t)-p_t)^2\mathbb{I}(t\in\mathcal{M}^c,\mathcal{A})\right]\\&+
C_r\sum_{k=0}^{\Upsilon_T}\mathbb{E}\left[ \sum_{t=\widehat{\tau}_{k}+1}^{{\tau}_{k+1}}(\varphi(\theta_t, z_t)-p_t)^2 %\varphi(\widehat{\theta}_t,x_t)
\mathbb{I}(t\in\mathcal{M}^c,\mathcal{A})\right]+  C_r\mathbb{E}\left[\sum_{t=1}^T(\varphi(\theta_t, z_t)-p_t)^2\mathbb{I}(\mathcal{A}^c)\right]
\\=&  R_{T,\mathrm{I}}+ R_{T,\mathrm{II}}+ R_{T,\mathrm{III}}+ R_{T,\mathrm{IV}},
\end{flalign*}
where $R_{T,i}$, $i\in\{\mathrm{I,II,III,IV}\}$ corresponds to regret due to (I)-(IV) respectively. \Cref{thm:regret} directly follows from Lemmas \ref{lem_regret_experiment}-\ref{lem_regret_bad} below, which give the upper bounds for $R_{T,\mathrm{I}}, R_{T,\mathrm{II}}, R_{T,\mathrm{III}}$, and $R_{T,\mathrm{IV}}$, respectively.

We remark that in view of Sections \ref{sec:GLM}, \ref{sec:GLM_mix} and \ref{sec:proof_prop}, all the constants such as $c_2,c_5$, $c_1',c_2',c_3'$ that appear in Lemmas \ref{lem_regret_experiment}-\ref{lem_regret_bad} below are absolute positive constants that depend only on Assumptions \ref{assum_model}, \ref{assum_moment} and \ref{ass_spacing}, and  $c_m$, $c_{\lambda}$ and $c_{\gamma}$. 

\begin{lemma}[Regret due to price experimentation] \label{lem_regret_experiment}
There exists an absolute constant $ C_{R,\mathrm{I}}$ depending only on $c_m$, $C_r$ and $C_b$, such that
$$R_{T,\mathrm{I}}\leq C_{R,\mathrm{I}}  s\log(Td)\sqrt{\Upsilon_T T}.$$  
\end{lemma}

\begin{lemma}[Regret due to detection delay] \label{lem_regret_delay}
There exists an absolute constant $ C_{R,\mathrm{II}}$ depending only on $c_m$, $c_\lambda$, $c_5$, $C_r$ and $C_b$ such that 
$$
R_{T,\mathrm{II}}\leq 2C_{R,\mathrm{II}}\kappa_{\min}^{-2}\sqrt{\Upsilon_TT}.
$$
\end{lemma}

\begin{lemma} [Regret due to estimation error] \label{lem_regret_error}
There exists an absolute constant $C_{R,\mathrm{III}}$ depending only on $c_m,  c_{\lambda}, c_2, C_r$, $C_b$, $C_{\varphi}$, $C_\theta$ and $\lambda_{\max}(\Sigma)$ such that, 
$$R_{T,\mathrm{III}}\leq 2C_{R,\mathrm{III}}\sqrt{\Upsilon_T T}\log T. $$
\end{lemma}

\begin{lemma} [Regret due to failed change-point detection] \label{lem_regret_bad}
There exists a constant $C_{R,\mathrm{IV}}$ depending only on $C_r$ and $C_b$, such that 
$$R_{T,\mathrm{IV}}\leq C_{R,\mathrm{IV}}. $$
\end{lemma}

Summarizing the results from Lemmas \ref{lem_regret_experiment}-\ref{lem_regret_bad} above, we have that, for some absolute positive constants $C_R$ and $C$,
\begin{flalign*}
    R_T\leq&  C_R \left[s\log(Td)+\kappa_{\min}^{-2}+\log T\right]\sqrt{\Upsilon_T T}\leq C\sqrt{\Upsilon_T T}(s\log(Td)\vee \kappa_{\min}^{-2}).
\end{flalign*}
This concludes the proof. \qed

\medskip
\noindent\textbf{Proof of \Cref{cor_regret}}:
The proof follows the same argument as that in the proof of \Cref{thm:regret} by replacing \Cref{lem_regret_delay} with \Cref{lem_regret_delay2}. In the following, suppose $c_m\geq c_1'$, with $c_1'$ being the absolute constant taken from \Cref{lem_mixture}, all other settings are the same as that in the proof of \Cref{thm:regret}.

\begin{thmbis}{lem_regret_delay}[Regret due to detection delay under \Cref{assum:mixture}]\label{lem_regret_delay2}
    Suppose in addition Assumption \ref{assum:mixture} holds, there exists an absolute constant $C_{R,\mathrm{II}}'$ depending only on  $c_m, c_\lambda, c_{5}, c_2',c_3',C_M$, $C_r$, $C_{\varphi}$, $C_b$, $C_\theta$ and $\lambda_{\max}(\Sigma)$  such that
    $$
   R_{T,\mathrm{II}} \leq 2C_{R,\mathrm{II}}' \sqrt{\Upsilon_T T}(\log T+s).
   %\left[\sum_{k=1}^{\Upsilon_T}n_{k-1}(\log T +s)\right].
    $$
\end{thmbis}
\qed

\subsection{Proof of Lemmas \ref{lem_regret_experiment}-\ref{lem_regret_bad} and \Cref{lem_regret_delay2}}

\noindent\textbf{Proof of Lemma \ref{lem_regret_experiment}}: When $\mathcal{A}$ holds, we have $\widehat \Upsilon_T=\Upsilon_T$. Note that $|p_t^*-p_t|=|\varphi(\theta_t,z_t)-p_t|\leq p_u\leq  C_b$ by \Cref{assum_model} (iv). Hence, we have 
\begin{flalign*}
    R_{T,\mathrm{I}}\leq& \mathbb E \left\{C_r \sum_{k=0}^{{\Upsilon}_T}  C_b^2 |\mathcal{M}_k|\mathbb{I}(\mathcal{A})\right\}
    \\\stackrel{(a)}{\leq} & C_r C_b^2  \mathbb E \left\{\sum_{k=0}^{\Upsilon_T} m\left[\frac{(\widehat{\tau}_{k+1}-\widehat{\tau}_{k}-m)}{l_k}+1\right]\mathbb{I}(\mathcal{A})\right\}
    \\\leq & C_r C_b^2  \mathbb E \left\{\left[\max_{0\leq k\leq \Upsilon_T}\frac{m}{l_k}\sum_{k=0}^{\Upsilon_T}  (\widehat{\tau}_{k+1}-\widehat{\tau}_{k})+   \sum_{k=0}^{\Upsilon_T}m\right]\mathbb{I}(\mathcal{A})\right\}
    \\\stackrel{(b)}{\leq }&  C_r C_b^2  \left(\max_{0\leq k\leq \Upsilon_T}\frac{m}{l_k}T+({\Upsilon_T}+1)m\right),
\end{flalign*}
where (a) holds by noting that the cycle containing $\widehat{\tau}_k$ has only $m$ data points, and that between $\widehat{\tau}_{k}$ and $\widehat{\tau}_{k+1}$, there are exactly $({\widehat{\tau}_{k+1}-\widehat{\tau}_k-m})/{l_k}$ cycles of length $l_k$ and one cycle of length $m$;   (b) holds by noting that $\widehat{\tau}_{\Upsilon_T+1}=T$ when $\mathcal{A}$ holds. 

Note that $\max_{0\leq k\leq \Upsilon_T}(T/l_k)=T/{l_{\Upsilon_T}}\leq 2\sqrt{T\Upsilon_T}$, and $\Upsilon_T\leq \sqrt{T\Upsilon_T}$, we thus obtain that $
R_{T,I}\leq 4C_rC_b^2 m\sqrt{T\Upsilon_T}.$ The result then follows. \qed 

\medskip 

\noindent\textbf{Proof of Lemma \ref{lem_regret_delay}}: By \Cref{assum_model} (iv),   $|\varphi(\theta_t,z_t)-p_t|\leq C_b$. We have that 
\begin{flalign*}
R_{T,\mathrm{II}}\leq  C_rC_b^2\sum_{k=1}^{\Upsilon_T}\mathbb{E}\left[ \sum_{t={\tau}_k+1}^{\widehat{\tau}_k
}\mathbb{I}(t\in\mathcal{M}^c,\mathcal{A})\right]\leq C_rC_b^2\sum_{k=1}^{\Upsilon_T}\mathbb{E}\left[ \lceil\frac{\widehat{\tau}_k-\tau_k}{l_{k-1}}\rceil n_{k-1}\mathbb{I}(\mathcal{A})\right].
\end{flalign*}
Note that on event $\mathcal{A}$, $\tau_k\leq \widehat{\tau}_k\leq \tau_k+d_k$, and $d_k=\lceil c_{\mathrm{snr}}/(c_m\kappa_k^2)\rceil l_{k-1}$, with $c_{\mathrm{snr}}=c_5c_{\lambda}^2$ defined in \Cref{lem-one-change_alter},  we have
\begin{flalign*}
    R_{T,\mathrm{II}}\leq 2C_rC_b^2\sum_{k=1}^{\Upsilon_T} \frac{d_k}{l_{k-1}}n_{k-1}=2C_rC_b^2c_\lambda^2\sum_{k=1}^{\Upsilon_T}\lceil c_{5}/(c_m\kappa_k^2)\rceil \sqrt{T/k}\leq C_{R,\mathrm{II}}\kappa_{\min}^{-2}\sum_{k=1}^{\Upsilon_T}\sqrt{T/k}.
\end{flalign*}
\qed

\medskip 

\noindent\textbf{Proof of Lemma \ref{lem_regret_error}}: Define ${\mathcal{A}}_k^{\circ}=\{\widehat{\tau}_k\in[\tau_k,\tau_k+d_k]\}$ and $\mathcal B_{t,k}=\{\widehat{\Upsilon}_t=k\}$, where $\widehat{\Upsilon}_t$ denotes the number of detected change-points up to time $t.$ Note that by definition, we have $\mathcal{A}\subseteq {\mathcal{A}}_k^{\circ}$ and $\mathbb{I}(\mathcal{A}_k^{\circ})=\sum_{i=\tau_k}^{\tau_k+d_k} \mathbb{I}(\widehat{\tau}_k=i)$. We have that  
\begin{flalign}
\notag R_{T,\mathrm{III}}=&C_r\sum_{k=0}^{\Upsilon_T}\mathbb{E}\left[ \sum_{t=\widehat{\tau}_{k}+1}^{{\tau}_{k+1}}(\varphi(\theta_t, z_t)-\varphi(\widehat{\theta}_t,z_t))^2\mathbb{I}(t\in\mathcal{M}^c,\mathcal{A})\right]  
\\\notag \stackrel{(a)}{\leq} & C_rC_{\varphi}^2\sum_{k=0}^{\Upsilon_T}\mathbb{E}\left[ \sum_{t=\widehat{\tau}_{k}+1}^{{\tau}_{k+1}} \{|(\alpha_t-\widehat{\alpha}_t)^{\top}z_t|+|\beta_t-\widehat{\beta}_t|\}^2\mathbb{I}(\mathcal{A})\mathbb{I}(t\in\mathcal{M}_k^c)\right]  
\\\notag {=} & 
C_rC_{\varphi}^2\sum_{k=0}^{\Upsilon_T}\mathbb{E}\left[ \sum_{i=\tau_k}^{\tau_k+d_k}\sum_{t=i+1}^{{\tau}_{k+1}}\{|(\alpha_t-\widehat{\alpha}_t)^{\top}z_t|+|\beta_t-\widehat{\beta}_t|\}^2\mathbb{I}(\mathcal{A})\mathbb{I}(t\in\mathcal{M}_k^c) \mathbb{I}(\widehat{\tau}_k=i)\right]
\\\notag \stackrel{(b)}{\leq }&2C_rC_{\varphi}^2\sum_{k=0}^{\Upsilon_T}\mathbb{E}\left[ \sum_{i=\tau_k}^{\tau_k+d_k}\sum_{t=i+1}^{{\tau}_{k+1}}\{|(\alpha_t-\widehat{\alpha}_t)^{\top}z_t|^2+|\beta_t-\widehat{\beta}_t|^2\}\mathbb{I}(\mathcal{B}_{t,k})\mathbb{I}(t\in\mathcal{M}_k^c) \mathbb{I}(\widehat{\tau}_k=i)\right]
\\\notag \stackrel{(c)}{\leq }&2C_rC_{\varphi}^2\sum_{k=0}^{\Upsilon_T}\mathbb{E}\left[ \sum_{i=\tau_k}^{\tau_k+d_k}\sum_{t=i+1}^{{\tau}_{k+1}}(\lambda_{\max}(\Sigma)\vee 1)\|\theta_t-\widehat{\theta}_t\|_2^2\mathbb{I}(\mathcal{B}_{t,k})\mathbb{I}(t\in\mathcal{M}_k^c) \mathbb{I}(\widehat{\tau}_k=i)\right]
\\\label{RT3_bound}= &2C_rC_{\varphi}^2(\lambda_{\max}(\Sigma)\vee 1)\sum_{k=0}^{\Upsilon_T}\mathbb{E}\left[ \sum_{i=\tau_k}^{\tau_k+d_k}\mathbb{I}(\widehat{\tau}_k=i)\mathbb{E}\left(\sum_{t=i+1}^{{\tau}_{k+1}}\|\theta_t-\widehat{\theta}_t\|_2^2\mathbb{I}(\mathcal{B}_{t,k})\mathbb{I}(t\in\mathcal{M}_k^c) \bigg\vert\widehat{\tau}_k=i \right)\right],
\end{flalign}
where $(a)$ holds by \eqref{C_phi}, and $(b)$ holds by the definition of $\mathcal{A}$ and $\mathcal{B}_{t,k}$, and Cauchy-Schwarz inequality, and (c) holds due to tower property of conditional expectation, by noting that due to temporal independence and design of the policy, $z_t$ is independent from $\widehat{\theta}_t,\mathbb{I}(\mathcal{B}_{t,k}),\mathbb{I}(t\in\mathcal{M}_k^c),$ and $\mathbb{I}(\widehat{\tau}_k=i)$ for $t\geq i+1.$

Note that $\sum_{i=\tau_k}^{\tau_k+d_k}\mathbb{I}(\widehat{\tau}_k=i)\leq 1$. Thus, we can control $R_{T,\mathrm{III}}$ if an upper bound on 
$$\mathbb{E}\left(\sum_{t=i+1}^{{\tau}_{k+1}}\|\theta_t-\widehat{\theta}_t\|_2^2\mathbb{I}(\mathcal{B}_{t,k})\mathbb{I}(t\in\mathcal{M}_k^c) \bigg\vert\widehat{\tau}_k=i \right)$$
can be established for $i\in[\tau_k,\tau_k+d_k]$ uniformly.  
% \textcolor{red}{$[a, b]$ and $\{a, \ldots, b\}$ are both used.  we should make them consistent.}

Given $\widehat{\tau}_k=i$, on the event $\mathcal{B}_{t,k}$, we note that we can partition  $[i+1,\tau_{k+1}]\cap \mathcal{M}_k^c$ into $J_k$ groups of size at most $n_k$.  That is,  $[i+1,\tau_{k+1}]\cap \mathcal{M}_k^c=\bigcup_{j=1}^{J_k} G_j$ such that $G_j=\{ t\in [i+1,\tau_{k+1}]\cap\mathcal{M}_k^c: \lceil \frac{t-i}{l_k}\rceil= j \}$ and $|G_j|\leq n_k$.
Furthermore, for the periods in the $j$th group, the Lasso estimators are the same, i.e., $\widehat{\theta}_t\equiv \widehat{\theta}^j(k)$ for $t\in G_j$.  Let $$
L_t=c_2c_{\lambda}\sqrt{\frac{s\log (Td)}{mj}}, ~ t\in G_j%\mathbb{I}(t\in [i+1,\tau_{k+1}]\cap\mathcal{M}_k^c),
$$
with $c_2$ defined in \Cref{lem:est}. Define $\mathcal{U}_t=\left\{\|\theta_t-\widehat{\theta}_t\|_2\leq L_t\right\}$. We have that
\begin{flalign}
&\mathbb{E}\left(\sum_{t=i+1}^{{\tau}_{k+1}}\|\theta_t-\widehat{\theta}_t\|_2^2\mathbb{I}(\mathcal{B}_{t,k})\mathbb{I}(t\in\mathcal{M}_k^c) \bigg\vert\widehat{\tau}_k=i \right)\nonumber\\
\notag= & \mathbb{E}\left(\sum_{t=i+1}^{{\tau}_{k+1}}\|\theta_t-\widehat{\theta}_t\|_2^2\mathbb{I}(\mathcal{B}_{t,k})\mathbb{I}(t\in\mathcal{M}_k^c) \mathbb{I}(\mathcal{U}_t)\bigg\vert\widehat{\tau}_k=i \right) + \mathbb{E}\left(\sum_{t=i+1}^{{\tau}_{k+1}}\|\theta_t-\widehat{\theta}_t\|_2^2\mathbb{I}(\mathcal{B}_{t,k})\mathbb{I}(t\in\mathcal{M}_k^c) \mathbb{I}(\mathcal{U}_t^c) \bigg\vert\widehat{\tau}_k=i \right)
\\\label{RT3}\stackrel{(d)}{\leq} &\mathbb{E}\left(\sum_{t=i+1}^{{\tau}_{k+1}}L_t^2\mathbb{I}(\mathcal{B}_{t,k})\mathbb{I}(t\in\mathcal{M}_k^c)\bigg\vert\widehat{\tau}_k=i \right)+4C_{\theta}^2\mathbb{E}\left(\sum_{t=i+1}^{{\tau}_{k+1}}\mathbb{I}(\mathcal{B}_{t,k})\mathbb{I}(t\in\mathcal{M}_k^c) \mathbb{I}(\mathcal{U}_t^c)\bigg\vert\widehat{\tau}_k=i \right),
\end{flalign}
where (d) holds by definition of $\mathcal{U}_t$ and that $\|\theta_t-\widehat{\theta}_t\|^2_2\leq (\|\theta_t\|_2+\|\widehat{\theta}_t\|_2)^2\leq 4C_{\theta}^2.$

For the first term in \eqref{RT3}, we have 
\begin{flalign*}
\mathbb{E}\left(\sum_{t=i+1}^{{\tau}_{k+1}}L_t^2\mathbb{I}(\mathcal{B}_{t,k})\mathbb{I}(t\in\mathcal{M}_k^c)\bigg\vert\widehat{\tau}_k=i \right)\leq c_2^2c_{\lambda}^2 \sum_{j=1}^{J_k} n_{k} \frac{s\log(Td)}{mj} 
\stackrel{(e)}{\leq}& c_2^2c_{\lambda}^2 c_m^{-1}C_L n_k\log(J_k)
\\{\leq}&c_2^2c_{\lambda}^2 c_m^{-1}C_L n_k\log T, 
\end{flalign*}
where $(e)$ holds by elementary inequality that for some  $C_L>0$, $\sum_{i=1}^{k}i^{-1}\leq C_L\log(k)$ and that $m=c_ms\log(Td)$. %and (f) holds   under  \Cref{ass_spacing} with $c_{\mathrm{snr}}^*>2c_{\mathrm{snr}}/c_m$, we have $\tau_{k+1}>\tau_k+d_k$, and hence $J_k\leq (\tau_{k+1}-i)/l_k<\tau_{k+1}-\tau_k$.

For the second term in \eqref{RT3}, we have that for  $t\in\{i+1,\cdots,\tau_{k+1}\}\cap \mathcal{M}_k^c$,  given $\widehat{\tau}_k=i$, on the event $\mathcal{B}_{t,k}$, there are at least $m>c_1s\log(Td)$ periods used for estimation, hence by \Cref{lem:est},
$$
\mathbb{E}\left(\sum_{t=i+1}^{{\tau}_{k+1}}\mathbb{I}(\mathcal{B}_{t,k})\mathbb{I}(t\in\mathcal{M}_k^c) \mathbb{I}(\mathcal{U}_t^c)\bigg\vert\widehat{\tau}_k=i \right)\leq \sum_{t=i+1}^{{\tau}_{k+1}} T^{-4}\leq T^{-3}.
$$

Together, we have\begin{flalign*}
    R_{T,\mathrm{III}}\leq& 2C_rC_{\varphi}^2(\lambda_{max}(\Sigma)\vee 1) [c_2^2c_{\lambda}^2 c_m^{-1}C_L\sum_{k=0}^{\Upsilon_T}{n_k}\log T+4C_{\theta}^2(\Upsilon_T+1)T^{-3}]
    \\\leq& C_{R,\mathrm{III}}[\sum_{k=0}^{\Upsilon_T}{n_k}\log T]\leq 2 C_{R,\mathrm{III}} \sqrt{\Upsilon_T T}\log T.
\end{flalign*}
\qed

\medskip
\noindent\textbf{Proof of Lemma \ref{lem_regret_bad}}: It is clear that $$R_{T,\mathrm{IV}}=\mathbb{E}\left[\sum_{t=1}^T C_r(\varphi(\theta_t, z_t)-p_t)^2\mathbb{I}(\mathcal{A}^c)\right]\leq C_rC_b^2T\mathbb{P}(\mathcal{A}^c).$$

Thus, we only need to bound $\mathbb{P}(\mathcal{A}^c)$. Denote $\mathcal{A}_k$ as the event that the first $k$ change-points have been detected within $d_j$ time points for $j\leq k$, i.e., $\mathcal{A}_{k}=\{\text{for all } 1\leq j\leq k, \tau_j\leq \widehat{\tau}_j\leq \tau_j+d_j \}$.   Recall  $\widehat{\tau}_0=\tau_0=0$ and let $\mathcal{A}_0=\varnothing$. We have
\begin{flalign*}
\mathbb{P}(\mathcal{A}^c)\leq \sum_{k=1}^{\Upsilon_T+1}\mathbb{P}(\widehat{\tau}_k< \tau_k|\mathcal{A}_{k-1})+\sum_{k=1}^{\Upsilon_T}\mathbb{P}(\widehat{\tau}_k>\tau_k+d_k|\mathcal{A}_{k-1}),
\end{flalign*}
where the first term corresponds to false alarm, and the second term corresponds to large delay. 

%{\color{red}should we include $\widehat{\tau}_{\Upsilon_T+1}$ in $\mathcal{A}$?}

{\bf (I). Controlling the false alarm:} 
%Note that there are at least $3$ cycles of price experimentation between $\widehat{\tau}_k$ and $\widehat{\tau}_{k-1}$ by the change-point detection algorithm (with the last cycle of length $2m_{k-1}$), we thus have $\widehat{\tau}_k\geq \widehat{\tau}_{k-1}+(2l_{k-1}+2m_{k-1})$. 
Fix $k\in\{1,2,\ldots,{\Upsilon_{T}+1}\}$. Conditional on $\mathcal{A}_{k-1}$~(in fact $\widehat{\tau}_{k-1}$), we declare a change-point before $\tau_k$ if for some $r\in[\widehat{\tau}_{k-1}+1, \tau_k-1]$, $\text{CPT}([\widehat{\tau}_{k-1}+1, r]\cap \mathcal{M}_{k-1},\lambda,\gamma,m)$ declares a change-point. We  denote this event as $\text{CPT}([\widehat{\tau}_{k-1}+1, r]\cap \mathcal{M}_{k-1})$ for simplicity. 
Thus, we have that
\begin{flalign*}
    &\mathbb{P}(\widehat{\tau}_k< \tau_k|\mathcal{A}_{k-1})
    \\=&\mathbb{P}\left(\bigcup_{r=\widehat{\tau}_{k-1}+1}^{\tau_k-1} \text{CPT}([\widehat{\tau}_{k-1}+1, r]\cap \mathcal{M}_{k-1})\bigg\vert\mathcal{A}_{k-1}\right)    
    %\\\leq&\mathbb{P}\left(\bigcup_{r=\widehat{\tau}_{k-1}+1}^{\tau_k}\bigcup_{q=\widehat{\tau}_{k-1}+1}^{r} \mathcal{D}(\widehat{\tau}_{k-1}+1,q,r)|\mathcal{A}_{k-1}\right)
    \\ \stackrel{(a)}{=} &  \mathbb{E}\left(\sum_{s=\tau_{k-1}}^{\tau_{k-1}+d_{k-1}}\mathbb{I}(\widehat{\tau}_{k-1}=s)\mathbb{P}\left(\bigcup_{\substack{r=s+1,\\r\in\mathcal{M}_{k-1}}}^{\tau_k-1}\text{CPT}([s+1, r]\cap \mathcal{M}_{k-1})  \bigg\vert\widehat{\tau}_{k-1}=s\right)\bigg\vert\mathcal{A}_{k-1}\right),
\end{flalign*} 
where $(a)$ holds by the tower property of conditional expectation and the fact that conditionally on $\mathcal{A}_{k-1}$, $\widehat{\tau}_{k-1}\in [\tau_{k-1}+1,\tau_{k-1}+d_{k-1}]$.

When $\text{CPT}([s+1, r]\cap \mathcal{M}_{k-1})$ holds, we have for some
$q\in [s+1, r]$, the following event holds, i.e.\
\begin{align*}
\widetilde{\mathcal{D}}(s+1,q,r)=\left\{ L(\widehat{\theta}_I, I) - L(\widehat\theta_{I_1}, I_1) - L(\widehat{\theta}_{I_2}, I_2) + \lambda \sqrt{|I_1|} \|\widehat{\theta}_I-\widehat{\theta}_{I_1}\|_1 + \lambda \sqrt{|I_2|} \|\widehat{\theta}_I-\widehat{\theta}_{I_2}\|_1>\gamma\right\},
\end{align*}
where $I = [s+1, r]\cap \mathcal{M}_{k-1}$, $I_1 = [s+1, q]\cap \mathcal{M}_{k-1}$ and $I_2 = [q+1, r]\cap \mathcal{M}_{k-1}$.

Denote $\mathbb{I}(\mathcal{D}^*(s+1,q,r))=\mathbb{I}(|[s+1,q]\cap \mathcal{M}_{k-1}| \geq m) \mathbb{I}(|[q+1,r]\cap \mathcal{M}_{k-1}| \geq m)$.
For each $s$, we further have that
\begin{align*}
&\mathbb{P}\left(\bigcup_{\substack{r=s+1,\\r\in\mathcal{M}_{k-1}}}^{\tau_k-1}\text{CPT}([s+1, r]\cap \mathcal{M}_{k-1})  \bigg\vert\widehat{\tau}_{k-1}=s\right)
\\=&\mathbb{E}\left(\bigcup_{\substack{r=s+1,\\r\in\mathcal{M}_{k-1}}}^{\tau_k-1}\bigcup_{\substack{q=s+1,\\q\in\mathcal{M}_{k-1}}}^{r}\mathbb{I}(\widetilde{\mathcal{D}}(s+1,q,r))\mathbb{I}(\mathcal{D}^*(s+1,q,r)) \bigg\vert\widehat{\tau}_{k-1}=s\right)
\\\stackrel{(b)}{\leq}  &\sum_{r=s+1,r\in\mathcal{M}_{k-1}}^{\tau_k-1}\sum_{q=s+1,q\in\mathcal{M}_{k-1}}^{r}\mathbb{I}(\mathcal{D}^*(s+1,q,r))\mathbb{P}( \widetilde{\mathcal{D}}(s+1,q,r)|\widehat{\tau}_{k-1}=s)
\\\stackrel{(c)}{\leq}  & \left(\frac{mT}{l_{k-1}}\right)^23T^{-4}= 3m^2l_{k-1}^{-2}T^{-2}\leq 3T^{-2},
\end{align*}
where (b) holds by union bound and (c) follows by \Cref{lem-no-change}, and that the total number of tuples $(q,r)$ is smaller than $[(\tau_k-\tau_{k-1})m/l_{k-1}]^2\leq [mT/l_{k-1}]^2.$  Thus, recalling $\sum_{s=\tau_{k-1}}^{\tau_{k-1}+d_{k-1}}\mathbb{I}(\widehat{\tau}_{k-1}=s)\leq 1$ conditional on $\mathcal{A}_{k-1}$, we have that $\mathbb{P}(\widehat{\tau}_k< \tau_k|\mathcal{A}_{k-1})\leq 3T^{-2}$.

{\bf (II). Controlling large delay:}
Conditional on $\mathcal{A}_{k-1}$~(in fact $\widehat{\tau}_{k-1}$), if $\widehat{\tau}_k>\tau_k+d_k$, then  for all $r\in\{\widehat{\tau}_{k-1}+1,\cdots,\tau_k+1,\cdots,\tau_k+d_k\}\cap\mathcal{M}_{k-1}$,  no change-points are declared between $\widehat{\tau}_{k-1}$  and $r$. Hence, for all such $r$, the event $\text{CPT}^c([\widehat{\tau}_{k-1}+1, r]\cap \mathcal{M}_{k-1})$ holds.

%for all such $r$ and $q\in\{\widehat{\tau}_{k-1}+1,\cdots, r\}\cap\mathcal{M}_{k-1}$,  events $\mathcal{D}^c(\widehat{\tau}_{k-1}+1,q,r)$ hold. Denote $\mathcal{N}(t)=\max_{s}\{s\leq t, s\in\mathcal{M}\}$ as the earliest time point before  (or including) $t$ that conducts the price experiment. We then have ,\\r\in\mathcal{M}_{k-1}
Thus, we have
\begin{flalign*}
    &\mathbb{P}(\widehat{\tau}_k> \tau_k+d_k|\mathcal{A}_{k-1})
    \\=&\mathbb E\left(\sum_{s=\tau_{k-1}}^{\tau_{k-1}+d_{k-1}}\mathbb{I}(\widehat{\tau}_{k-1}=s)\mathbb P \left( \bigcap_{\substack{r=s+1}}^{\tau_k+d_k}
\text{CPT}^c([s+1, r]\cap \mathcal{M}_{k-1})\bigg\vert\widehat{\tau}_{k-1}=s\right)\bigg\vert\mathcal{A}_{k-1}\right).
\end{flalign*}
Furthermore, for each $s$, we have
\begin{flalign*}    
&\mathbb P \left( \bigcap_{\substack{r=s+1}}^{\tau_k+d_k}\text{CPT}^c([s+1, r]\cap \mathcal{M}_{k-1})\bigg\vert\widehat{\tau}_{k-1}=s\right)
\\=&\mathbb E \left( \bigcap_{\substack{r=s+1}}^{\tau_k+d_k}\bigcap_{\substack{q=s+1}}^{r} \mathbb{I}(\widetilde{\mathcal{D}}^c(s+1,q,r))\mathbb{I}(\mathcal{D}^*(s+1,q,r))\bigg\vert\widehat{\tau}_{k-1}=s\right)
\\\stackrel{(a)}{\leq} & \mathbb{P}( \widetilde{\mathcal{D}}^c(s+1, \tau_k, \tau_k+d_{k})|\widehat{\tau}_{k-1}=s)
\end{flalign*}
where $(a)$ holds by monotonicity of probability and the fact that $\mathbb{I}(\mathcal{D}^*(s+1,\tau_k,\tau_k+d_k))=1$. To see this, note that by Assumption \ref{ass_spacing} with $c_{\mathrm{snr}}^*>2c_{\mathrm{snr}}/c_m$, for all $s\in[\tau_{k-1},\tau_{k-1}+d_{k-1}]$, we have $$\tau_k-s \geq \tau_k-\tau_{k-1}-d_{k-1} {\geq} d_{k}.$$

%(b) bolds by the fact that conditional on $\mathcal{A}_{k-1}$, $\widehat{\tau}_{k-1}\in\{\tau_{k-1}+1,\cdots,\tau_{k-1}+d_{k-1}\}$; and (c) by union bound.
%Note that we choose $r= \tau_k+d_k $, $q= \tau_k$ such that 

Therefore,  there exist at least $\lceil c_{\mathrm{snr}}/(c_m{\kappa_k^2})\rceil$ cycles between $s+1$ and $\tau_k$, and between $\tau_k+1$ and $\tau_k+d_k$, respectively, suggesting at least $\lceil c_{\mathrm{snr}}/(c_m{\kappa_k^2})\rceil m$ data points from both segments used for change-point detection. In addition, note that the data points in $[s+1, \tau_k]\cap\mathcal{M}$ and $[\tau_k+1,\tau_k+d_k]\cap\mathcal{M}$  form two separate stationary segments with a change size $\kappa_k.$

Thus, by Lemma \ref{lem-one-change_alter},  we have that 
$$
\mathbb{P}( \widetilde{\mathcal{D}}^c(s+1, \tau_k, \tau_k+d_{k})|\widehat{\tau}_{k-1}=s)\leq 3T^{-4}.
$$
Recalling $\sum_{s=\tau_{k-1}}^{\tau_{k-1}+d_{k-1}}\mathbb{I}(\widehat{\tau}_{k-1}=s)\leq 1$ conditional on $\mathcal{A}_{k-1}$, we have that $\mathbb{P}(\widehat{\tau}_k>\tau_k+d_k|\mathcal{A}_{k-1})\leq 3T^{-4}$.

Summarizing the above results, we have $R_{T,\mathrm{IV}}\leq 3C_rC_b^2T[({\Upsilon_T+1}) T^{-2}+ {\Upsilon_T}T^{-4}]\leq C_{R,\mathrm{IV}}$.
\qed

\medskip
\noindent\textbf{Proof of Lemma \ref{lem_regret_delay2}}: Recall  
\begin{flalign*}
    R_{T,\mathrm{II}}=&C_r\sum_{k=1 }^{\Upsilon_T}\mathbb{E}\left[ \sum_{t={\tau}_k+1}^{\widehat{\tau}_k
}(\varphi(\theta_t, z_t)-p_t)^2\mathbb{I}(t\in\mathcal{M}^c,\mathcal{A})\right]\\\leq &C_r\sum_{k=1 }^{\Upsilon_T}\mathbb{E}\left[ \sum_{t={\tau}_k+1}^{(\tau_k+l_{k-1})\wedge \widehat{\tau}_k}(\varphi(\theta_t, z_t)-p_t)^2\mathbb{I}(t\in\mathcal{M}^c,\mathcal{A})\right]
    \\&+C_r\sum_{k=1 }^{\Upsilon_T}\mathbb{E}\left[ \sum_{t=\tau_k+l_{k-1}+1}^{\widehat{\tau}_k}(\varphi(\theta_t, z_t)-p_t)^2\mathbb{I}(t\in\mathcal{M}^c,\mathcal{A},\widehat{\tau}_k>\tau_k+l_{k-1})\right]
    \\=&R_{T,\mathrm{II}}^{(1)}+R_{T,\mathrm{II}}^{(2)}.
\end{flalign*}

Clearly, we have 
\begin{equation}\label{RT21}
    R_{T,\mathrm{II}}^{(1)}\leq C_rC_b^2\sum_{k=1 }^{\Upsilon_T}n_{k-1}.
\end{equation}

Recall $\mathcal{A}_k^{\circ}=\{\widehat{\tau}_k\in[\tau_k,\tau_k+d_k]\}$ and $\mathcal B_{t,k-1}=\{\widehat{\Upsilon}_t=k-1\}$, where $\widehat{\Upsilon}_t$ denotes the number of detected change-points up to time $t.$
By similar arguments as \eqref{RT3_bound} used in the proof of Lemma \ref{lem_regret_error}, we have that 
\begin{flalign*}&R_{T,\mathrm{II}}^{(2)}\leq 
% & C_rC_{\varphi}^2\sum_{k=1}^{\Upsilon_T}\mathbb{E}\left[ \mathbb{E}\left(\sum_{t=\tau_k+l_{k-1}+1}^{{\tau}_{k}+d_k}\{(\beta_t-\widehat{\beta}_t)^{\top}z_t\}^2\mathbb{I}(\mathcal{A}_{k-1}^{\circ})\mathbb{I}(t\in\mathcal{M}_{k-1}^c) \mathbb{I}(\mathcal B_{t,k-1})\right)\right]
% \\\leq & 
2C_rC_{\varphi}^2(\lambda_{max}(\Sigma)\vee 1)\\&\times\sum_{k=1}^{\Upsilon_T}\mathbb{E}\left[ \sum_{i=\tau_k+l_{k-1}+1}^{\tau_k+d_k}\mathbb{I}(\widehat{\tau}_k=i)\mathbb{E}\left(\sum_{t=\tau_k+l_{k-1}+1}^{i}\|\widehat{\theta}_t-\theta_t\|_2^2\mathbb{I}(\mathcal{A}_{k-1}^{\circ})\mathbb{I}(t\in\mathcal{M}_{k-1}^c) \mathbb{I}(\mathcal B_{t,k-1})\bigg\vert\widehat{\tau}_k=i\right)\right].
\end{flalign*}

 When $\mathcal{B}_{t,k-1}\cap \mathcal{A}_{k-1}^{\circ}$ holds,   for $\widehat{\tau}_k=i>\tau_k+l_{k-1}$, and $t\in [\tau_k+l_{k-1}+1,i]\cap \mathcal{M}_{k-1}^c$,  the parameter $\widehat{\theta}_t$ is estimated based on the mixture of two stationary segments $[\widehat{\tau}_{k-1}+1, \tau_k]\cap\mathcal{M}_{k-1}$ and $[\tau_{k}+1,t]\cap\mathcal{M}_{k-1}$. Furthermore, we have that $\left|[\widehat{\tau}_{k-1}+1,\tau_k]\cap\mathcal{M}_{k-1}\right|\geq m$ and 
$|[\tau_{k}+1,t]\cap\mathcal{M}_{k-1}|\geq m$.

We then  partition $[\tau_{k}+l_{k-1}+1,i]\cap\mathcal{M}_{k-1}^c$ into $J_{k}^*$ groups of size at most $n_{k-1}$ such that $[\tau_{k}+l_{k-1}+1,i]\cap\mathcal{M}_{k-1}^c=\bigcup_{j}G_j^*$, with $G_j^*=\{t\in [\tau_{k}+l_{k-1}+1,i]\cap\mathcal{M}_{k-1}^c:  \lceil\frac{(t-\widehat{\tau}_{k-1})}{l_{k-1}}\rceil 
=j\}$. Note here $j$ does not start with $j=1$, but  we can ensure the enumeration of $j$ is consecutive and $J_{k}^*\leq (d_k-l_{k-1})/{l_{k-1}}=\lceil c_{\mathrm{snr}}/(c_m \kappa_k^2)\rceil-1$.
Also note that the Lasso estimators are the same for the periods in $G_j^*$, i.e., $\widehat{\theta}_t\equiv \widehat{\theta}^{j*}(k-1)$ for $t\in G_j^*$.

Define $$L_t^*=\sqrt{\frac{c_2'c_{\lambda}^2 s\log(Td)}{mj}+c_3'\kappa_{k}^2}+C_M\sqrt{2s}\kappa_k,~t\in G_j^*$$
and denote $\mathcal{U}_t^*=\left\{\|\theta_t-\widehat{\theta}_t\|_2\leq L_t^*\right\}$, where $c_2',c_3'$ are taken from \Cref{lem_mixture}.

To proceed, for $t=\tau_k+l_{k-1}+1,\ldots, i$,  we have, 
\begin{flalign*}
    &\mathbb{E}\left(\|\widehat{\theta}_t-\theta_t\|_2^2\mathbb{I}(\mathcal{A}_{k-1}^{\circ})\mathbb{I}(t\in\mathcal{M}_{k-1}^c) \mathbb{I}(\mathcal B_{t,k-1})\bigg\vert\widehat{\tau}_k=i\right)
    \\=&\mathbb{E}\left(\|\widehat{\theta}_t-\theta_t\|_2^2\mathbb{I}(\mathcal{A}_{k-1}^{\circ})\mathbb{I}(t\in\mathcal{M}_{k-1}^c) \mathbb{I}(\mathcal B_{t,k-1}) \mathbb{I}( \mathcal{U}_t^*)\bigg\vert\widehat{\tau}_k=i\right)\\&+\mathbb{E}\left(\|\widehat{\theta}_t-\theta_t\|_2^2\mathbb{I}(\mathcal{A}_{k-1}^{\circ})\mathbb{I}(t\in\mathcal{M}_{k-1}^c) \mathbb{I}(\mathcal B_{t,k-1}) \mathbb{I}(\mathcal{U}_t^{*c})\bigg\vert\widehat{\tau}_k=i\right)
    \\\leq& (L_t^*)^2 + 4C_{\theta}^2 \mathbb{E}\left(\mathbb{I}(\mathcal{A}_{k-1}^{\circ})\mathbf \mathbb{I}(\mathcal B_{t,k-1}) \mathbb{I}(\mathcal{U}_t^{*c})\bigg\vert\widehat{\tau}_k=i\right).
\end{flalign*}
For $t\in G_j^*$, denote $\theta_t^*$ as the pseudo true value of based on mixture distributions, then  
\begin{flalign*}
   &\mathbb{E}\left(\mathbb{I}(\mathcal{A}_{k-1})\mathbf \mathbb{I}(\mathcal B_{t,k-1}) \mathbb{I}(\mathcal{U}_t^{*c})\vert\widehat{\tau}_k=i\right)\leq \mathbb{P}(\mathcal{U}_t^{*c}\vert\widehat{\tau}_k=i)
   \leq  \mathbb{P}( \|\theta_t-\theta_t^*\|_2+\|\widehat{\theta}_t-\theta_t^*\|_2> L_t^* \vert\widehat{\tau}_k=i)
   \\\leq &\mathbb{P}( \|\theta_t-\theta_t^*\|_2>C_M\sqrt{2s}\kappa_k \vert\widehat{\tau}_k=i)+  \mathbb{P}\left( \|\widehat{\theta}_t-\theta_t^*\|_2^2\geq {\frac{c_2'c_{\lambda}^2 s\log(Td)}{mj}+c_3'\kappa_{k}^2} \vert\widehat{\tau}_k=i\right)\leq 0+T^{-4},
\end{flalign*}
where the last inequality holds  by \Cref{lem_mixture} and by Assumption \ref{assum:mixture} where $\|\theta_t-\theta_t^*\|_2\leq \|\theta_t-\theta_t^*\|_1\leq C_M\|\theta_{\tau_k}-\theta_{\tau_k+1}\|_1\leq C_M\sqrt{2s}\kappa_k$ using that the sparsity level of $\theta_{\tau_k}-\theta_{\tau_k+1}$ is at most $2s$.

Therefore,  we have that 
\begin{flalign*}
  &\mathbb{E}\left(\sum_{t=\tau_k+l_{k-1}+1}^{i}\|\widehat{\theta}_t-\theta_t\|_2^2\mathbb{I}(\mathcal{A}_{k-1}^{\circ})\mathbb{I}(t\in\mathcal{M}_{k-1}^c) \mathbb{I}(\mathcal B_{t,k-1})\bigg\vert\widehat{\tau}_k=i\right) 
  \leq  \sum_j \sum_{t\in G_j^*} [L_t^{*2}+4C_{\theta}^2T^{-4}]
  \\\leq & 2n_{k-1} \sum_j  \left(\frac{c_2'c_{\lambda}^2}{c_m j}+c_3'\kappa_{k}^2+2C_M^2{s}\kappa_k^2+2C_{\theta}^2 T^{-4}\right)
  \\\leq & 2n_{k-1}\sum_{j=1}^{\lceil c_{\mathrm{snr}}/(c_m \kappa_k^2)\rceil-1}  \left(\frac{c_2'c_{\lambda}^2}{c_m j}+c_3'\kappa_{k}^2+2C_M^2{s}\kappa_k^2+2C_{\theta}^2 T^{-4}\right) 
  \\\leq & 2n_{k-1}\left(\frac{c_2'c_{\lambda}^2}{c_m }C_L\log T+(c_3'+2C_M^2{s}) c_{\mathrm{snr}}/c_m+2C_{\theta}^2 T^{-3}\right),
\end{flalign*}
where the last inequality uses the fact $\lceil c_{\mathrm{snr}}/(c_m \kappa_k^2)\rceil=d_k/l_{k-1}<d_k<T$.

Using the fact that $\sum_{i=\tau_k+l_{k-1}+1}^{\tau_k+d_k}\mathbb{I}(\widehat{\tau}_k=i)\leq 1$, and recall that $c_{\mathrm{snr}}=c_5c_{\lambda}^2$,  we thus obtain that 
\begin{align*}
    R_{T,\mathrm{II}}^{(2)}\leq &  2C_rC_{\varphi}^2(\lambda_{max}(\Sigma)\vee 1)\sum_{k=1}^{\Upsilon_T}2n_{k-1}\left(\frac{c_2'c_{\lambda}^2}{c_m }C_L\log T+(c_3'+2C_M^2{s}) c_{\mathrm{snr}}/c_m+2C_{\theta}^2 T^{-3}\right).
\\\leq & C_{R,\mathrm{II}}'' \sum_{k=1}^{\Upsilon_T} n_{k-1} (\log T +s)
\end{align*}
Combined with \eqref{RT21}, we thus have that $$ R_{T,\mathrm{II}}^{(2)}\leq 2C_{R,\mathrm{II}}' \sqrt{\Upsilon_T T}(\log T +s).$$
\qed

\section[]{Proof of \Cref{thm:lowerbound}} \label{sec:proo_lower}

In this section, we give the proof of \Cref{thm:lowerbound}, which provides the regret lower bound. In particular, we consider the following linear demand model.

Let $y_t=\alpha_t+\beta_tp_t+\epsilon_t$ with $\epsilon_t\stackrel{i.i.d.}{\sim}\mathcal{N}(0,1)$ and $\theta_t=(\alpha_t,\beta_t).$ Given $\theta_t$, the optimal price is $\varphi(\theta_t)=-\alpha_t/(2\beta_t).$ Consider the setting where there are $\Upsilon_T$ number of change-points and thus partition $\{1,2,\cdots,T\}$ into $\Upsilon_T+1$ stationary segments. For $k=0,\ldots, \Upsilon_{T}$, the $k$th stationary segment is of length $$N_k=\frac{T}{(k+1) C_{\Upsilon_T}},$$ where $C_{\Upsilon_T}=\sum_{k=0}^{\Upsilon_T}(k+1)^{-1}$. For $k=1,\dots, \Upsilon_{T}$, denote the change-points as $\tau_k=\sum_{i=0}^{k-1}N_i$. Furthermore, define $\tau_0=0$ and $\tau_{\Upsilon_T+1}=T$. Denote $\theta^{(k)}=\theta_{\tau_k+1}=\cdots=\theta_{\tau_{k+1}}$ as the parameter between $\tau_k+1$ and $\tau_{k+1}$. Denote a generic model parameter sequence as $\widetilde{\theta}=\{\theta^{(0)},\cdots,\theta^{(\Upsilon_T)}\}$. 

Define $\theta^0=(\alpha^0,\beta^0)^{\top}=\left(2,-1\right)^{\top}$. For $k=0,\ldots, \Upsilon_{T}$, the parameter $\theta^{(k)}$ can only take either of the two values in $\{\theta^0,\theta^0+(-1)^k\Delta_k\}=\{\theta^{0,(k)},\theta^{1,(k)}\}$, where $\Delta_k=C_{\Delta}C_{\Upsilon_T}^{1/4}N_k^{-1/4}(-1,1)^{\top}$ for some constant $C_{\Delta}$. We remark that by construction, it is easy to verify that \Cref{ass_spacing} is satisfied for a sufficiently large $C_{\Delta}$. In the above setting, there are a total of $2^{\Upsilon_{T}+1}$ possible model parameter sequences $\widetilde\theta=\{\theta^{(0)}, \theta^{(1)},\cdots,\theta^{(\Upsilon_T)}\}$. However, to examine the role of $\Upsilon_T$, we only consider the sequences where there are \textit{exactly} $\Upsilon_T$ change-points. Denote this set as
$$\Gamma(\Upsilon_T)=\left\{\widetilde\theta=\{\theta^{(k)}\}_{k=0}^{\Upsilon_T}: \theta^{(k)}\neq \theta^{(k+1)} \text{ for all } k\right\}.$$

For $\Upsilon_T=n$, collect all $\widetilde\theta$ in $\Gamma(n)$ that starts with $\theta^{(0)}=\theta^{0,(0)}$ as into a set $\Pi_0(n)$, and collect all $\widetilde\theta$ in $\Gamma(n)$ that starts with $\theta^{(0)}=\theta^{0}+\Delta_0=\theta^{1,(0)}$ into a set $\Pi_1(n)$. Thus, we have $\Gamma(\Upsilon_T)=\Pi_0(\Upsilon_T)\cup\Pi_1(\Upsilon_T)$. By elementary induction, we have  $|\Pi_0(n)|=|\Pi_1(n-1)|$, and $|\Pi_1(n)|=|\Pi_0(n-1)|+|\Pi_1(n-1)|$. This implies that $|\Pi_1(n)|=|\Pi_1(n-1)|+|\Pi_1(n-2)|$, which forms a Fibonacci sequence with $|\Pi_1(0)|=1$ and $|\Pi_1(1)|=2$. In the following, for notational simplicity, we drop $\Upsilon_T$ in $\Gamma(\Upsilon_T)$ when no confusion arises.

% Note in above settings,  there are total of $2^{\Upsilon_{T}+1}$ possible combinations for model parameters.  %Denote the set of all the enumerations of model parameters as  $\Gamma$. we let  be equal to with the coordinate $k$ empty and  by filling the empty coordinate by  $\theta^{(k)}$.  For $\widetilde{\theta}\in  \Gamma$, i.e., the probability measures with exact $\Upsilon_T$ change-points, 

Given $\widetilde{\theta}\in \Gamma$, denote $\widetilde{\theta}^{(-k)}$ as the corresponding model parameter sequence with the $k$th parameter set as $\varnothing$, while keeping the rest parameters in $\widetilde{\theta}$ unchanged. Furthermore, denote $(\widetilde{\theta}^{(-k)},\theta^{(k)})$ as the model parameter sequence by replacing the $k$th empty parameter of $\widetilde{\theta}^{(-k)}$ with $\theta^{(k)}.$ Denote $\Gamma_{-k}$ as the collection of all $\widetilde{\theta}^{(-k)}$ such that both $(\widetilde{\theta}^{(-k)},\theta^{0,(k)})\in \Gamma$ and $(\widetilde{\theta}^{(-k)},\theta^{1,(k)})\in \Gamma$. An important observation is that for $\Upsilon_T\geq 1$, we have $|\Gamma_{-k}|=|\Pi_1(\Upsilon_T-k-1)||\Pi_1(k-1)|$ with the convention $\Pi_1(-1)=1$. Together with the fact that $|\Gamma|=|\Pi_0(\Upsilon_T)|+|\Pi_1(\Upsilon_T)|=|\Pi_1(\Upsilon_T+1)|$, we show later that although $|\Gamma_{-k}|\leq |\Gamma|$, they are in the same order.

%Define   and $\theta^{(k)}=\theta^{0,(k)}$, similarly define  $\widetilde{\theta}^{(-k)}\in \Gamma_{-k}$ and $\theta^{(k)}=\theta^{1,(k)}$.  
% \begin{flalign*}
%     R_T(\mu)= &\sum_{t=1}^T\mathbf{E}_{\mu}^{\pi}\left\{[r(p_t^*,\theta_t,x_t)-r(p_t^{\pi},\theta_t,x_t)] |x_t\right\}
%     \\=& \sum_{k=0}^{\Upsilon_T} \sum_{t=\tau_k+1}^{\tau_{k+1}} E_{\mu}^{\pi}\left\{[r(\varphi(\theta_t,x_t),\theta_t,x_t)-r(\varphi(\hat{\theta_t},x_t),\theta_t,x_t)]|x_t \right\}
%     \\:=& \sum_{k=0}^{\Upsilon_T}\sum_{t=\tau_k+1}^{\tau_{k+1}}\mathrm{Regret}_t({\mu},x_t)
% \end{flalign*}
% where the second last equality holds by the fact that $\sum_{t=\tau_k+1}^{\tau_{k+1}}[r(\varphi(\theta_t,x_t),\theta_t,x_t)-r(\varphi(\hat{\theta_t},x_t),\theta_t,x_t)]$ is $\mathcal{F}_{\tau_{k+1}}$ measurable, and that the law of $\{y_t\}_{t=0}^{\tau_{k+1}}$  remains  the same under  ${\mu}_0(\cdot)$ and $\mu_{k+1}(\cdot)$.

For a given $\widetilde{\theta}^{(-k)}\in \Gamma_{-k}$, denote $\mu_{\widetilde{\theta}^{(-k)}}(0)$ as the probability measure given by $(\widetilde{\theta}^{(-k)},\theta^{0,(k)})$ and denote $\mu_{\widetilde{\theta}^{(-k)}}(1)$ as the probability measure given by $(\widetilde{\theta}^{(-k)},\theta^{1,(k)})$. Note that the two probability measures are the same except for the parameter $\theta^{(k)}$ between $\tau_k$ and $\tau_{k+1}$

For a non-anticipating policy $\pi$, we compute the KL divergence between the two probability measures $\mu_{\widetilde{\theta}^{(-k)}}(0)$ and $\mu_{\widetilde{\theta}^{(-k)}}(1)$. By definition, we have 
\begin{flalign*}
\mathbb{KL}(\mu_{\widetilde{\theta}^{(-k)}}(0),\mu_{\widetilde{\theta}^{(-k)}}(1))=& \mathbb{E}^\pi_{\mu_{\widetilde{\theta}^{(-k)}}(0)}\left[ \log \frac{\mathbb{P}^{\pi}_{\mu_{\widetilde{\theta}^{(-k)}}(0)}(\{y_t\}_{t=1}^T)}{\mathbb{P}^{\pi}_{\mu_{\widetilde{\theta}^{(-k)}}(1)}(\{y_t\}_{t=1}^T)}\right]\\=&\mathbb{E}^\pi_{\mu_{\widetilde{\theta}^{(-k)}}(0)}\left[ \log \prod_{t=1}^T \frac{\mathbb{P}^{\pi}_{\mu_{\widetilde{\theta}^{(-k)}}(0)}(y_t|\mathcal{F}_{t-1})}{\mathbb{P}^{\pi}_{\mu_{\widetilde{\theta}^{(-k)}}(1)}(y_t|\mathcal{F}_{t-1})}\right]
\\=&  \mathbb{E}^\pi_{\mu_{\widetilde{\theta}^{(-k)}}(0)}\left[ \sum_{i=0}^{\Upsilon_T}\sum_{t=\tau_i+1}^{\tau_{i+1}} \log \frac{\mathbb{P}^{\pi}_{\mu_{\widetilde{\theta}^{(-k)}}(0)}(y_t|\mathcal{F}_{t-1})}{\mathbb{P}^{\pi}_{\mu_{\widetilde{\theta}^{(-k)}}(1)}(y_t|\mathcal{F}_{t-1})}\right]\\
=&  \sum_{t=\tau_{k}+1}^{\tau_{k+1}}\mathbb{E}^\pi_{\mu_{\widetilde{\theta}^{(-k)}}(0)}\left[ \log \frac{\mathbb{P}^{\pi}_{\mu_{\widetilde{\theta}^{(-k)}}(0)}(y_t|\mathcal{F}_{t-1})}{\mathbb{P}^{\pi}_{\mu_{\widetilde{\theta}^{(-k)}}(1)}(y_t|\mathcal{F}_{t-1})}\right]
\end{flalign*}
where the last equality holds by the fact that the law of $\{y_t\}_{t=1}^{\tau_k}$ and $\{y_t\}_{t=\tau_{k+1}+1}^{T}$ are  the same  under $\mu_{\widetilde{\theta}^{(-k)}}(0)$ and $\mu_{\widetilde{\theta}^{(-k)}}(1)$ for the given $\pi$.

Furthermore, when $\epsilon_t$ is i.i.d. normally distributed, and recall 
$z_t=(1,p_t)^{\top}$, we further have 
\begin{flalign}
  \notag  \mathbb{KL}(\mu_{\widetilde{\theta}^{(-k)}}(0),\mu_{\widetilde{\theta}^{(-k)}}(1))=& -\frac{1}{2}\sum_{t=\tau_{k}+1}^{\tau_{k+1}}\mathbb{E}^\pi_{\mu_{\widetilde{\theta}^{(-k)}}(0)}[\epsilon_t^2-(\epsilon_t+(-1)^k\Delta_k^{\top}z_t)^2]
    \\  \notag\stackrel{(a)}{=}& \frac{1}{2}\sum_{t=\tau_{k}+1}^{\tau_{k+1}}\mathbb{E}^\pi_{\mu_{\widetilde{\theta}^{(-k)}}(0)}[\Delta_k^{\top}z_t]^2
    \\  \notag\stackrel{(b)}{=}& \frac{C_{\Delta}^2C_{\Upsilon_T}^{1/2}}{2N_k^{1/2}} \sum_{t=\tau_{k}+1}^{\tau_{k+1}}\mathbb{E}^\pi_{\mu_{\widetilde{\theta}^{(-k)}}(0)}[p_t-1]^2
 %\\  \notag\stackrel{(c)}{=}& \frac{C_{\Delta}^2C_{\Upsilon_T}^{1/2}}{2N_k^{1/2}} \sum_{t=\tau_{k}+1}^{\tau_{k+1}}\mathbb{E}^\pi_{\mu_{\widetilde{\theta}^{(-k)}}(0)}[p_t-\varphi(\theta^0)]^2,
  \\\label{KL1}\stackrel{(c)}{=}& \frac{C_{\Delta}^2C_{\Upsilon_T}^{1/2}}{2N_k^{1/2}} \sum_{t=\tau_{k}+1}^{\tau_{k+1}}\mathbb{E}^\pi_{\mu_{\widetilde{\theta}^{(-k)}}(0)} \mathrm{Regret}_t(\mu_{\widetilde{\theta}^{(-k)}}(0)),
\end{flalign}
where $(a)$ follows using the independence between $\epsilon_t$, $(b)$ follows using $\Delta_k=(-1)^kC_{\Delta}C_{\Upsilon_T}^{1/4}N_k^{-1/4}(-1,1)^{\top}$, $(c)$ follows as under $\mu_{\widetilde{\theta}^{(-k)}}(0)$ we have $\theta^{0,(k)}=\theta^0=(2,-1)$, and thus the optimal price $p_t^*=1$ and $(p_t-1)^2$ is the regret of the policy $\pi$.

%and $(d)$ follows as 
%$(p_t-1)^2=p_t^* (1,p_t^*)^{\top}\theta^{0,(k)} -p_t (1,p_t)^{\top}\theta^{0,(k)}$ 

Now, fix a constant $\eta>0$.

If $\mathbb{KL}(\mu_{\widetilde{\theta}^{(-k)}}(0),\mu_{\widetilde{\theta}^{(-k)}}(1))>\eta$, by \eqref{KL1}, we have that  \begin{flalign}\label{reg_bound1n}
    \mathbb{E}^\pi_{\mu_{\widetilde{\theta}^{(-k)}}(0)}\sum_{t=\tau_{k}+1}^{\tau_{k+1}} \mathrm{Regret}_t(\mu_{\widetilde{\theta}^{(-k)}}(0))\geq 2 \eta \frac{N_k^{1/2}}{C_{\Delta}^2{C^{1/2}_{\Upsilon_T}}}=\frac{2 \eta}{C_{\Delta}^2{C_{\Upsilon_T}}}\sqrt{T/(k+1)}.
\end{flalign}

If $\mathbb{KL}(\mu_{\widetilde{\theta}^{(-k)}}(0),\mu_{\widetilde{\theta}^{(-k)}}(1))\leq \eta$, define $I_{i,k}=[\varphi(\theta^{i,(k)})- {C_{\Delta}C_{\Upsilon_T}^{1/4}}{N_k^{-1/4}}/4,\varphi(\theta^{i,(k)})+{C_{\Delta}C_{\Upsilon_T}^{1/4}}{N_k^{-1/4}}/4]$, $i=0,1$. For each $t \in  [\tau_{k}+1,\tau_{k+1}]$, consider the following hypothesis:
$$
H_0: p_t \in I_{0,k},\quad H_1: p_t\in I_{1,k}.
$$
We have that
\begin{flalign}
    \notag &\mathbb{E}^\pi_{\mu_{\widetilde{\theta}^{(-k)}}(0)}\sum_{t=\tau_{k}+1}^{\tau_{k+1}} \mathrm{Regret}_t(\mu_{\widetilde{\theta}^{(-k)}}(0))+ \mathbb{E}^\pi_{\mu_{\widetilde{\theta}^{(-k)}}(1)}\sum_{t=\tau_{k}+1}^{\tau_{k+1}} \mathrm{Regret}_t(\mu_{\widetilde{\theta}^{(-k)}}(1))\\
     \notag\stackrel{(a)}{=}& \mathbb{E}^\pi_{\mu_{\widetilde{\theta}^{(-k)}}(0)}\sum_{t=\tau_{k}+1}^{\tau_{k+1}} -\beta^{0,(k)}[p_t+\frac{\alpha^{0,(k)}}{2\beta^{0,(k)}}]^2+ \mathbb{E}^\pi_{\mu_{\widetilde{\theta}^{(-k)}}(1)}\sum_{t=\tau_{k}+1}^{\tau_{k+1}} -\beta^{1,(k)}[p_t+\frac{\alpha^{1,(k)}}{2\beta^{1,(k)}}]^2
    \\ \notag \stackrel{(b)}{\geq}&  \frac{1}{2}\sum_{t=\tau_{k}+1}^{\tau_{k+1}} \left\{\mathbb{E}^\pi_{\mu_{\widetilde{\theta}^{(-k)}}(0)}[p_t-\varphi(\theta^{0,(k)})]^2+\mathbb{E}^\pi_{\mu_{\widetilde{\theta}^{(-k)}}(1)}[p_t-\varphi(\theta^{1,(k)})]^2\right\}
    \\ \notag\stackrel{(c)}{\geq}& \frac{1}{2}\sum_{t=\tau_{k}+1}^{\tau_{k+1}} \left\{\mathbb{E}^\pi_{\mu_{\widetilde{\theta}^{(-k)}}(0)}[p_t-\varphi(\theta^{0,(k)})]^2\mathbbm{1}(p_t\not \in I_{0,k})+\mathbb{E}^\pi_{\mu_{\widetilde{\theta}^{(-k)}}(1)}[p_t-\varphi(\theta^{1,(k)})]^2\mathbbm{1}(p_t\not \in I_{1,k})\right\}
        \\ \notag\stackrel{(d)}{\geq}& \frac{C_{\Delta}^2C_{\Upsilon_T}^{1/2}}{32N_k^{1/2}} \sum_{t=\tau_{k}+1}^{\tau_{k+1}} \mathbb{P}_{\mu_{\widetilde{\theta}^{(-k)}}(0)}(p_t\not \in I_{0,k})+\mathbb{P}_{\mu_{\widetilde{\theta}^{(-k)}}(1)}(p_t\not\in I_{1,k})
    \\ \notag\stackrel{(e)}{\geq}& \frac{C_{\Delta}^2C_{\Upsilon_T}^{1/2}}{32N_k^{1/2}}   \sum_{t=\tau_{k}+1}^{\tau_{k+1}} \mathbb{P}_{\mu_{\widetilde{\theta}^{(-k)}}(0)}(p_t\in I_{1,k})+\mathbb{P}_{\mu_{\widetilde{\theta}^{(-k)}}(1)}(p_t\in I_{0,k})
    \\\label{reg_bound2n}\stackrel{(f)}{\geq} & \frac{C_{\Delta}^2C_{\Upsilon_T}^{1/2}}{128N_k^{1/2}} \frac{T}{(k+1)C_{\Upsilon_T}} \exp(-\eta)=\frac{C_{\Delta}^2}{128}\sqrt{T/(k+1)}\exp(-\eta),
\end{flalign}
where $(a)$ holds using the definition of regret, $(b)$ holds by noting $\varphi(\theta^{i,(k)})=-\alpha^{i,(k)}/(2\beta^{i,(k)})$ and $-\beta^{i,(k)}\geq 1-C_{\Delta}C_{\Upsilon_T}^{1/4} N_k^{-1/4}> 1/2$ for large $T$, $(c)$ holds by  writing $1=\mathbbm{1}(p_t\in I_{i,k})+\mathbbm{1}(p_t\not\in I_{i,k})$, $(d)$ holds by definition of $I_{i,k}$, $(e)$ holds by noting that $I_{0,k}$ and $I_{1,k}$ are disjoint, and $p_t\not\in I_{i,k}$ is implied by $p_t\in I_{1-i,k}$, for $i=0,1$, and $(f)$ holds by  Theorem 2.2 in \cite{tsybakov2009} for the minimax probability error of testing two hypothesis using observations $\mathcal{F}_{t-1}$ and the monotonicity of the KL divergence. 

Therefore, in view of \eqref{reg_bound1n} and \eqref{reg_bound2n}, we have that 
$$
\mathbb{E}^\pi_{\mu_{\widetilde{\theta}^{(-k)}}(0)}\sum_{t=\tau_{k}+1}^{\tau_{k+1}} \mathrm{Regret}_t(\mu_{\widetilde{\theta}^{(-k)}}(0))+ \mathbb{E}^\pi_{\mu_{\widetilde{\theta}^{(-k)}}(1)}\sum_{t=\tau_{k}+1}^{\tau_{k+1}} \mathrm{Regret}_t(\mu_{\widetilde{\theta}^{(-k)}}(1)) \geq C_{LB}  [T/(k+1)]^{1/2},
$$
and  $C_{LB}=\min\{C_{\Delta}^2/128\exp(-\eta),2\eta/(C_{\Delta}^2C_{\Upsilon_T})\}$.

% Therefore, consider the average regret for $\tilde{\theta}\in \Gamma$, we have 
% \begin{flalign*}
%     &2^{-(\Upsilon_T+1)}\sum_{\tilde{\theta}\in\Gamma} R_T(\mu_{\tilde{\theta}})
%     \\=& 2^{-(\Upsilon_T+1)}\sum_{k=0}^{\Upsilon_T}\sum_{\tilde{\theta}\in\Gamma}  \mathbb{E}^{\pi}_{\mu_{\tilde{\theta}}}\sum_{t=\tau_k+1}^{\tau_{l+1}} \mathrm{Regret}_t(\mu_{\tilde{\theta}})
%     \\=&2^{-(\Upsilon_T+1)}\sum_{k=0}^{\Upsilon_T}\sum_{\tilde{\theta}^{(-k)}\in\Gamma_{-k}}\sum_{t=\tau_k+1}^{\tau_{k+1}}\left[\mathbb{E}^{\pi}_{\mu_{\tilde{\theta}^{(-k)}}(0)}\mathrm{Regret}_t(\mu_{\tilde{\theta}^{(-k)}}(0))+\mathbb{E}^{\pi}_{\mu_{\tilde{\theta}^{(-k)}}(1)}\mathrm{Regret}_t(\mu_{\tilde{\theta}^{(-k)}}(1))\right]    \\\geq& 2^{-(\Upsilon_T+1)}\sum_{k=0}^{\Upsilon_T}\left[\sum_{\tilde{\theta}^{(-k)}\in\Gamma_{-k}}C_{LB} [T/(k+1)]^{1/2}\right]\geq 2^{-1}C_{LB}\sqrt{T(\Upsilon_T+1)}.
% \end{flalign*}

Therefore, consider the average regret for $\widetilde{\theta}\in  \Gamma=\Pi_0(\Upsilon_T)\cup \Pi_1(\Upsilon_T)$, i.e., the probability measures with exact $\Upsilon_T$ change-points,  we have 
\begin{flalign*}
    &|\Gamma|^{-1}\sum_{\widetilde{\theta}\in\Gamma} R_T^\pi(\mu_{\widetilde{\theta}})
    \\=& |\Gamma|^{-1}\sum_{k=0}^{\Upsilon_T}\sum_{\widetilde{\theta}\in\Gamma}  \mathbb{E}^{\pi}_{\mu_{\widetilde{\theta}}}\sum_{t=\tau_k+1}^{\tau_{l+1}} \mathrm{Regret}_t(\mu_{\widetilde{\theta}})
\\\geq &|\Gamma|^{-1}\sum_{k=0}^{\Upsilon_T}\sum_{\widetilde{\theta}^{(-k)}\in\Gamma_{-k}}\sum_{t=\tau_k+1}^{\tau_{k+1}}\left[\mathbb{E}^{\pi}_{\mu_{\widetilde{\theta}^{(-k)}}(0)}\mathrm{Regret}_t(\mu_{\widetilde{\theta}^{(-k)}}(0))+\mathbb{E}^{\pi}_{\mu_{\widetilde{\theta}^{(-k)}}(1)}\mathrm{Regret}_t(\mu_{\widetilde{\theta}^{(-k)}}(1))\right]    \\\geq& |\Gamma|^{-1}\sum_{k=0}^{\Upsilon_T}\sum_{\widetilde{\theta}^{(-k)}\in\Gamma_{-k}}C_{LB}  [T/(k+1)]^{1/2}\geq  |\Gamma|^{-1}\inf_{0\leq k\leq \Upsilon_T}|\Gamma_{-k}| C_{LB}\sum_{k=0}^{\Upsilon_T}\sqrt{T/(k+1)}.
\end{flalign*}
Note that  for $\Upsilon_T\geq 1$, we have $|\Gamma_{-k}|=
    |\Pi_1(\Upsilon_T-k-1)||\Pi_1(k-1)|$ with the convention $\Pi_1(-1)=1$, and that 
 $|\Gamma|=|\Pi_0(\Upsilon_T)|+|\Pi_1(\Upsilon_T)|=|\Pi_1(\Upsilon_T+1)|$. 
 By the property of a Fibonacci sequence, we have  $$|\Pi_1(n)|=\frac{\phi^{n+2}-(1-\phi)^{n+2}}{\phi-(1-\phi)}$$ with $\phi=(1+\sqrt{5})/2$, and thus
\begin{flalign*}
   |\Gamma|^{-1}\inf_{0\leq k\leq \Upsilon_T}|\Gamma_{-k}| =&\inf_{0\leq k\leq \Upsilon_T} \frac{[\phi^{\Upsilon_T-k+1}-(1-\phi)^{\Upsilon_T-k+1}][\phi^{k+1}-(1-\phi)^{k+1}]}{[\phi^{\Upsilon_T+3}-(1-\phi)^{\Upsilon_T+3}][\phi-(1-\phi)]}\geq 1/5,
\end{flalign*}
for all $\Upsilon_T\geq 1$. This implies that $|\Gamma|^{-1}\sum_{\widetilde{\theta}\in\Gamma} R_T^\pi(\mu_{\widetilde{\theta}})\gtrsim C_{LB}\sqrt{T\Upsilon_T}.$

Finally, note that $C_{\Upsilon_T}\asymp \log(\Upsilon_T)$, we obtain $C_{LB}\gtrsim 1/\log(\Upsilon_T)$, and hence 
$$
\inf_{\pi}\sup_{\widetilde{\theta}\in\Gamma}R_T^\pi(\mu_{\tilde{\theta}})\gtrsim \sqrt{T\Upsilon_T}/\log(\Upsilon_T).
$$ 
\qed

% \zifeng{Roadmap of Proof: \Cref{lem:DB} is deviation bound. \Cref{lem:RE} is RSC. \Cref{lem:est} and \Cref{lem_mixture} are built upon \Cref{lem:DB} and \Cref{lem:RE}. Change-points \Cref{lem-no-change} and \Cref{lem-one-change_alter} are built upon \Cref{lem:est}. Regret is built upon \Cref{lem:est} and \Cref{lem-no-change} and \Cref{lem-one-change_alter}.}

\section{Additional Numerical Results}\label{sec:add_simu}

\subsection{Performance of SW-DP and DF-DP w.r.t.\ tuning parameters}\label{subsec:swdf}
\begin{figure}[ht]
\vspace{-5mm}
\hspace*{-6mm} 
	\begin{subfigure}{0.32\textwidth}
		\includegraphics[angle=270, width=1.2\textwidth]{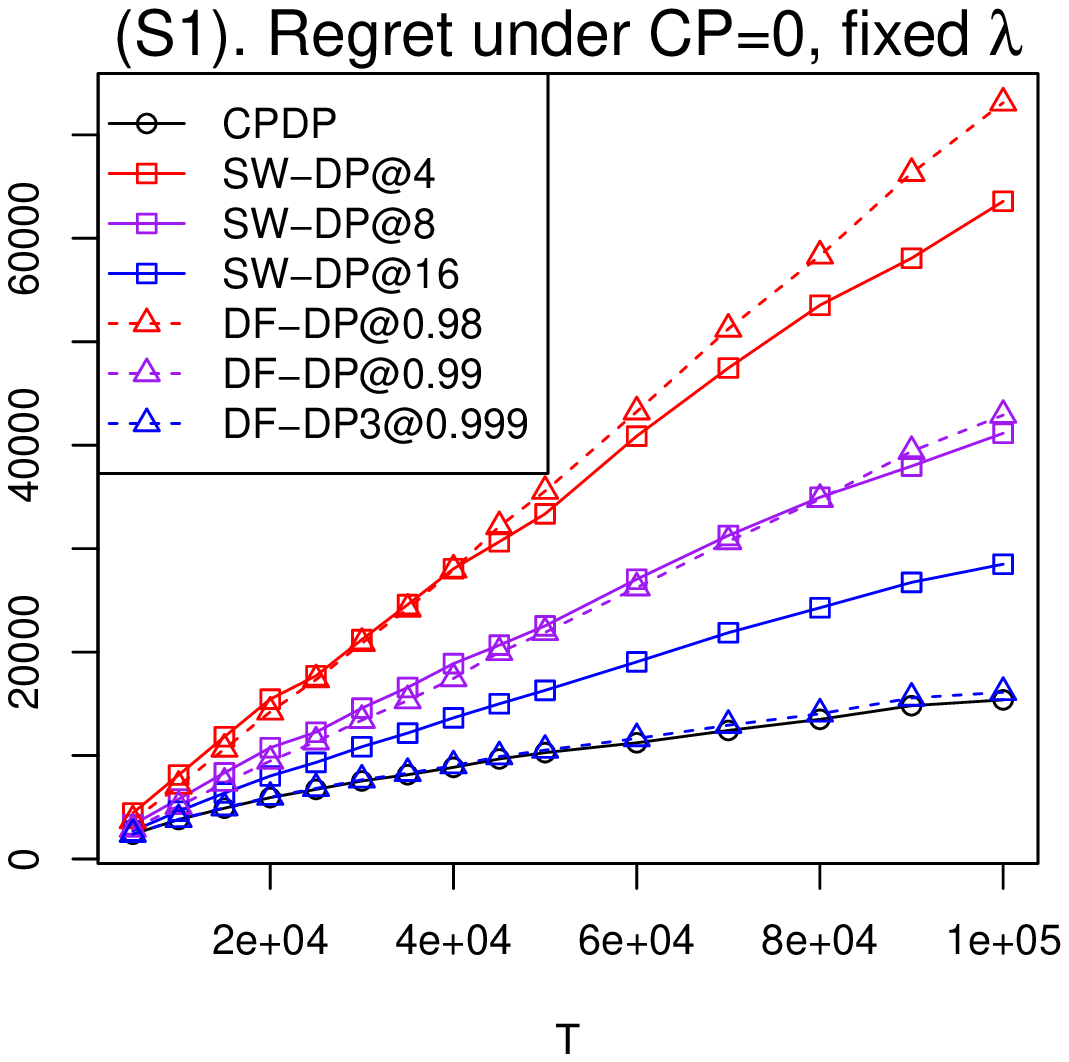}
		\vspace{-0.5cm}
	\end{subfigure}
	~
	\begin{subfigure}{0.32\textwidth}
		\includegraphics[angle=270, width=1.2\textwidth]{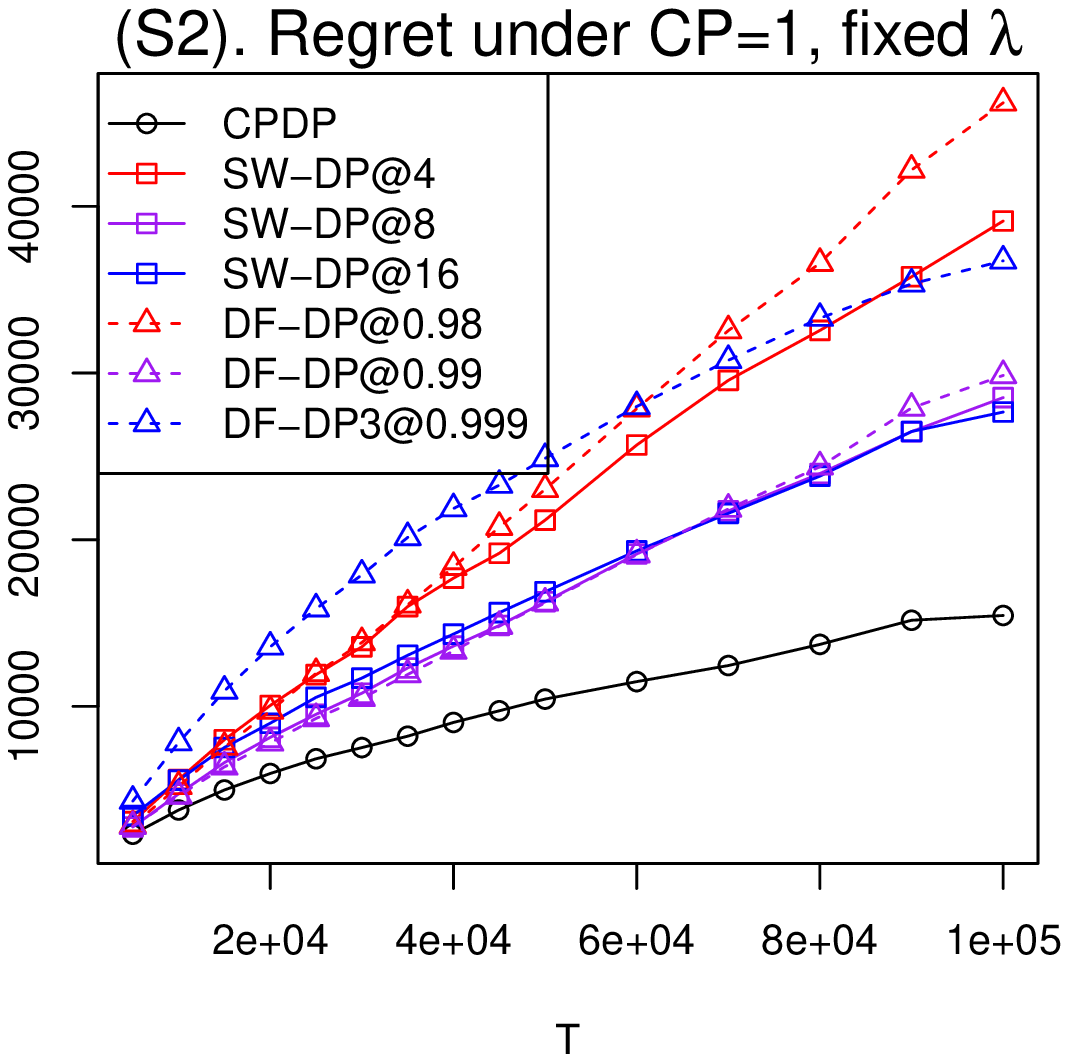}
		\vspace{-0.5cm}
	\end{subfigure}
	~
	\begin{subfigure}{0.32\textwidth}
		\includegraphics[angle=270, width=1.2\textwidth]{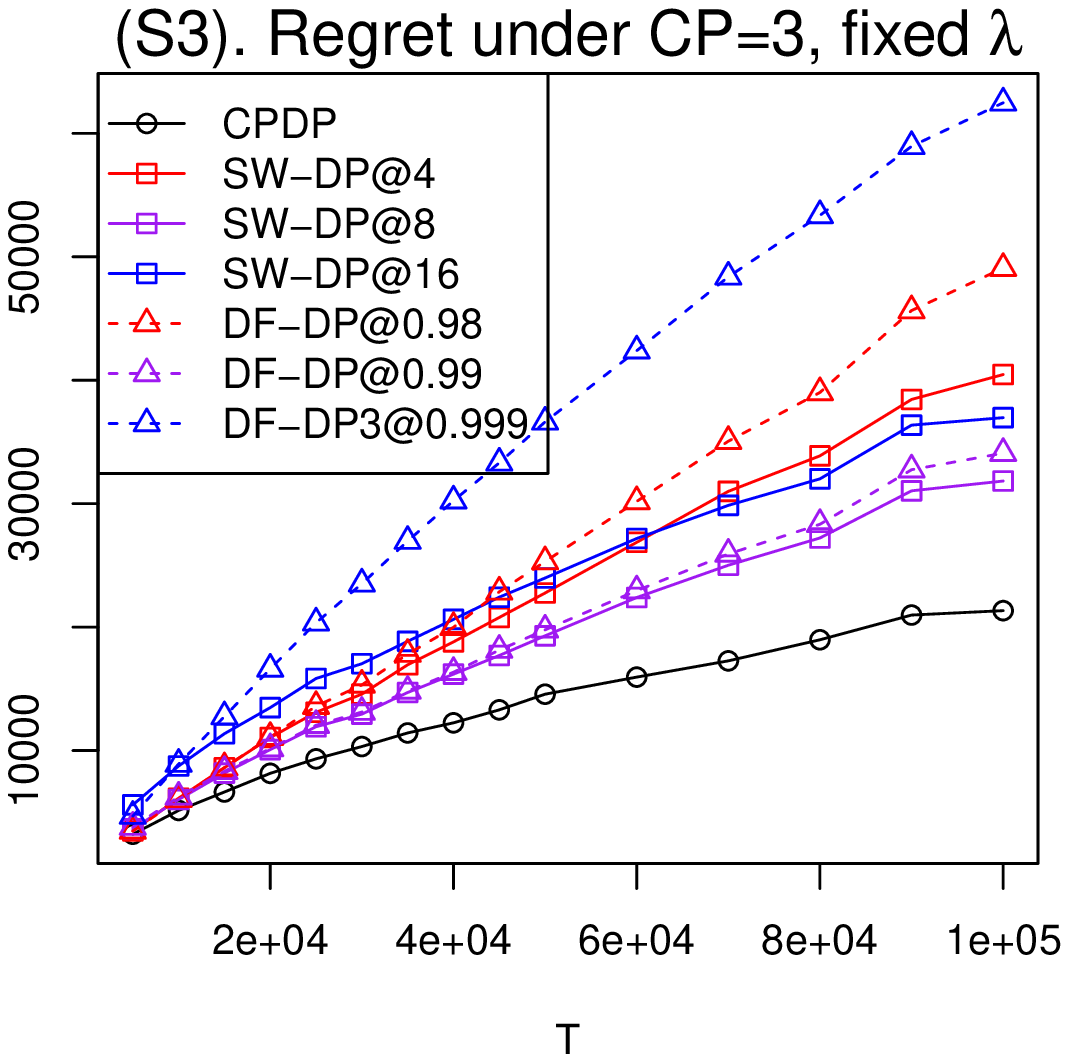}
		\vspace{-0.5cm}
	\end{subfigure}
 	\caption{ Mean regret vs.\ $T$ under (S1) [left], (S2) [middle], (S3) [right] with $\lambda_{\mathrm{fix}}=0.2\sqrt{\log(Td)}$.}
	\label{fig:swdf}
\end{figure}

\Cref{fig:swdf} gives the mean regret of CPDP, SW-DP and DF-DP under scenarios (S1), (S2) and (S3), where for SW-DP, we set $\eta_{\mathrm{sw}}=4,8,16$ and for DF-DP, we set $\gamma_{\mathrm{df}}=0.98,0.99,0.999$. As can be seen, the performance of SW-DP and DF-DP vary across tuning parameters $\eta_{\mathrm{sw}}$ and $\gamma_{\mathrm{df}}.$ For (S1) where there is no change-point, as expected, larger $\eta_{\mathrm{sw}}$ and $\gamma_{\mathrm{df}}$ lead to better performance. On the other hand, with more change-points as in (S2) and (S3), $\eta_{\mathrm{sw}}=8$ and $\gamma_{\mathrm{df}}={0.99}$ provide the best overall performance.

\subsection{Performance of CPDP w.r.t.\ change size}\label{subsec:halfchange}
In this section, we provide additional simulation result to investigate the impact of change size on the performance of each algorithm. In particular, we modify the simulation scenarios (S1), (S2) and (S3) in \Cref{subsec:valueCPD} of the main text by replacing $\theta^{(1)}=(0,1,1,2,2,\mathbf{0}_{44}^\top,-0.25)$ with $\theta^{(3)}=\theta^{(1)}/2$ and replacing $\theta^{(2)}=(0,1,1,1,1,\mathbf{0}_{44}^\top,-0.5)$ with $\theta^{(4)}=\theta^{(2)}/2$. Denote the new scenarios as (S4), (S5) and (S6). Note that the change size measured in $\ell_2$-norm reduces by 50\% for the new simulation settings. Indeed, for $p_t\sim \text{uniform}[\widetilde{p}_l,\widetilde{p}_u]=[1,15]$, the KL divergence between $\theta^{(3)}$ and $\theta^{(4)}$ is $\text{KL}(\theta^{(3)},\theta^{(4)})=0.271$, much smaller than $\text{KL}(\theta^{(1)},\theta^{(2)})=0.814.$

\begin{figure}[ht]
\vspace{-5mm}
\hspace*{-6mm} 
	\begin{subfigure}{0.32\textwidth}
		\includegraphics[angle=270, width=1.2\textwidth]{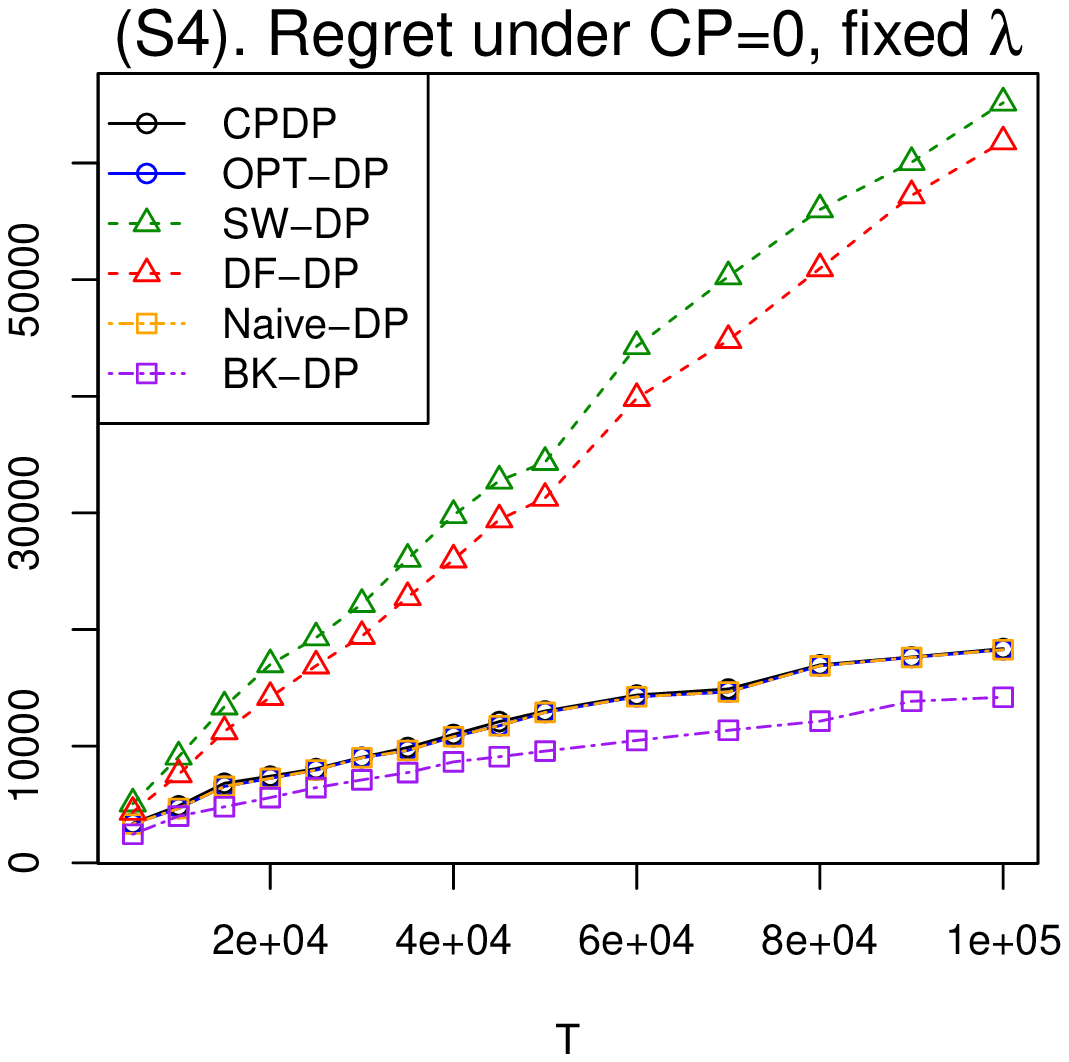}
		\vspace{-0.5cm}
	\end{subfigure}
	~
	\begin{subfigure}{0.32\textwidth}
		\includegraphics[angle=270, width=1.2\textwidth]{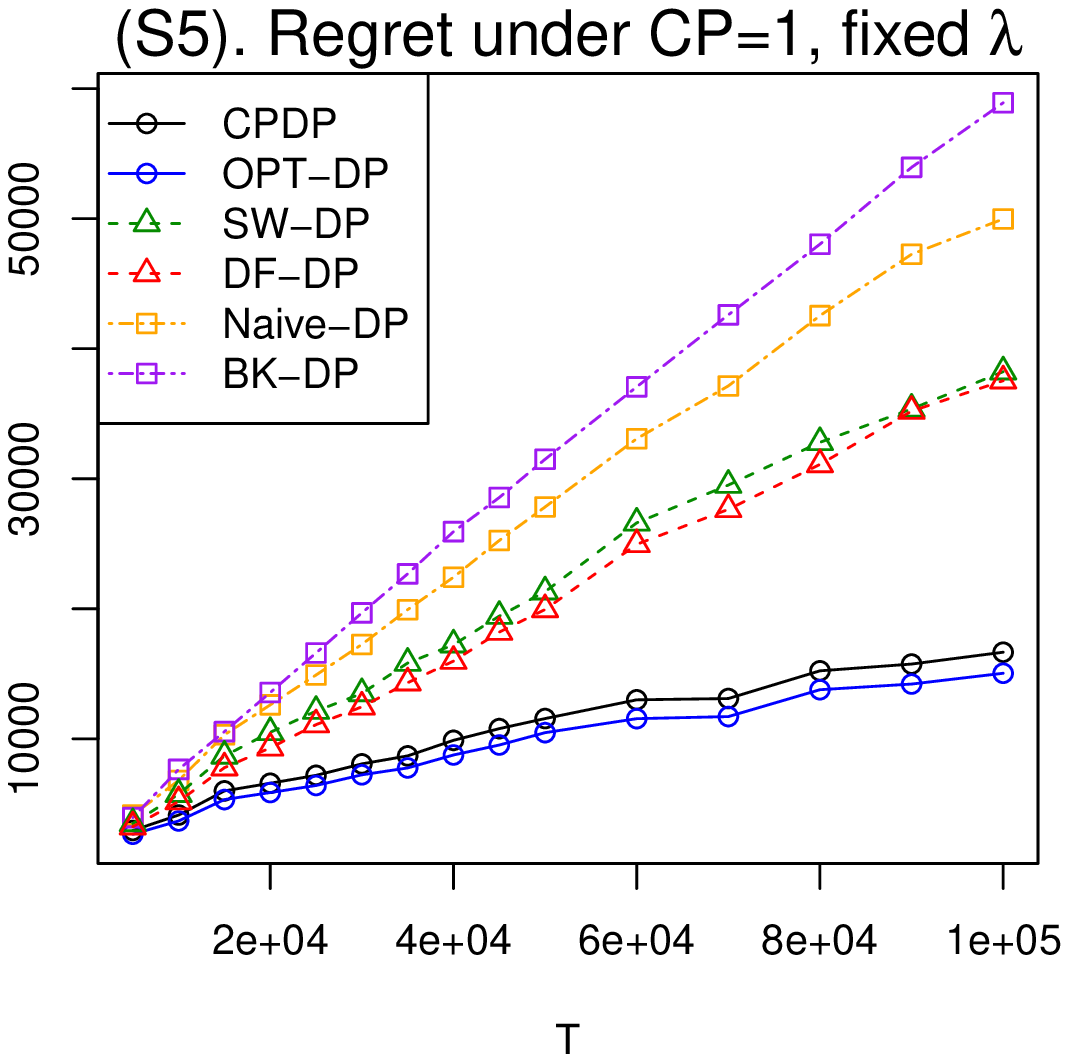}
		\vspace{-0.5cm}
	\end{subfigure}
	~
	\begin{subfigure}{0.32\textwidth}
		\includegraphics[angle=270, width=1.2\textwidth]{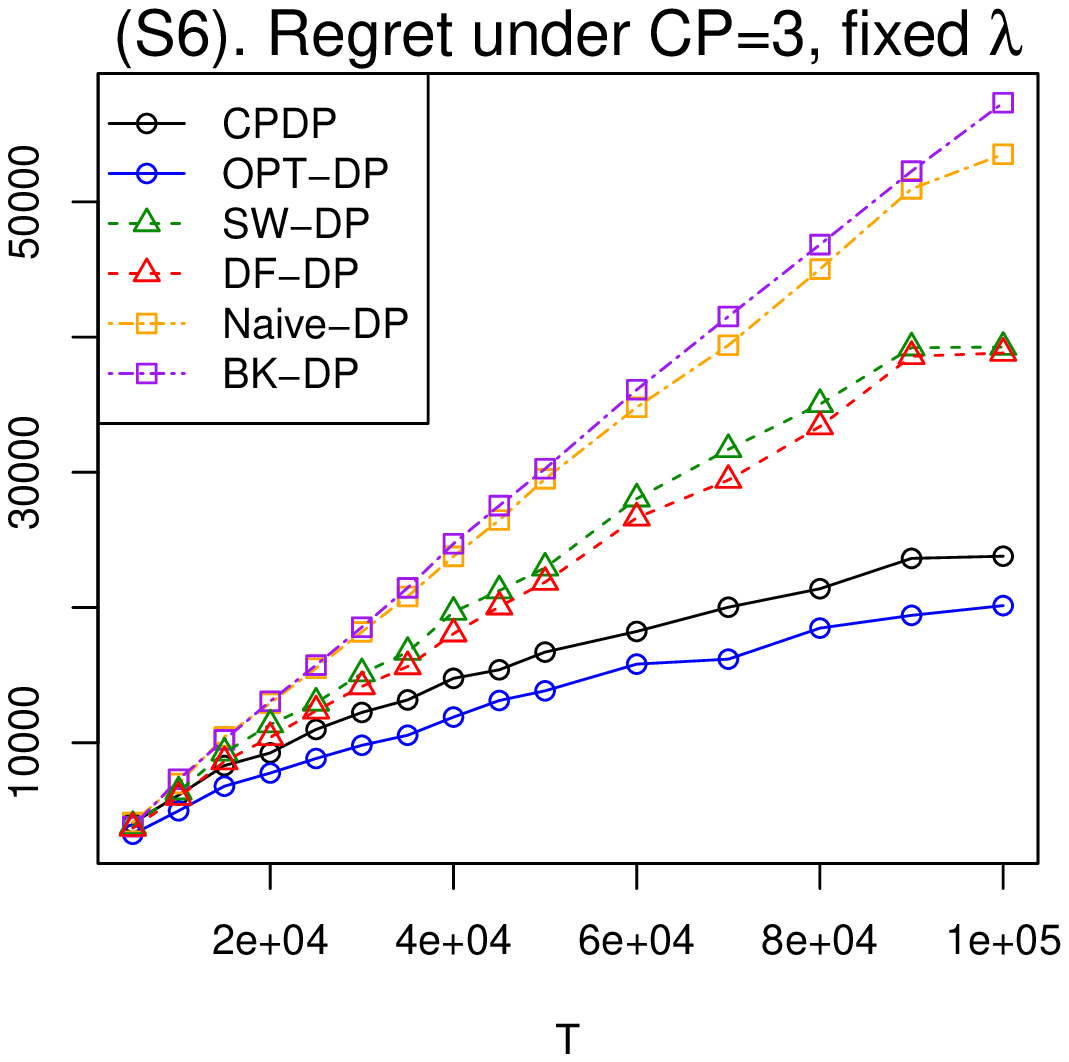}
		\vspace{-0.5cm}
	\end{subfigure}
 	\caption{ Mean regret vs.\ $T$ under (S4) [left], (S5) [middle], (S6) [right] with $\lambda_{\mathrm{fix}}=0.2\sqrt{\log(Td)}$.}
	\label{fig:halfchange}
\end{figure}

\Cref{fig:halfchange} reports the mean regret by each algorithm under simulation scenarios (S4), (S5) and (S6). Compared to \Cref{fig:fixedlambda}, the notable difference is that the performance gaps between CPDP and Naive-DP are smaller. For example, at $T=50000$, for (S3), the ratio of regret by Naive-DP w.r.t.\ CPDP is 2.85, while for (S6), the corresponding ratio is 1.77. In addition, due to larger detection delay, the gaps between CPDP and OPT-DP are wider. Interestingly, the performance gap between CPDP and SW-DP (with $\eta_{\mathrm{sw}}=8$) and DF-DP (with $\gamma_{\mathrm{df}}=0.99$) stays relatively stable: where at $T=50000$, the ratios of regret by SW-DP and DF-DP w.r.t.\ CPDP are 1.32, 1.36 for (S3), and are 1.37 and 1.31 for (S6).

\subsection{Additional results for the auto loan dataset}\label{subsec:addrealdata}
\textbf{Offline change-point detection algorithm}: We run a standard model selection based offline change-point detection algorithm for logistic regression~\citep[e.g.,][]{davis2006structural,bai2003computation} on $\{(y_t,z_t^*,p_t)\}_{t=1}^{50000}$, which returns three change-points at $t=12916, 24986, 37054.$ In the following, we describe the algorithm, which estimates change-points based on optimizing a penalized likelihood function.

Given $\theta=(\alpha,\beta),$ the logistic regression specifies that
$$\mathbb E (y_t|z_t^*,p_t)=\psi'(\alpha^\top z_t^*+\beta p_t),$$
where $\psi'(\cdot)$ is the logistic function.

Define $\mathcal T=\{\boldsymbol\tau: 0=\tau_0<\tau_1<\tau_2<\cdots <\tau_K<\tau_{K+1}=T, K\in \mathbb N\}$ as the candidate set of all possible vectors of change-points. For a given change-point candidate $\boldsymbol{\tau}=(\tau_1,\cdots,\tau_K)$, the penalized likelihood function is defined as
\begin{align*}
    F(\boldsymbol{\tau})=-2\sum_{k=0}^K \left[\max_{\theta} \sum_{t=\tau_k+1}^{\tau_{k+1}}\log(f(y_t|z_t^*,p_t,\theta)) \right] + K(d+1)\log T 
\end{align*}
where $d$ is dimension of $(z_t^*,p_t)$ and $f(y_t|z_t^*,p_t,\theta)$ is the likelihood function of $y_t$ given $z_t^*,p_t$ and $\theta.$ In other words, $F(\boldsymbol{\tau})$ is the BIC-based model selection criterion, where the first component is the summation of negative log-likelihood of each stationary segment and the second component is the BIC penalty for model complexity.

The change-point estimator is defined as
\begin{align*}
    \widehat{\boldsymbol{\tau}} =(\widehat\tau_1,\cdots,\widehat\tau_{\widehat K})= \argmin_{\boldsymbol\tau \in \mathcal{T}} F(\boldsymbol{\tau}),
\end{align*}
which can be solved in $O(T^2)$ operations via dynamic programming. We refer to \cite{bai2003computation} for more details.

\begin{table}[ht]
\caption{Summary of the auto loan dataset used in \Cref{subsec:autoloan}}
\begin{tabular}{lll}
\hline\hline
Variable            & Type        & Description                                                                                                               \\\hline
apply               & Binary      & Indicator for eventual contract (dependent variable)                                                                      \\
Price               & Continuous  & Price of the loan                                                                                                         \\
Primary\_FICO       & Continuous  & FICO score                                                                                                                \\
Competition\_rate   & Continuous  & Competitor’s rate                                                                                                         \\
Amount\_Approved    & Continuous  & Loan amount approved                                                                                                      \\
onemonth            & Continuous  & Prime rate                                                                                                                \\
Term                & Continuous  & Approved term in months                                                                                                   \\
Tier                & Categorical & Segmentation (1–7) based on FICO scores (defined by company) \\
rate                & Continuous  & Customer rate                                                                                                             \\
CarType           & Categorical & Type of car (new and used)                                                                                               \\
rate1               & Continuous  & Rate relative to the prime rate                                                                                           \\
rel\_compet\_rate   & Continuous  & Rate relative to the competitor’s rate                                                                                    \\
mp                  & Continuous  & Monthly payment                                                                                                           \\
mp\_rto\_amtfinance & Continuous  & Monthly payment over amount financed                                                                                      \\
partnerbin          & Categorical & Segmentation based on loan partners                                                                                       \\
States              & Categorical & Customer state \\\hline\hline
\end{tabular}
\end{table}

\end{document}